\documentclass[preprint,a4paper,12pt,3p]{elsarticle} 
\usepackage{graphicx}
\usepackage{graphics}
\usepackage[mathscr]{eucal}
\usepackage{amssymb}
\usepackage{amsmath}
\usepackage{tabls} 
\usepackage{multirow}
\usepackage{wasysym}
\usepackage{float}

\usepackage{listings}

\newcommand{\sla}[1]{\ooalign{\hfil/\hfil\crcr$#1$}} 
\newcommand{\Rea}{\widetilde{\text{Re}}}

\newcommand{\Nge}{\mathcal{N}}
\newcommand{\Mass}{\mathcal{M}}

\newcommand{\LogA}{K_1}
\newcommand{\LogB}{K_2}
\newcommand{\LogC}{K_3}
\newcommand{\prog}{{\tt CNNDecays} }

\begin{document}

\begin{frontmatter}

\title{Electroweak corrections to Neutralino and Chargino decays into a
$W$-boson in the (N)MSSM }

\author[wue]{Stefan Liebler}
\ead{sliebler@physik.uni-wuerzburg.de}

\author[wue]{Werner Porod}
\ead{porod@physik.uni-wuerzburg.de}

\address[wue]{
Institut f\"ur Theoretische Physik und Astrophysik, Universit\"at W\"urzburg,\\
D-97074  W\"urzburg, Germany
}

\begin{abstract}
We present the complete electroweak one-loop corrections to
the partial widths for  two-body decays of a chargino (neutralino)
into a $W$-boson and  a neutralino (chargino). We perform
the calculation for the minimal and the next-to-minimal supersymmetric
standard model using an  on-shell 
renormalization scheme. 
Particular attention is paid to the question
of gauge invariance which is achieved using the so-called pinch technique.
Furthermore we show that these corrections show a strong parameter 
dependence and usually are in the range of $1$-$10$ percent if the neutralino
involved is a higgsino or wino like state. However, in case of a bino-like
or a singlino-like neutralino the corrections can go up to $50$\% and more. 
Moreover we present the public program \prog performing these calculations.
\end{abstract}

\end{frontmatter}

\tableofcontents

\section{Introduction}
\label{sec:introduction}

A new energy frontier has been opened with the start of the LHC, namely the
exploration of the tera-scale. An important part of the experimental program
of the LHC is the search for supersymmetric particles
 \cite{Weiglein:2004hn,Ball:2007zza,Aad:2009wy}. If they are indeed
found, the determination of the underlying parameters will then be among the
main tasks. Here the properties of neutralinos and charginos will play an
important role as they occur in various steps in the cascade decays of
squarks and gluinos which can be copiously produced at the LHC. 
These investigations will then most likely be complemented and continued
at a prospective future linear collider  (ILC) 
\cite{AguilarSaavedra:2001rg,Weiglein:2004hn} and/or at
 a multi-TeV collider such
as CLIC \cite{Accomando:2004sz}.

The two-body 
decay of a chargino (neutralino) into a $W$-boson and a neutralino (chargino) 
is one of the most important modes if kinematically allowed
\cite{Bartl:1985fk,Baer:1988kx,Feng:1995zd}.
The determination
of the corresponding branching ratio can give important information on the nature
of the chargino and neutralino involved and therefore one will need at least
an ILC for its precise value. This is one of the reasons to investigate the
one-loop corrections to this decay mode. We will see below that the corrections
can be as large as the LO decay width in certain regions of parameters space. A second
important aspect is the question how to perform the complete electroweak
corrections in a gauge invariant way. These decay modes offer an ideal
play ground to tackle the technical questions involved.

We will use an on-shell renormalization scheme for masses and couplings.
The on-shell renormalization of the mass matrices of charginos and neutralinos
in the minimal supersymmetric standard model (MSSM) was first discussed in \cite{Eberl:2001eu,Fritzsche:2002bi}
and extensively addressed in \cite{Baro:2008bg}. Moreover, 
electroweak corrections for these decays taking only into account third generation
squarks and quarks within the MSSM have been calculated in \cite{Zhang:2001rd}.
Corrections of up to $10$\% have been found there. Note, that for these corrections
gauge invariance is not an issue. 

We will work  in the general linear $R_\xi$-gauge. An important
aspect will be  the question how to renormalize the mixing
matrices of charginos and neutralinos, since it is already known from
the Standard Model that the on-shell renormalization prescription
according to \cite{Denner:1990yz,Denner:1991kt} leads to a gauge
dependent CKM-matrix  \cite{Gambino:1998ec} which in turn
implies a gauge dependence of processes like $t\rightarrow W b$ at
 next-to-leading order level \cite{Yamada:2001px}.

Since then, some papers proposed different solutions to the problem
addressed above: whereas \cite{Espriu:2002xv} argued that missing
absorptive parts due to the unstable nature of the external particles
have to be included in the calculation, \cite{Yamada:2001px} proposed
a method on how to construct a gauge invariant counterterm for the mixing
matrices inspired by the pinch technique \cite{pinch}, which defines
gauge independent form factors for gauge bosons.
Another perspective is presented in \cite{Pilaftsis:2002nc},
where the gauge-variant on-shell renormalized mixing matrix is related
to a gauge independent one in a generalized $\overline{\text{MS}}$
scheme of renormalization.

For the special case of the CKM matrix different methods were
discussed in the literature: \cite{Gambino:1998ec} proposed to set the
momenta of the non-diagonal entries of the quark self-energies to
zero, \cite{Denner:2004bm,Kniehl:2006bs,Kniehl:2006rc} suggested new
variants of renormalization in particular for the mixing matrices
themselves partially based on physical processes which allows one to get
gauge independent decay widths.  For lepton or neutrino mass matrices
also \cite{Diener:2001qt,Almasy:2009kn} proposed useful
renormalization conditions allowing gauge independent results.

In this paper we will apply the technique presented in \cite{Yamada:2001px}.
The technical features are the same for the 
MSSM and the next-to-minimal one (NMSSM). As for the processes
under consideration the MSSM is a trivial sub-class of the NMSSM in the
limit of heavy singlet states we will perform
the calculation right from the start in the NMSSM. 
The paper is organized as follows: in Section~\ref{sec:NMSSM} 
we fix our notation for both models and in Section~\ref{sec:treelevel} we
summarize the tree-level results for the mass matrices of charginos
and  neutralinos as well as 
the tree-level decay widths for 
$\tilde{\chi}_i^{\pm}\rightarrow \tilde{\chi}_k^{0}W^{\pm}$ 
and $\tilde{\chi}_l^{0}\rightarrow \tilde{\chi}_j^{\pm} W^{\mp}$. 
The virtual one-loop contributions and real corrections including
the question of gauge invariance are discussed in
Section~\ref{sec:oneloopdecaywidth}.
We present our numerical result in Section~\ref{sec:results}
and in Section~\ref{sec:conclusions} our conclusions.

~\ref{sec:vertexcorr} contains the generic formulas for the
matrix element contributions of the vertex corrections.
In ~\ref{sec:photemission} we give the formulas
for the  the real corrections which are non-factorizable
as the higgsinos form a vector-like $SU(2)$ gauge state.
The usage of the $R_\xi$-gauge implies that  all possible derivatives of 
the two-point one-loop functions
show up in the calculation. Some of them are to our knowledge not available
in analytical form in the literature and, thus, we provide them
for completeness in ~\ref{sec:oneandtwopointfunctions}.
In ~\ref{sec:genericfortran} we present
the program \prog for the numerical evaluation of these corrections.

\section{The models}
\label{sec:NMSSM}

In this section we fix the notation for the models considered, the MSSM 
\cite{Chung:2003fi} and the NMSSM \cite{Ellwanger:2009dp}.
Detailed formulas for mass-matrices and couplings can be found
in \cite{Chung:2003fi,Staub:2010ty}.
Both model have in common the Yukawa part of the matter superfields coupled to the $SU(2)$-doublet
Higgs fields
\begin{align}
{\cal W}_{Y} =  \epsilon_{ab} \left(Y_U^{ij}  \widehat Q_i^a \widehat U_j \widehat H_u^b
+ Y_D^{ij} \widehat Q_i^b   \widehat D_j \widehat H_d^a
+  Y_e^{ij} \widehat L_i^b \widehat E_j \widehat H_d^a \right)\qquad,
\label{eq:WsuppotY}
\end{align}
where $\epsilon_{ab}$ is the complete antisymmetric $SU(2)$ tensor with $\epsilon_{12}=1$.
In case of the MSSM the $\mu$-term is added and the total superpotential reads as 
\begin{equation}
{\cal W}_{\mathrm{MSSM}} = {\cal W}_{Y} - \mu \widehat H_d \widehat H_u\qquad,
\label{eq:WsuppotMSSM}
\end{equation} 
whereas in the NMSSM the Higgs doublets are coupled to a gauge singlet
superfield $\hat{S}$
\begin{align}
{\cal W}_{\rm{NMSSM}} = {\cal W}_{Y}  - \lambda \widehat S \widehat H_d \widehat H_u
+\frac{1}{3}\kappa \widehat S \widehat S \widehat S\qquad.
\label{eq:Wsuppot}
\end{align}
If the scalar component of the gauge singlet $\hat S$ gets a vacuum
expectation value $\frac{1}{\sqrt{2}}v_S$
an effective $\mu$-term is generated 
\begin{align}
 \mu = \frac{1}{\sqrt{2}} \lambda v_S \qquad.
\label{eq:mueff}
\end{align}
In this way one gets $\mu$ naturally of the order of the  electroweak scale
\cite{Ellwanger:2009dp}.
To complete the models one has to add  the soft-SUSY breaking terms, where
the part common to both models reads as\footnote{We use the notation of
\cite{Skands:2003cj,Allanach:2008qq}.}:
\begin{align}\nonumber
 V_{soft}^{1} = &{m_Q^{ij}}^2 \tilde{Q}_i^{a \ast}
\tilde{Q}_j^a + {m_U^{ij}}^2 \tilde{U}_i \tilde{U}_j^{\ast} +
{m_D^{ij}}^2 \tilde{D}_i \tilde{D}_j^{\ast} + {m_L^{ij}}^2
\tilde{L}_i^{a \ast} \tilde{L}_j^a + {m_E^{ij}}^2 \tilde{E}_i
\tilde{E}_j^{\ast} \\\nonumber
& + m_{H_d}^2 H_d^{a \ast} H_d^a +
m_{H_u}^2 H_u^{a \ast} H_u^a \\\nonumber
& + \frac{1}{2} \left[ M_1 \tilde{B}^0
\tilde{B}^0 + M_2 \tilde{W}^c \tilde{W}^c + M_3 \tilde{g}^d
\tilde{g}^d + \text{h.c.} \right]\\ 
& + \epsilon_{ab}
\left[ T_U^{ij} \tilde{Q}_i^a \tilde{U}_j^\ast H_u^b + T_D^{ij}
\tilde{Q}_i^b \tilde{D}_j^\ast H_d^a + T_E^{ij} \tilde{L}_i^b
\tilde{E}_j^\ast H_d^a + \text{h.c.} \right] \qquad ,
\label{eq:softmssm1} 
\end{align}
with $a,b=1,2$ as summation indices. The complete soft-SUSY breaking potential of the MSSM is then given by
\begin{equation}
 V_{soft}^{\mathrm{MSSM}} = V_{soft}^{1}  - B \mu \epsilon_{ab}H_d^aH_u^b
\end{equation}
and the NMSSM one by
\begin{equation}
 V_{soft}^{\mathrm{NMSSM}} = V_{soft}^{1} + m_{S}^2 S
 S^{\ast} - \left[\epsilon_{ab} T_\lambda S H_d^a H_u^b+ \text{h.c.} \right]
 + \left[\frac{1}{3}T_{\kappa} SSS + \text{h.c.}\right]\qquad.
\label{eq:softsing}
\end{equation}

\section{Tree-level results}
\label{sec:treelevel}

Here we collect the tree-level results which form the basis for the 
one-loop corrections considered in the subsequent sections.

\subsection{Masses of Charginos and Neutralinos}
\label{subsec:neutralinoscharginos}

Using Weyl spinors in the basis
$(\psi^-)^T=(\tilde{W}^-,\tilde{H}_d^-)$, $(\psi^+)^T=(\tilde{W}^+,\tilde{H}_u^+)$
the chargino mass matrix is given in both models by
\begin{align}
\Mass_c = \left( \begin{array}{c c}
M_2 & \frac{1}{\sqrt{2}}gv_u \\
\frac{1}{\sqrt{2}}gv_d & \mu  \end{array} \right)\quad.
\label{eq:chimimasstree}
\end{align}
The mass eigenstates are obtained by two unitary matrices $U$ and $V$ 
\begin{align}
F_i^+=V_{it}\psi_t^+ \quad\text{and}\quad F_i^-=U_{it}\psi_t^-\quad,
\end{align}
such that the mass eigenvalues are  given by:
\begin{eqnarray}
\text{Diag}\left(m_{\tilde{\chi}_1^+},m_{\tilde{\chi}_2^+}\right)&=&U^*\mathcal{M}_c V^{-1}
\\
\nonumber
m_{\tilde{\chi}_1^+} \le m_{\tilde{\chi}_2^+}
\end{eqnarray}
and $\tilde{\chi}_i^+$ is the Dirac spinor given by
\begin{equation}
\tilde{\chi}_i^+ = \left( \begin{array}{c}
F_i^+ \\ \overline{F^-_i}
\end{array} \right)\qquad.
\end{equation}
For the neutralinos we obtain in the basis\footnote{$\tilde{S}$ is of course only present
in the NMSSM.} $( \psi^0 )^T = ( {\tilde B}^0, {\tilde W}_3^0, {\tilde
H}_d^0, {\tilde H}_u^0, \tilde{S} )$
the  mass matrix 
\begin{align}
\Mass_n = \left( \begin{array}{c c c c c}
M_1 & 0 & -\frac{1}{2}g' v_d & \frac{1}{2}g' v_u & 0\\
0 & M_2 & \frac{1}{2}g v_d & -\frac{1}{2}g v_u & 0\\
-\frac{1}{2}g' v_d & \frac{1}{2}g v_d & 0 & -\mu & -\frac{1}{\sqrt{2}} \lambda v_u\\
\frac{1}{2}g' v_u & -\frac{1}{2}g v_u & -\mu & 0 & -\frac{1}{\sqrt{2}}\lambda v_d\\
0&0&-\frac{1}{\sqrt{2}} \lambda v_u&-\frac{1}{\sqrt{2}}\lambda v_d & \sqrt{2} \kappa v_S \end{array} \right)\qquad,
\label{eq:chi0masstree}
\end{align}
where in the NMSSM $\mu$ is given by \eqref{eq:mueff}.
In case of the MSSM one has to omit the last column and the last row in 
Equation~\eqref{eq:chi0masstree}.
The mass eigenstates $F_i^0$ are obtained via
\begin{equation}
F_i^0=\mathcal{N}_{is}\psi_s^0 
\end{equation}
from the gauge eigenstates $\psi_s^0$, where the unitary matrix $\Nge$ diagonalizes 
the mass matrix $\Mass_n$ as
\begin{equation}
\text{Diag}\left(m_{\tilde{\chi}_1^0},\ldots,m_{\tilde{\chi}_j^0}\right)=\mathcal{N}^*\mathcal{M}_n\mathcal{N}^\dagger\qquad,
\end{equation}
where $j=4(5)$ in case of the MSSM (NMSSM). The masses are ordered
as $m_{\tilde{\chi}_i^0} \le m_{\tilde{\chi}_j^0}$ for $i<j$ and we have
chosen $\Nge$ such that all mass eigenvalues are positive. This implies that
in general the matrix $\Nge$ is complex.
Useful checks can be performed if one goes in case of real parameters to a basis
where $\Nge$ is real. Then one or more of the mass eigenvalues get negative
and the mass ordering applies to their moduli.
For completeness we note that $\tilde{\chi}_i^0$ is the $4$-component spinor given by
\begin{equation}
\tilde{\chi}_i^0 = \left( \begin{array}{c}
F_i^0 \\ \overline{F^0_i}
\end{array} \right)\qquad.
\end{equation}

\subsection{Decay widths}
\label{sec:treelevelwidth}

The partial widths for the decays
$\tilde{\chi}_j^{0}\rightarrow \tilde{\chi}_l^{\pm}W^{\mp}$ and
$\tilde{\chi}_l^{\mp}\rightarrow \tilde{\chi}_j^{0}W^{\mp}$ are obtained
from the following interaction Lagrangian:
\begin{align}
 \mathcal{L}=\overline{\tilde{\chi}_l^-}\gamma^\mu\left(P_LO_{Llj}+P_RO_{Rlj}\right)\tilde{\chi}_j^0 W^-_\mu + \mathrm{h.c.}\qquad.
\label{eq:treelevellagrangian}
\end{align}
In both models the  couplings  are given by:
\begin{align}
O_{Llj} = &- g \Nge^*_{j2}U_{l1} - \frac{1}{\sqrt{2}} g \Nge^*_{j3}U_{l2},\qquad
O_{Rlj} = - gV_{l1}^*\Nge_{j2} + \frac{1}{\sqrt{2}} gV_{l2}^*\Nge_{j4} \,.
\label{eq:treelevelcouplings}
\end{align}
The widths have the form 
\begin{equation}
\Gamma^{0} = \frac{1}{16\pi m_i^3}\sqrt{\kappa(m_i^2,m_o^2,m_W^2)} \frac{1}{2}\sum_{pol}|M_T|^2\qquad,
\label{eq:leadingorderwidth}
\end{equation}
where $m_i$ ($m_o$) is the mass of the mother (daughter) particle and 
$M_T$ is the tree-level matrix element. Explicitly they 
 are given by
\begin{eqnarray}
\label{eq:neutdecaytree}
\Gamma^{\mathrm{0}}\left(\tilde{\chi}_j^{0}\rightarrow \tilde{\chi}_l^{+}W^{-}\right)
&=& \frac{1}{16\pi m_j^3}\sqrt{\kappa(m_j^2,m_l^2,m_W^2)} \nonumber  \\ 
 \times \Big( \big(
|O_{Llj}|^2&+&|O_{Rlj}|^2 \big) f(m_j^2,m_l^2,m_W^2)
-6 \mathrm{Re}(O_{Llj}O_{Rlj}^*)m_jm_l \Big) \\
\Gamma^{\mathrm{0}}\left(\tilde{\chi}_i^{+}\rightarrow \tilde{\chi}_k^{0}W^{+}\right)
&=& \frac{1}{16\pi m_i^3}\sqrt{\kappa(m_i^2,m_k^2,m_W^2)} \nonumber  \\ 
 \times \Big( \big(
|O_{Lik}|^2&+&|O_{Rik}|^2 \big) f(m_i^2,m_k^2,m_W^2)
-6 \mathrm{Re}(O_{Lik}O_{Rik}^*)m_im_k \Big) 
\label{eq:chardecaytree}
\end{eqnarray}
with
\begin{eqnarray}
f(x,y,z) &=& \frac{1}{2}(x+y)- z+\frac{(x-y)^2}{2 z} \\
\kappa(x,y,z)&=&x^2+y^2+z^2-2xy-2xz-2yz\qquad.
\label{eq:kaellenfunction}
\end{eqnarray}

\section{One loop results}
\label{sec:oneloopdecaywidth}

We discuss the one-loop corrections in the context of the NMSSM
as it contains the MSSM in the limit of very heavy gauge singlet
states. There are no technical differences for the decays considered
as the gauge parts of both models coincides.
The one-loop corrected (renormalized) amplitude(s) $M$ can be expressed as
\begin{equation}
M = M_T + \Delta M = M_T + M_V + M_{WV}\qquad,
\end{equation}
where $M_{V}$ contains the vertex corrections and $M_{WV}$ is the sum of
the wave-function corrections and counterterms. Moreover we consider
finite mass shifts 
$m_{\tilde{\chi}_i^{\pm 0}}\rightarrow m^{1L}_{\tilde{\chi}_i^{\pm 0}}$
for the charginos and neutralinos at the one-loop level,
since the number of physical parameters is smaller than the number of masses.

\subsection{Wave-function and coupling renormalization}

\begin{figure}[t]
\begin{center}
\includegraphics[width=0.7\textwidth]{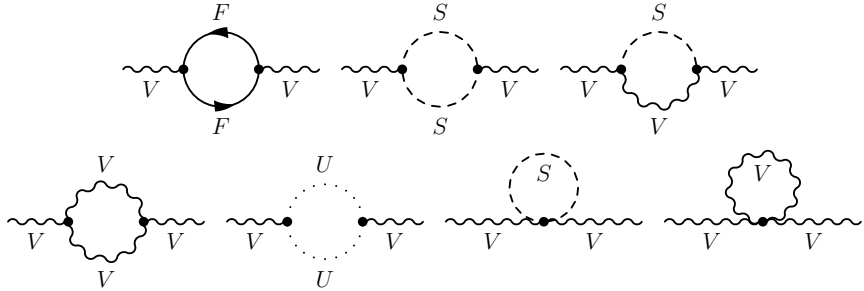}
\caption{Generic self-energy diagrams for gauge bosons.}
\label{fig:selfenergy1}
\end{center}
\end{figure}
\begin{figure}[t]
\begin{center}
\includegraphics[width=0.4\textwidth]{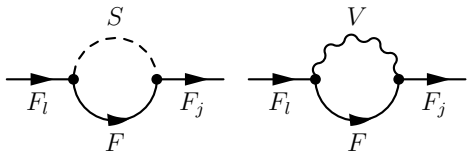}
\caption{Generic self-energy diagrams for fermions.}
\label{fig:selfenergy2}
\end{center}
\end{figure}
General formulas for the wave-function renormalization of
supersymmetric particles can be found e.g.~in
\cite{Pierce:1997wu}. Here we summarize the most important points
related to the decay under consideration and discuss in particular the
problem of gauge invariance. We 
work in a linear $R_\xi$ gauge given by
\begin{align}
\mathcal{L}=-\frac{1}{2\xi_\gamma}(F^\gamma)^2-\frac{1}{2\xi_Z}(F^Z)^2-\frac{1}{\xi_W}F^+F^-
\end{align}
with 
\begin{align}
 F^\pm &= \partial^\mu W_\mu^\pm \mp i M_W\xi_W G^\pm\\
 F^Z &= \partial^\mu Z_\mu - M_Z\xi_ZG^0\\
 F^\gamma &= \partial^\mu A_\mu \qquad .
\end{align}
The inverse propagator of the vector bosons is given at  tree-level by
\begin{align}
\label{eq:mgaugebosonprop}
-ig^{\mu\nu}(k^2-m_V^2)+i\left(1-\frac{1}{\xi_V}\right)k^\mu k^\nu
=-ig_T^{\mu\nu}(k^2-m_V^2)-ig_L^{\mu\nu}\left(m_V^2-\frac{k^2}{\xi_V}\right)
\end{align}
with $V=W,Z$ and the transverse and longitudinal projectors are defined as:
\begin{align}
g_T^{\mu\nu}=g^{\mu\nu}-\frac{k^\mu k^\nu}{k^2},\qquad g_L^{\mu\nu}=\frac{k^\mu k^\nu}{k^2}
\end{align}
The multiplicative renormalization of the bare parameters is done in
the following form:
\begin{align}\nonumber
m_{V0}^2 &\rightarrow Z_m m_V^2=(1+\delta Z_m)m_V^2=m_V^2+\delta m_V^2 \qquad\text{with}\qquad \delta m_V^2=m_V^2 \delta Z_m\\\nonumber
V_\mu^0 &\rightarrow Z_V V_\mu=(1+\tfrac{1}{2}\delta Z_V)V_\mu\\
\xi_{V0} &\rightarrow Z_{\xi_V}^{-1}\xi_V=(1+\delta Z_{\xi_V})^{-1}\xi_V
\label{multrenormwboson}
\end{align}
The $Z_\xi$s are fixed by the conditions that propagators mixing the
Goldstone bosons with the vector bosons vanish
\cite{Chankowski:1992er} and will not be discussed further here as they are not
needed in the subsequent discussion.
The vector boson propagator then reads at the one-loop level
\begin{align}\label{propwboson}
 i\hat{\Gamma}^{V,\mu\nu}(k^2)=-ig_T^{\mu\nu}(k^2-m_V^2)+ig_L^{\mu\nu}\left(m_V^2-\frac{k^2}{\xi}\right)
-ig_T^{\mu\nu}\hat{\Sigma}_T^V(k^2)-\frac{i}{\xi}g_L^{\mu\nu}\hat{\Sigma}_L^V(k^2)
\end{align}
with
\begin{eqnarray}\nonumber
\label{renormwbosontransverse}
 \hat{\Sigma}_T^V(k^2)&=&\Sigma_T^V(k^2)+k^2\delta Z_V-m_V^2-\delta m_V^2\\
 \hat{\Sigma}_L^V(k^2)&=&\Sigma_L^V(k^2)+\xi m_V^2\delta Z_V
 +\xi \delta m_V^2-k^2(\delta Z_\xi + \delta Z_V) \qquad,
\label{renormwbosonlongitudinal}
\end{eqnarray}
where $\Sigma_T^V(k^2)$ and $\Sigma_L^V(k^2)$ are the contributions obtained
from the one-loop graphs whose generic structures are shown in
 Figure~\ref{fig:selfenergy1}.

We require an on-shell renormalization for the vector bosons,
e.g.~that the pole of the propagator occurs at the physical mass and
that its residuum is one. This yields the conditions
\begin{align}
\label{wbosonphysicalmass} 
\Rea \hat{\Sigma}_T^V(m_V^2)&=0\\
\Rea \hat{\Sigma'}_T^V(m_V^2):=\left. \frac{\partial \Rea \hat{\Sigma}_T^V(k^2)}{\partial k^2}\right|_{k^2=m_V^2}&=0\qquad.
\label{wbosonresidat1}
\end{align}
Note, that the $\Rea$ applies only to the one-loop functions but
not to the couplings. 
Finally one gets:
\begin{align}
 \delta m_V^2=\Rea \Sigma_T^V(m_V^2), \qquad \delta Z_V=-\Rea {\Sigma'}_T^V(m_V^2)
\end{align}
Moreover one has to fix the renormalization of the
couplings $e$ and $g$ and the $\cos$ of the Weinberg angle $\cos\theta_W$:
\begin{align}
e&\rightarrow \delta Z_ee=(1+\delta Z_e)e=e+\delta e\\
g&\rightarrow \delta Z_gg=(1+\delta Z_g)g=g+\delta g\\
\cos\theta_W &\rightarrow \cos\theta_W+\delta\cos\theta_W
\end{align}
Not all of these quantities are independent of the mass renormalization
and one obtains \cite{Denner:1991kt}:
\begin{align}
 \delta \cos\theta_W &= \frac{1}{2}\cos\theta_W\left(\frac{\delta m_W^2}{m_W^2}-\frac{\delta m_Z^2}{m_Z^2}\right) \quad\Longrightarrow\quad
 \delta \sin\theta_W = - \frac{1}{\tan\theta_W}\delta\cos\theta_W \\
 \delta Z_e &= \frac{1}{2}{\Sigma'}_T^{\gamma\gamma}(0)-\frac{\tan\theta_W}{m_Z^2}\Sigma_T^{Z\gamma}(0) \quad\Longrightarrow\quad
 \delta g = \left(\delta Z_e - \frac{\delta \sin \theta_W}{\sin\theta_W}\right)g
\end{align}

In case of charginos and neutralinos the mixing effects have to be taken
into account. The fact, that  both sectors together
have less parameters than mass eigenstates,
leads to additional finite shifts and has been discussed
in detail for the MSSM in \cite{Eberl:2001eu,Fritzsche:2002bi,Oller:2003ge}
being relevant for the calculation of NLO masses. Before addressing this
topic we first present the principal idea of the wave-function renormalization for
mixed fermions and discuss gauge invariance in all detail.
For the wave-function and mass renormalization one needs
\begin{align}
\tilde{\chi}_i^0&\rightarrow \left(\delta_{ij}+\tfrac{1}{2}\delta Z_{ij}^{0L}P_L+\tfrac{1}{2}\delta Z_{ij}^{0R}P_R\right)\tilde{\chi}_j^0\\
\tilde{\chi}_i^\pm&\rightarrow \left(\delta_{ij}+\tfrac{1}{2}\delta Z_{ij}^{\pm L}P_L+\tfrac{1}{2}\delta Z_{ij}^{\pm R}P_R\right)\tilde{\chi}_j^\pm \\
m_{fi} &\rightarrow m_{fi} + \delta m_{fi} \qquad.
\label{eq:masseigenstatesmassct}
\end{align}
It is well known that the wave-function renormalization constants of the
neutralinos and charginos satisfy the following relations
\begin{align}
\delta Z_{ij}^{0L}=\delta Z_{ij}^{0R*},\qquad \delta Z_{ij}^{+L}=\delta Z_{ij}^{-R*},\qquad \delta Z_{ij}^{-L}=\delta Z_{ij}^{+R*} \qquad.
\end{align}
As already pointed out in \cite{Denner:1990yz} the combination of
tree-level mixing with non-diagonal field renormalization gives
anti-hermitian parts, which however can be canceled by introducing
counterterms for the mixing matrices themselves:
\begin{align}
\Nge_{ij}&\rightarrow \Nge_{ij}+\delta \Nge_{ij} \quad\text{with}\quad \delta \Nge_{ij}=\frac{1}{4}\left(\delta Z_{ik}^{0L}-\delta Z_{ki}^{0R}\right)\Nge_{kj}\\
U_{ij} &\rightarrow U_{ij}+\delta U_{ij}\quad\text{with}\quad \delta U_{ij}=\frac{1}{4}\left(\delta Z_{ik}^{+R*}-\delta Z_{ki}^{+R}\right)U_{kj}\\
V_{ij} &\rightarrow V_{ij}+\delta V_{ij}\quad\text{with}\quad \delta V_{ij}=\frac{1}{4}\left(\delta Z_{ik}^{+L}-\delta Z_{ki}^{+L*}\right)V_{kj}
\end{align}
Here we recall the basic steps to renormalize a system of Dirac
fermions. The case of Majorana fermions can be obtained from this by
noting that the Majorana nature induces relationships between
$Z_{ij}^{L}$ and $Z_{ij}^{R}$. As in case of the vector bosons one
starts with the inverse propagator given by
\begin{align}\nonumber
i\hat{\Gamma}^f_{ij}\left(p\right)=i\delta_{ij}(\sla{p}-m_{fi})+i&\left[\sla{p}\left(P_L\hat{\Sigma}_{ij}^{fL}\left(p^2\right)
+P_R\hat{\Sigma}_{ij}^{fR}\left(p^2\right)\right)\right.\\
&\quad\left.+P_L\hat{\Sigma}_{ij}^{fSL}(p^2)+P_R\hat{\Sigma}_{ij}^{fSR}(p^2)\right]
\label{eq:fermionselfenergy} 
\end{align}
with
\begin{align}\label{diracfermionrenorm1}
\hat{\Sigma}_{ij}^{fL}(p^2)&=\Sigma_{ij}^{fL}(p^2)+\tfrac{1}{2}\left(\delta Z_{ij}^{fL}+\delta Z_{ij}^{fL\dagger}\right)\\\label{diracfermionrenorm2}
\hat{\Sigma}_{ij}^{fR}(p^2)&=\Sigma_{ij}^{fR}(p^2)+\tfrac{1}{2}\left(\delta Z_{ij}^{fR}+\delta Z_{ij}^{fR\dagger}\right)\\\label{diracfermionrenorm3}
\hat{\Sigma}_{ij}^{fSL}(p^2)&=\Sigma_{ij}^{fSL}(p^2)-\tfrac{1}{2}\left(m_{fi}\delta Z_{ij}^{fL}+m_{fj}\delta Z_{ij}^{fR\dagger}\right)-\delta_{ij}\delta m_{fi}\\\label{diracfermionrenorm4}
\hat{\Sigma}_{ij}^{fSR}(p^2)&=\Sigma_{ij}^{fSR}(p^2)-\tfrac{1}{2}\left(m_{fi}\delta Z_{ij}^{fR}+m_{fj}\delta Z_{ij}^{fL\dagger}\right)-\delta_{ij}\delta m_{fi}\qquad.
\end{align}
The corresponding generic one-loop contributions are shown in Figure~\ref{fig:selfenergy2}.
The on-shell conditions read as 
\begin{align}\label{diracfermioncondition1}
\left.\overline{u}_i(p)\Rea \hat{\Gamma}_{ij}^f(p)\right|_{p^2=m_{fi}^2}=0 \qquad
& \lim_{p^2\rightarrow m_{fi}^2} \overline{u}_i(p)\Rea \hat{\Gamma}_{ii}^f(p)\frac{\sla{p}+m_{fi}}{p^2-m_{fi}^2}=\overline{u}_i(p)\\
\left.\Rea \hat{\Gamma}_{ij}^f(p)u_j(p)\right|_{p^2=m_{fj}^2}=0 \qquad
& \lim_{p^2\rightarrow m_{fi}^2} \frac{\sla{p}+m_{fi}}{p^2-m_{fi}^2}\Rea \hat{\Gamma}_{ii}^f(p)u_i(p)=u_i(p)\qquad,
\label{diracfermioncondition2}
\end{align}
which in turn lead to the counterterm for the masses $m_{fi}$
\begin{align}
\delta m_{fi}=\frac{1}{2}\left[m_{fi}\Rea \Sigma_{ii}^{fL}(m_{fi}^2)
+m_{fi}\Rea \Sigma_{ii}^{fR}(m_{fi}^2)+\Rea \Sigma_{ii}^{fSL}(m_{fi}^2)+\Rea \Sigma_{ii}^{fSR}(m_{fi}^2)\right]
\end{align}
and the following wave-function renormalization constants
\begin{align}
\label{eq:diagonalwv1}\nonumber
\delta Z_{ii}^{fL}=-\Rea&\left[\Sigma_{ii}^{fL}(m_{fi}^2)+m_{fi}^2\left({\Sigma'}_{ii}^{fL}(m_{fi}^2)+ {\Sigma'}_{ii}^{fR}(m_{fi}^2)\right)\right.\\
&\left.\quad+m_{fi}\left({\Sigma'}_{ii}^{fSL}(m_{fi}^2)+{\Sigma'}_{ii}^{fSR}(m_{fi}^2)\right)\right]\\\nonumber
\delta Z_{ii}^{fR}=-\Rea&\left[\Sigma_{ii}^{fR}(m_{fi}^2)+m_{fi}^2\left({\Sigma'}_{ii}^{fL}(m_{fi}^2)+ {\Sigma'}_{ii}^{fR}(m_{fi}^2)\right)\right.\\
&\left.\quad+m_{fi}\left({\Sigma'}_{ii}^{fSL}(m_{fi}^2)+{\Sigma'}_{ii}^{fSR}(m_{fi}^2)\right)\right]
\label{eq:diagonalwv2}
\end{align}
for the diagonal entries, whereas for the off-diagonal ones one finds:
\begin{align}\nonumber
\delta Z_{ij}^{fL}=\frac{2}{m_{fi}^2-m_{fj}^2}&\left[m_{fj}^2\Rea \Sigma_{ij}^{fL}(m_{fj}^2)
+m_{fi}m_{fj}\Rea \Sigma_{ij}^{fR}(m_{fj}^2)\right.\\&\left.\qquad+m_{fi}\Rea \Sigma_{ij}^{fSL}(m_{fj}^2)+m_{fj}\Rea \Sigma_{ij}^{fSR}(m_{fj}^2)\right]
\label{eq:nondiagonalwv1}\\
\nonumber
 \delta Z_{ij}^{fR}=\frac{2}{m_{fi}^2-m_{fj}^2}&\left[m_{fi}m_{fj}\Rea \Sigma_{ij}^{fL}(m_{fj}^2)
+m_{fj}^2\Rea \Sigma_{ij}^{fR}(m_{fj}^2)\right.\\&\left.\qquad+m_{fj}\Rea \Sigma_{ij}^{fSL}(m_{fj}^2)+m_{fi}\Rea \Sigma_{ij}^{fSR}(m_{fj}^2)\right]
\label{eq:nondiagonalwv2}
\end{align}

\subsection{Gauge invariance}

Before discussing the finite mass shifts for neutralinos and charginos
and the question how to construct $M_{WV}$ in the next subsections
we have to clarify a subtle
point with respect to gauge invariance.
The counterterms  given in
Equations~\eqref{eq:diagonalwv1}-\eqref{eq:nondiagonalwv2} cancel the
UV divergences and avoid anti-hermitian parts in the
Lagrangian. However, this does not necessarily imply
gauge invariance for the process as well as the one-loop masses.
To achieve this we use the method proposed in
\cite{Yamada:2001px} which is inspired by the pinch technique
\cite{pinch}, which defines gauge independent form factors for gauge
bosons.
We are aware that it has one weak point, namely a dependence on the
choice of the gauge fixing for the mixing matrix counterterm. Its
advantage is that it is model independent and does not rely
on the details of the renormalization of physical parameters.

\begin{figure}[t]
\begin{center}
\includegraphics[width=0.5\textwidth]{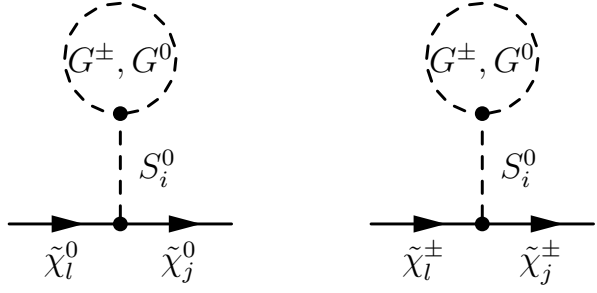}
\caption{Tadpole contributions induced by Goldstone bosons
$G^\pm$, $G^0$ ensuring gauge invariance.}
\label{fig:mixingtadpole}
\end{center}
\end{figure}

The basic idea of this method is to treat the gauge dependence of the
counterterms for the wave-function renormalization constants differently from 
the ones for the mixing matrices. Therefore we calculate two
variants of wave-function renormalization constants, namely in case of
the neutralinos $\delta Z_{ij}^{0L}, \delta Z_{ij}^{0R}$ for arbitrary
values of $\xi_V$ and $\delta \breve{Z}_{ij}^{0L}, \delta
\breve{Z}_{ij}^{0R}$ for $\xi_V=1$ ('t~Hooft-Feynman gauge).  The same
is done for the wave-function renormalization constants of the
charginos which we denote by
$\delta Z_{ij}^{\pm L}, \delta Z_{ij}^{\pm R}$ and $\delta
\breve{Z}_{ij}^{\pm L}, \delta \breve{Z}_{ij}^{\pm R}$.  The
counterterms for the mixing matrices are calculated via the
wave-function renormalization constants in the 't~Hooft-Feynman gauge
\begin{align}
\delta \Nge_{ij}&=\frac{1}{4}\left(\delta \breve{Z}_{ik}^{0L}-\delta \breve{Z}_{ki}^{0R}\right)\Nge_{kj}\\
\delta U_{ij}=\frac{1}{4}\left(\delta \breve{Z}_{ik}^{+R*}-\delta \breve{Z}_{ki}^{+R}\right)U_{kj},&\qquad
\delta V_{ij}=\frac{1}{4}\left(\delta \breve{Z}_{ik}^{+L}-\delta \breve{Z}_{ki}^{+L*}\right)V_{kj}\qquad,
\end{align}
whereas for the 
wave-function renormalization constants as given in general form in
Equations~\eqref{eq:diagonalwv1}-\eqref{eq:nondiagonalwv2}
we use the full gauge dependent form.
This splitting forces one to include the
additional contributions from the tadpole graphs with the Goldstone
bosons as shown in Figure~\ref{fig:mixingtadpole}.
A few remarks are in order here:
for $\xi_V=1$ the tadpole graph contributions to the wave-function
renormalization exactly cancel their contributions to $\delta \Nge$,
$\delta U$ and $\delta V$ in case of the considered processes.
For  $\xi_V\neq 1$ they cancel also $\xi_V$-dependent parts of other one-loop 
contributions.
Moreover, they only contribute to the ultraviolet divergence for $\xi_V\neq 1$
but of course the rearrangement is such that the total ultraviolet divergence
cancels. 
The infrared divergence in case of the
processes is not affected 
at all by this mechanism, since the photon
only contributes to the diagonal entries of fermionic self-energies.
Last but not least we note that all these statements hold
as well for the calculation of gauge invariant, ultraviolet and infrared
finite masses $m^{1L}_{\tilde{\chi}_i^{\pm 0}}$ in the next section, implying that
we use gauge invariant counterterms for the mixing matrices including the
described Goldstone contributions.

\subsection{On-shell masses of Neutralinos and Charginos}

A special feature of the chargino/neutralino sector is, that the number of
 parameters is lower than the number of imposed on-shell conditions. This
implies finite corrections to the tree-level masses of neutralinos and charginos.
We closely follow \cite{Eberl:2001eu} and start with the one-loop contributions to
 neutralino masses
$\delta \Mass_n^\circledast$ and chargino masses $\delta\Mass_c^\circledast$
\begin{align}
\label{eq:onshellmasses1}
&\left(\delta \Mass_n^\circledast\right)_{ij}=\delta\left(\Nge^T\Mass_{n,dia.}^\circledast \Nge\right)_{ij}\\\nonumber
&\quad=\sum_{n,l}\left[\delta\Nge_{ni}\left(\Mass_{n,dia.}^\circledast\right)_{nl}\Nge_{lj}+\Nge_{ni}\left(\delta\Mass_{n,dia.}^\circledast\right)_{nl}\Nge_{lj}+\Nge_{ni}\left(\Mass_{n,dia.}^\circledast\right)_{nl}\delta\Nge_{lj}\right]\\
\label{eq:onshellmasses2}
&\left(\delta \Mass_c^\circledast\right)_{ij}=\delta\left(U^T\Mass_{c,dia.}^\circledast V\right)_{ij}\\\nonumber
&\quad=\sum_{n,l}\left[\delta U_{ni}\left(\Mass_{c,dia.}^\circledast\right)_{nl}V_{lj}+U_{ni}\left(\delta\Mass_{c,dia.}^\circledast\right)_{nl}V_{lj}+U_{ni}\left(\Mass_{c,dia.}^\circledast\right)_{nl}\delta V_{lj}\right]
\end{align}
with the diagonalized mass matrices $\Mass_{dia.}^\circledast$ and their counterterms $\delta \Mass_{dia.}^\circledast$:
\begin{align}
\left(\Mass_{n,dia.}^\circledast\right)_{nl}=\delta_{nl}m_{\tilde{\chi}_l^0}&,\quad \left(\delta \Mass_{n,dia.}^\circledast\right)_{nl}
=\delta_{nl}\delta m_{\tilde{\chi}_l^0}\\ \left(\Mass_{c,dia.}^\circledast\right)_{nl}=\delta_{nl}m_{\tilde{\chi}_l^\pm}&,
\quad \left(\delta \Mass_{c,dia.}^\circledast\right)_{nl}=\delta_{nl}\delta m_{\tilde{\chi}_l^\pm}
\end{align}
The bare mass matrices of the neutralinos and charginos can now be expressed as the correct on-shell mass matrix
with the corrections \eqref{eq:onshellmasses1} and \eqref{eq:onshellmasses2} or via the tree-level mass matrix
expressed in physical parameters together with the renormalization constants of those:
\begin{align}
\Mass_{n,c}^0=\Mass_{n,c}^\circledast+\delta \Mass_{n,c}^\circledast=\Mass_{n,c}+\delta\Mass_{n,c}
\end{align}
Therefore the relations between tree-level and one-loop mass matrices take the form:
\begin{align}
\label{eq:oneloopmasses}
\Mass_{n,c}^\circledast&=\Mass_{n,c}+\delta\Mass_{n,c}-\delta\Mass_{n,c}^\circledast=:\Mass_{n,c}+\varDelta \Mass_{n,c}
\end{align}
In the following we will define the model-dependent physical parameters, write down the
renormalization of the mass matrices $\delta\Mass_{n,c}$ and identify the renormalization constants
of the physical parameters,
which should be fixed in the neutralino or chargino sector.
Some physical parameters, namely $\delta m_W,
\delta m_Z$ and thus $\delta\cos\theta_W$ are fixed in the gauge boson sector
and $\delta\tan\beta$ in the Higgs sector. In particular for
$\delta\tan\beta$ we take
the $\overline{\text{DR}}$ renormalization \cite{Freitas:2002um}, such that
UV divergences in the masses and the considered process cancel
\begin{align}
 \frac{\delta\tan\beta}{\tan\beta}=\frac{1}{32\pi^2}\Delta\left( 3\text{Tr}({Y_d}{Y_d}^\dagger)
 - 3\text{Tr}({Y_u}{Y_u}^\dagger) + \text{Tr}({Y_l}{Y_l}^\dagger) \right)
\end{align}
with $\Delta= \frac{1}{\epsilon}-\gamma +\ln(4\pi)$.
Note, that this choice maintains also the gauge invariance of masses and
the considered decay widths.

\subsubsection{MSSM}
In case of the MSSM we follow \cite{Eberl:2001eu}. The tree-level neutralino mass matrix
in \eqref{eq:chi0masstree} can be written as
\begin{align}
\Mass_n&=\begin{pmatrix}M_1&0&-m_Z\sin\theta_W\cos\beta&m_Z\sin\theta_W\sin\beta\\
&M_2&m_Z\cos\theta_W\cos\beta&-m_Z\cos\theta_W\sin\beta\\
&&0&-\mu\\
&\text{sym.}&&0\end{pmatrix}
\end{align}
and the chargino mass matrix in \eqref{eq:chimimasstree} as
\begin{align}
\Mass_c=\begin{pmatrix}M_2&\sqrt{2}m_W\sin\beta\\\sqrt{2}m_W\cos\beta&\mu\end{pmatrix}
\end{align}
The variation of all  given entries of the tree-level mass matrix leads to
\begin{align}
\label{eq:phfixneutralinosF}
\delta \Mass_n^{11}&=\delta M_1=\frac{\delta M_1}{M_1}\Mass_n^{11}\\
\delta \Mass_n^{13}&=-\delta(m_Z\sin\theta_W\cos\beta)
=\left(\frac{\delta m_Z}{m_Z}+\frac{\delta \sin\theta_W}{\sin\theta_W}+\frac{\delta \cos\beta}{\cos\beta}\right)\Mass_n^{13}\\
\delta \Mass_n^{14}&=\delta(m_Z\sin\theta_W\sin\beta)
=\left(\frac{\delta m_Z}{m_Z}+\frac{\delta \sin\theta_W}{\sin\theta_W}+\frac{\delta \sin\beta}{\sin\beta}\right)\Mass_n^{14}\\
\delta \Mass_n^{22}&=\delta M_2=\frac{\delta M_2}{M_2}\Mass_n^{22}\\
\delta \Mass_n^{23}&=\delta(m_Z\cos\theta_W\cos\beta)
=\left(\frac{\delta m_Z}{m_Z}+\frac{\delta \cos\theta_W}{\cos\theta_W}+\frac{\delta \cos\beta}{\cos\beta}\right)\Mass_n^{23}\\
\delta \Mass_n^{24}&=-\delta(m_Z\cos\theta_W\sin\beta)
=\left(\frac{\delta m_Z}{m_Z}+\frac{\delta \cos\theta_W}{\cos\theta_W}+\frac{\delta \sin\beta}{\sin\beta}\right)\Mass_n^{24}\\
\delta \Mass_n^{34}&=-\delta\mu=\frac{\delta\mu}{\mu}\Mass_n^{34}
\label{eq:phfixneutralinosL}
\end{align}
whereas all the other variations $\delta\Mass_n^{12}=\delta\Mass_n^{21}=\delta\Mass_n^{33}=\delta\Mass_n^{44}=0$ necessarily vanish.
The corrections in the chargino mass matrix read
\begin{align}
\label{eq:phfixcharginosF}
\delta \Mass_c^{11}&=\delta M_2=\frac{\delta M_2}{M_2}\Mass_n^{22}\\
\delta \Mass_c^{12}&=\sqrt{2}\delta(m_W\sin\beta)
=\left(\frac{\delta m_W}{m_W}+\frac{\delta \sin\beta}{\sin\beta}\right)\Mass_c^{12}\\
\delta \Mass_c^{21}&=\sqrt{2}\delta(m_W\cos\beta)
=\left(\frac{\delta m_W}{m_W}+\frac{\delta \cos\beta}{\cos\beta}\right)\Mass_c^{21}\\
\delta \Mass_c^{22}&=\delta \mu=\frac{\delta \mu}{\mu}\Mass_c^{22}\quad.
\label{eq:phfixcharginosL}
\end{align}
We will fix $M_2$ and $\mu$ in the chargino sector, whereas $M_1$ is
fixed in the neutralino sector by imposing the conditions
\begin{align}
\varDelta\Mass_c^{11}=\varDelta\Mass_c^{22}=\varDelta\Mass_n^{11}\stackrel{!}{=}0
\end{align}
resulting in:
\begin{align}
\delta M_1&=\delta \Mass_n^{\circledast 11},\qquad \delta M_2=\delta \Mass_c^{\circledast 11},\qquad \delta \mu=\delta \Mass_c^{\circledast 22}
\end{align}
For all the remaining entries of the neutralino and chargino mass matrices
finite shifts $\varDelta \Mass_{n,c}$ have to be taken into account.

\subsubsection{NMSSM}

Defining the additional  angles and parameters
\begin{align}
\tan \beta_S=\frac{v_S}{v_u},\qquad\mu=\frac{1}{\sqrt{2}}\lambda v_S,\qquad m_S=\sqrt{2}\kappa v_S
\end{align}
the tree-level mass matrix of the neutralinos is given by
\begin{align}
\Mass_n&=\begin{pmatrix}M_1&0&-m_Z\sin\theta_W\cos\beta&m_Z\sin\theta_W\sin\beta&0\\&M_2&m_Z\cos\theta_W\cos\beta&-m_Z\cos\theta_W\sin\beta&0\\&&0&-\mu&-\frac{\mu}{\tan\beta_S}\\&\text{sym.}&&0&\frac{-\mu}{\tan\beta\tan\beta_S}\\&&&&m_S\end{pmatrix}
\end{align}
and the chargino mass matrix is equal to the one in the MSSM.
Beside the variations \eqref{eq:phfixneutralinosF}-\eqref{eq:phfixneutralinosL} and
\eqref{eq:phfixcharginosF}-\eqref{eq:phfixcharginosL}
already present in the MSSM one has in addition                               
\begin{align}
\delta \Mass_n^{35}&=\delta\left(-\frac{\mu}{\tan\beta_S}\right)
=\left(\frac{\delta \mu}{\mu}-\frac{\delta \tan\beta_S}{\tan\beta_S}\right)\Mass_n^{35}\\
\delta\Mass_n^{45}&=\delta\left(\frac{-\mu}{\tan\beta\tan\beta_S}\right)=\left(\frac{\delta\mu}{\mu}
-\frac{\delta\tan\beta}{\tan\beta}
-\frac{\delta\tan\beta_S}{\tan\beta_S}\right)\Mass_n^{45}\\
\delta\Mass_n^{55}&=\frac{\delta m_S}{m_S}\Mass_n^{55}\quad,
\end{align}
whereas all the other variations $\delta\Mass_n^{12}=\delta\Mass_n^{15}=\delta\Mass_n^{21}=\delta\Mass_n^{25}=\delta\Mass_n^{33}=\delta\Mass_n^{44}=0$ necessarily vanish.
Similar to the MSSM we will fix $M_2$ and $\mu$ in the chargino sector, whereas $M_1,\tan\beta_S$ and $m_S$ are
fixed in the neutralino sector by imposing the following conditions
\begin{align}
\varDelta\Mass_c^{11}=\varDelta\Mass_c^{22}=\varDelta\Mass_n^{11}=\varDelta\Mass_n^{35}=\varDelta\Mass_n^{55}\stackrel{!}{=}0\quad,
\end{align}
which results in:
\begin{align}
\delta M_1&=\delta \Mass_n^{\circledast 11},\qquad \delta M_2=\delta \Mass_c^{\circledast 11},\qquad \delta \mu=\delta \Mass_c^{\circledast 22}\\
 \delta \tan\beta_S &=\frac{\tan^2\beta_S}{\mu}\left(\delta \Mass_n^{\circledast 35}-\frac{1}{\tan\beta_S} \delta \Mass_n^{\circledast 34}\right),
\qquad \delta m_S=\delta \Mass_n^{\circledast 55}
\end{align}
For all the remaining entries of the neutralino and chargino mass matrices
finite shifts $\varDelta \Mass_{n,c}$ have to be taken into account.
Note, that one could also fix $\delta\tan\beta_S$ in the Higgs sector.

\subsubsection{Effect  on the considered processes}
With the procedure introduced in the last sections we can calculate one-loop on-shell neutralino and chargino masses
for the MSSM and the NMSSM, namely by combining the full one loop corrections $\delta\Mass_{n,c}^\circledast$ 
with the counterterms $\delta \Mass_{n,c}$ obtained according to Equation~\eqref{eq:oneloopmasses}.
This results in the one-loop mass matrix $\Mass_{n,c}^\circledast$, which diagonalizations lead to one-loop
neutralino $m^{1L}_{\tilde{\chi}_i^0}$
and chargino masses $m^{1L}_{\tilde{\chi}_i^\pm}$ and mixing matrices at the
 one-loop level $\Nge^{1L},U^{1L},V^{1L}$.
Note, that these masses are UV and IR finite as well as gauge independent, if one takes into
account the gauge independent
renormalization of the mixing matrices \eqref{eq:onshellmasses1} and \eqref{eq:onshellmasses2}.

Instead of having the diagonal counterterm for the masses in \eqref{eq:masseigenstatesmassct}
we have to replace
\begin{align}
\delta_{ij} m_{fi0} &\rightarrow \delta_{ij} m_{fi}+\delta \tilde{\Mass}_{ij}P_L + \delta \tilde{\Mass}_{ji}^*P_R\quad,
\end{align}
where $\delta \tilde{\Mass}=D_R^*\delta \Mass D_L^\dagger$ with $\delta \Mass$ being the physical mass counterterm
and $D_L,D_R$ being the rotation matrices, which diagonalize the tree-level mass matrix $\Mass_{dia.}=D_R^* \Mass D_L^\dagger$
in the notation of \cite{Baro:2008bg}.
With this counterterm one gets contributions to the non-diagonal wave function renormalization constants:
\begin{align}\nonumber
\delta Z_{ij}^{fL}=&\frac{2}{m_{fi}^2-m_{fj}^2}\left[m_{fj}^2\Rea \Sigma_{ij}^{fL}(m_{fj}^2)
+m_{fi}m_{fj}\Rea \Sigma_{ij}^{fR}(m_{fj}^2)\right.\\&\left.\quad+m_{fi}\Rea \Sigma_{ij}^{fSL}(m_{fj}^2)
+m_{fj}\Rea \Sigma_{ij}^{fSR}(m_{fj}^2)-m_i\delta\tilde{\Mass}_{ij}-m_j\delta\tilde{\Mass}_{ji}^*\right]
\\\nonumber
\delta Z_{ij}^{fR}=&\frac{2}{m_{fi}^2-m_{fj}^2}\left[m_{fi}m_{fj}\Rea \Sigma_{ij}^{fL}(m_{fj}^2)
+m_{fj}^2\Rea \Sigma_{ij}^{fR}(m_{fj}^2)\right.\\&\left.\quad+m_{fj}\Rea \Sigma_{ij}^{fSL}(m_{fj}^2)
+m_{fi}\Rea \Sigma_{ij}^{fSR}(m_{fj}^2)-m_j\delta\tilde{\Mass}_{ij}-m_i\delta\tilde{\Mass}_{ji}^*\right]
\end{align}
However it turns out, that these additional contributions are cancelled by the contributions to $\delta \Nge,\delta U$
and $\delta V$, which also have to be calculated using the new wave-function renormalization constants. This implies that
the reduced number of physical parameters only has an impact on the calculated neutralino and chargino masses, but
not directly on the considered processes itself.

The fact, that one has finite shifts to the on-shell masses at the one-loop
level leads to a complication: to obtain infrared finite results one has
to take into account the emission of an additional photon as discussed in
Section~\ref{sect:realcorrections}. Using one-loop corrected masses in these processes
requires that one also has to use one-loop corrected masses in the virtual
corrections. Formally this results in differences which are of higher order
and indeed numerically these differences are rather small. However, the
use of one-loop masses in the virtual corrections leads to a spurious gauge
dependence which is of two-loop order and which eventually gets canceled
by taking into account two-loop corrections. We have checked numerically
in the examples
presented in Section~\ref{sec:results}
that indeed this residual gauge dependence is very small.

\subsection{Counterterm for the considered processes}

Plugging everything together we get from the tree-level interaction in Equation~\eqref{eq:treelevellagrangian}
the  counterterm  Lagrangian
\begin{align}
 \delta\mathcal{L}&\supset\overline{\tilde{\chi}_i^-}\gamma^\mu\left(P_L\left[\delta O_{Lij} + \frac{1}{2}O_{Lij}\delta Z_W 
 + \frac{1}{2}\sum_{k=1}^5 O_{Lik}\delta Z_{Lkj}^0 + \frac{1}{2}\sum_{k=1}^2\delta Z_{Lki}^{-*}O_{Lkj}\right]\right.\\\nonumber
&\left.\qquad+P_R\left[\delta O_{Rij} + \frac{1}{2}O_{Rij}\delta Z_W 
 + \frac{1}{2}\sum_{k=1}^5 O_{Rik}\delta Z_{Rkj}^0 + \frac{1}{2}\sum_{k=1}^2\delta Z_{Rki}^{-*}O_{Rkj}\right]\right)\tilde{\chi}_j^0 W^-_\mu
\end{align}
with
\begin{align}\nonumber
 \delta O_{Lij} = &- \left(\delta g \Nge^*_{j2}U_{i1} + g\delta\Nge^*_{j2}U_{i1}+g\Nge_{j2}^*\delta U_{i1}\right)\\
 & - \frac{1}{\sqrt{2}}\left(\delta g \Nge^*_{j3}U_{i2} + g\delta\Nge^*_{j3}U_{i2}+g\Nge_{j3}^*\delta U_{i2}\right)\\\nonumber
 \delta O_{Rij} = &- \left(\delta gV_{i1}^*\Nge_{j2}+g\delta V_{i1}^*\Nge_{j2} + gV_{i1}^*\delta \Nge_{j2}\right)\\
 & + \frac{1}{\sqrt{2}}\left(\delta gV_{i2}^*\Nge_{j4}+g\delta V_{i2}^*\Nge_{j4} + gV_{i2}^*\delta \Nge_{j4}\right)\qquad.
\end{align}
Therefore we define
\begin{align}
\delta A_{Lij}&=i\left(\delta O_{Lij} + \frac{1}{2}O_{Lij}\delta Z_W 
 + \frac{1}{2}\sum_{k=1}^5 O_{Lik}\delta Z_{Lkj}^0 + \frac{1}{2}\sum_{k=1}^2\delta Z_{Lki}^{-*}O_{Lkj}\right)\\
\delta A_{Rij}&=i\left(\delta O_{Rij} + \frac{1}{2}O_{Rij}\delta Z_W 
 + \frac{1}{2}\sum_{k=1}^5 O_{Rik}\delta Z_{Rkj}^0 + \frac{1}{2}\sum_{k=1}^2\delta Z_{Rki}^{-*}O_{Rkj}\right)
\end{align}
and obtain
\begin{align}
 M_{WV} = \overline{u}(p_1)\gamma^\mu\left(P_L \delta A_{Llj}+ P_R \delta A_{Rlj}\right)u(k)\epsilon_\mu^*(p_2)\qquad.
\end{align}

\subsection{Vertex corrections}

\begin{figure}[t]
\begin{center}
\includegraphics[width=0.75\textwidth]{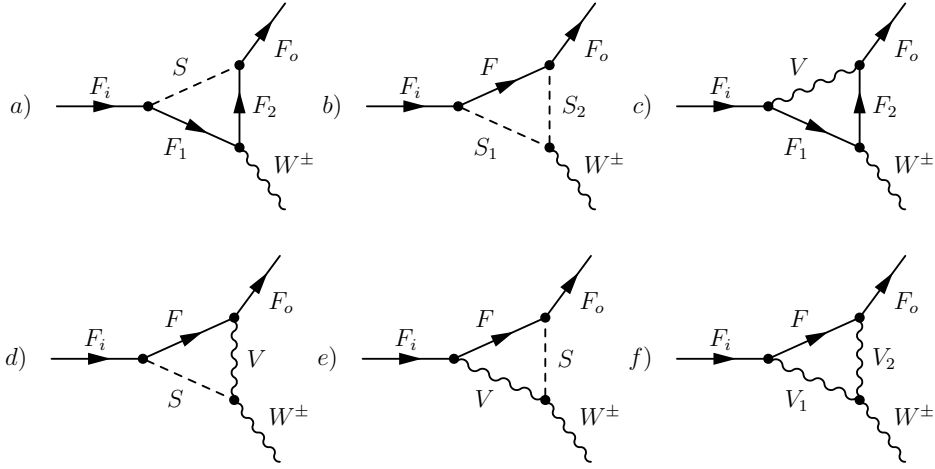}
\caption{Generic Vertex NLO corrections.}
\label{fig:vertexcorr}
\end{center}
\end{figure}

The next piece for the one-loop corrections are the vertex corrections.
In Figure~\ref{fig:vertexcorr} we depict the 6 generic contributions
to $M_V$.
In the Feynman diagrams a) and b) fermions and sfermions contribute as
well as charginos/neutralinos together with the neutral and charged
Higgs bosons. In diagrams c) and f) only charginos, neutralinos and
the vector bosons (including the photon) contribute whereas in
diagrams d) and e) there are in addition the charged and neutral Higgs
bosons as well as the Goldstone bosons. The individual contributions
from the diagrams in Figure~\ref{fig:vertexcorr} 
to the matrix element $M_V$ are given in \ref{sec:vertexcorr}
 for the 't~Hooft-Feynman gauge $\xi_V=1$. The general case $\xi_V\neq 1$ leads
 to rather lengthy formulas which are
 included in the program \prog \cite{Fortranonline}.
Putting everything together
we obtain for the width at the one-loop order including virtual corrections
only
\begin{equation}
\Gamma^{\mathrm{1L}} =\Gamma^{\mathrm{0}} +
\frac{\sqrt{\kappa(m^2_i,m^2_o,m^2_W)}}{16\pi m^3_i}
 \frac{1}{2}\sum_{pol} 2\mathrm{Re}\left( (M_{WV}M_T^*) +  (M_{V}M_T^*) \right)\qquad.
\end{equation}
This result is UV finite, gauge independent and also
not dependent on the renormalization scale $Q$,
which cancels between the contributions from $M_V$ and $M_{WV}$.
However we still have to address the IR infiniteness in the next section.

\subsection{Real corrections}
\label{sect:realcorrections}

\begin{figure}[t]
\begin{center}
\includegraphics[width=0.65\textwidth]{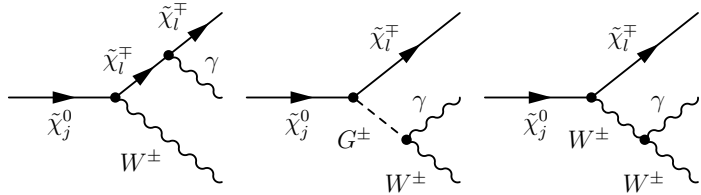}
\caption{Real corrections for $\tilde{\chi}^0_j\rightarrow \tilde{\chi}^\mp_l W^\pm \gamma$.}
\label{fig:realcorrtext1}
\end{center}
\vspace{-4mm}
\end{figure}

\begin{figure}[t]
\begin{center}
\includegraphics[width=0.65\textwidth]{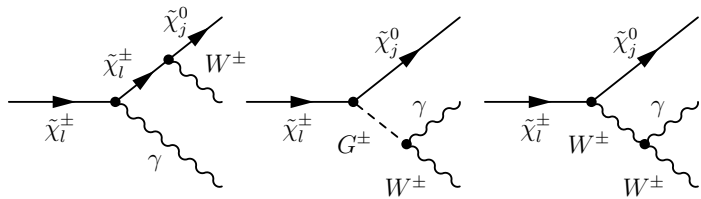}
\caption{Real corrections for $\tilde{\chi}^\pm_l\rightarrow \tilde{\chi}^0_j W^\pm \gamma$.}
\label{fig:realcorrtext2}
\end{center}
\end{figure}

Last but not least one has to take into account the real corrections
to cancel the infrared divergences occurring due the photon contributions
in the self-energy diagrams in Figures~\ref{fig:selfenergy1} and \ref{fig:selfenergy2}
or the vertex corrections in Figure~\ref{fig:vertexcorr}.
Technically this is achieved by introducing a small photon mass
for both, the virtual corrections and the real photon emission.
The corresponding graphs are shown in Figures~\ref{fig:realcorrtext1}
and \ref{fig:realcorrtext2}.
We calculate the real
bremsstrahlung corrections according to \cite{Denner:1991kt}
and collect the rather lengthy expression  in \ref{sec:photemission}.
We note, that (i) due to the presence of left and right couplings, see
\eqref{eq:treelevellagrangian}, these corrections do not factorize.
(ii) The gauge dependence of the graph with the Goldstone boson $G^+$ cancels
the one  with the virtual $W$-boson
implying that the real corrections are gauge independent.


Denoting the width due to real bremsstrahlung by $\Gamma^R$ we obtain
for the IR-finite decay width at next-to-leading order
\begin{equation}
\Gamma^{1} = \Gamma^{1L} + \Gamma^{R} \, .
\end{equation}

\section{Numerical results}
\label{sec:results}

Let us discuss a  general feature before  presenting our numerical results.
Inspecting the couplings given in \eqref{eq:treelevelcouplings} one sees
that only the wino and higgsino components of the neutralinos couple but
not the bino or the singlino one. This implies a small tree-level
width if the neutralino involved is either mainly bino or singlino which
potentially results in large corrections to the widths considered as we will
see below.

In the subsequent section we first present our results for 
the NMSSM benchmark points given in \cite{Djouadi:2008uw,Ellwanger:2008py}.
Afterwards we consider  cases where the neutralino
is nearly a  pure bino or singlino. There we demonstrate
the additional corrections not considered so far as important
as the  squark/quark contributions considered in the literature
\cite{Zhang:2001rd}.

\subsection{NLO masses and decay widths for benchmark scenarios}

In Table~\ref{tab:benchmarkpoints} we recall the most important parameters of the
benchmark scenarios  used in our analysis
based on \cite{Djouadi:2008uw} and \cite{Ellwanger:2008py}.
They are minimal supergravity  (mSUGRA) \cite{Djouadi:2008uw} and gauge mediated
SUSY breaking (GMSB)  \cite{Ellwanger:2008py} scenarios.
The parameter points can also be found within NMSSM-Tools \cite{NMSSMTools}.

\begin{table}[htb]
\begin{footnotesize}
\begin{center}
\begin{tabular}{| c p{2.5cm} || c | c | c |}
\hline
\multicolumn{2}{|c||}{} & mSUGRA $1$ & mSUGRA $3$ & mSUGRA $4$ \\ \hline\hline
GUT        & $M_0$ (in GeV)            & $180$ & $178$ & $780$ \\
scale      & $M_{1/2}$ (in GeV)        & $500$ & $500$ & $775$ \\ 
           & $A_{0}$ (in GeV)          & $-1500$ & $-1500$ & $-2250$ \\ \hline
SUSY       & $\tan\beta(m_Z)$          & $10$ & $10$ & $2.6$ \\ 
scale      & $\mu$ (in GeV)            & $969$ & $938$ & $-197$ \\ 
           & $\lambda,\kappa$          & $0.10,0.10$ & $0.40,0.30$ & $0.52,0.10$ \\ 
           & $A_\lambda,A_\kappa$ (in GeV)     & $-959,-1.6$ & $-616,-11$ & $-557,20$ \\   \hline\hline
\multicolumn{2}{|c||}{} & GMSB $1$ & GMSB $2$ & GMSB $5$ \\ \hline\hline
Mess.      & $M_{Mess}$ (in GeV)      & $10^{13}$ & $10^{13}$ & $5\cdot 10^{14}$ \\
scale      & $\Lambda$ (in GeV)       & $1.7\cdot 10^5$ & $1.7\cdot 10^5$ & $7.5\cdot 10^4$ \\ \hline
SUSY       & $\tan\beta(m_Z)$         & $8.5$ & $1.63$ & $50$ \\ 
scale      & $\mu$ (in GeV)           & $1404$ & $2351$ & $1376$ \\ 
           & $\lambda,\kappa$         & $0.020,0.004$ & $0.50,0.43$ & $0.010,-0.0007$ \\ 
           & $A_\lambda,A_\kappa$ (in GeV)     & $-52,-160$ & $-446,-2300$ & $118,4645$ \\  \hline
\end{tabular}
\end{center}
\end{footnotesize}
\caption{Most important parameters in the mSUGRA \cite{Djouadi:2008uw} and GMSB \cite{Ellwanger:2008py} benchmark scenarios.}
\label{tab:benchmarkpoints}
\end{table}

Note, that we calculate the soft breaking masses $m_{H_d}^2,m_{H_u}^2$
and in case of the NMSSM in addition $m_{S}^2$ from the tadpole equations, implying that
they are not input parameters in \prog. We do not give here the detailed soft SUSY breaking parameters
but refer for this to \cite{Djouadi:2008uw} and \cite{Ellwanger:2008py}
and/or the output of \prog discussed in the appendix.

\begin{table}[htb]
\begin{footnotesize}
\begin{center}
\begin{tabular}{| p{0.3cm} p{0.6cm} || c | c | c || c | c | c |}
\hline
\multicolumn{2}{|c||}{} & mSUGRA $1$ & mSUGRA $3$ & mSUGRA $4$ & GMSB $1$ & GMSB $2$ & GMSB $5$ \\ \hline\hline
$\tilde{\chi}^0_1:$ & $m$ & $210.79$ & $210.97$ & $89.08$ & $472.48$& $472.53$ & $203.30$\\
 & $m^{1L}$              & $210.61$ & $210.77$ & $91.45$& $472.38$& $472.39$ & $203.30$\\
 & $C$                   & $\tilde{B}$ & $\tilde{B}$ & $\tilde{S}$& $\tilde{B}$& $\tilde{B}$ & $\tilde{S}$\\
 \hline
$\tilde{\chi}^0_2:$ & $m$ & $387.18$ & $387.47$ & $215.38$ & $620.06$& $855.54$ & $496.87$\\ 
 & $m^{1L}$              & $387.10$ & $387.37$ & $215.58$& $620.06$& $855.53$ & $496.81$\\
 & $C$                   & $\tilde{W}$ & $\tilde{W}$ & $\tilde{H}$& $\tilde{S}$ & $\tilde{W}$ & $\tilde{B}$\\
  \hline
$\tilde{\chi}^0_3:$ & $m$ & $971.11$ & $942.27$ & $217.09$ & $854.13$& $2352.36$ & $899.60$\\ 
 & $m^{1L}$              & $971.75$ & $941.05$ & $217.51$& $854.43$& $2352.49$ & $899.98$\\ 
 & $C$                   & $\tilde{H}$ & $\tilde{H}$ & $\tilde{H}$& $\tilde{W}$& $\tilde{H}$ & $\tilde{W}$\\ 
 \hline
$\tilde{\chi}^0_4:$ & $m$ & $976.52$ & $943.16$ & $330.51$ & $1405.44$& $2355.92$ & $1377.67$\\ 
 & $m^{1L}$              & $975.14$ & $942.79$ & $331.01$& $1405.15$& $2354.84$ & $1377.45$\\ 
 & $C$                   & $\tilde{H}$ & $\tilde{H}$ & $\tilde{B}$& $\tilde{H}$& $\tilde{H}$ & $\tilde{H}$\\ 
 \hline
$\tilde{\chi}^0_5:$ & $m$ & $2101.57$ & $1421.70$ & $608.43$ & $1412.41$& $4062.82$& $1383.97$\\ 
 & $m^{1L}$              & $2101.57$ & $1421.67$ & $607.64$& $1411.46$& $4062.84$& $1383.04$\\ 
 & $C$                   & $\tilde{S}$ & $\tilde{S}$ & $\tilde{W}$& $\tilde{H}$& $\tilde{S}$& $\tilde{H}$\\ 
  \hline
$\tilde{\chi}^\pm_1:$ & $m$ & $387.16$ & $387.48$ & $201.36$& $854.11$& $855.53$ & $899.59$\\ 
 & $m^{1L}$                & $387.23$ & $387.53$ & $201.73$& $854.57$& $855.69$ & $900.14$\\ 
 & $C$                     & $\tilde{W}^\pm$ & $\tilde{W}^\pm$ & $\tilde{H}^\pm$& $\tilde{W}^\pm$& $ \tilde{W}^\pm$ & $\tilde{W}^\pm$\\ 
  \hline
$\tilde{\chi}^\pm_2:$ & $m$ & $977.07$ & $947.45$ & $608.40$ & $1412.29$& $2355.33$& $1384.24$\\ 
 & $m^{1L}$                & $976.69$ & $947.07$ & $607.80$& $1411.62$& $2355.06$& $1383.51$\\ 
 & $C$                     & $\tilde{H}^\pm$ & $\tilde{H}^\pm$ & $\tilde{W}^\pm$& $\tilde{H}^\pm$& $\tilde{H}^\pm$& $\tilde{H}^\pm$\\ 
  \hline
\end{tabular}
\end{center}
\end{footnotesize}
\caption{Neutralino and chargino masses $m$ on tree-level and $m^{1L}$ on one-loop-level in GeV and main particle character $C$
for the mSUGRA \cite{Djouadi:2008uw} and GMSB \cite{Ellwanger:2008py} benchmark scenarios.}
\label{tab:benchmarkmasses}
\end{table}

\begin{table}[]
\begin{center}
\begin{tabular}{| c || c | c | c || c | c || c |}
\hline

Sc. & Decay & $\Gamma^0$ (in GeV) & $\Gamma^1$ (in GeV) & $\delta_{1(\tilde{q},q)}$ & $\delta_2$ & $\delta_{1+2}$\\\hline\hline
 \multirow{7}{*}{$1$} 
  & $\tilde{\chi}^0_3\rightarrow \tilde{\chi}^-_1 W^+$  & $2.153$ & $2.234$ & $1.8\%$ & $2.0\%$ & $3.8\%$ \\\cline{2-7}
  & $\tilde{\chi}^0_4\rightarrow \tilde{\chi}^-_1 W^+$  & $2.181$ & $2.256$ & $1.8\%$ & $1.6\%$ & $3.4\%$ \\\cline{2-7}
  & $\tilde{\chi}^0_5\rightarrow \tilde{\chi}^-_1 W^+$  & $3.206\cdot 10^{-3}$ & $2.897\cdot 10^{-3}$ & $-2.6\%$ & $-7.0\%$ & $-9.6\%$ \\\cline{2-7}
  & $\tilde{\chi}^0_5\rightarrow \tilde{\chi}^-_2 W^+$  & $1.542\cdot 10^{-1}$ & $1.521\cdot 10^{-1}$ & $-1.9\%$ & $0.5\%$& $-1.4\%$\\\cline{2-7}
  & $\tilde{\chi}^-_1\rightarrow \tilde{\chi}^0_1 W^-$  & $2.575\cdot 10^{-3}$ & $2.561\cdot 10^{-3}$ & $0.8\%$ & $-1.3\%$ & $-0.5\%$\\\cline{2-7}
  & $\tilde{\chi}^-_2\rightarrow \tilde{\chi}^0_1 W^-$  & $5.860\cdot 10^{-1}$ & $5.766\cdot 10^{-1}$ & $0.2\%$ & $-1.8\%$ & $-1.6\%$ \\\cline{2-7}
  & $\tilde{\chi}^-_2\rightarrow \tilde{\chi}^0_2 W^-$  & $2.201$ & $2.222$ & $-0.3\%$ & $1.2\%$ & $0.9\%$\\\hline\hline

 \multirow{7}{*}{$3$} 
  & $\tilde{\chi}^0_3\rightarrow \tilde{\chi}^-_1 W^+$  & $2.085$ & $2.153$ & $1.8\%$ & $1.5\%$ & $3.3\%$ \\\cline{2-7}
  & $\tilde{\chi}^0_4\rightarrow \tilde{\chi}^-_1 W^+$  & $2.121$ & $2.181$ & $1.8\%$ & $1.0\%$ & $2.8\%$ \\\cline{2-7}
  & $\tilde{\chi}^0_5\rightarrow \tilde{\chi}^-_1 W^+$  & $2.937\cdot 10^{-2}$ & $2.755\cdot 10^{-2}$ & $-1.7\%$ & $-4.5\%$ & $-6.2\%$ \\\cline{2-7}
  & $\tilde{\chi}^0_5\rightarrow \tilde{\chi}^-_2 W^+$  & $1.302$ & $1.352$ & $-0.6\%$ & $4.5\%$ & $3.9\%$ \\\cline{2-7}
  & $\tilde{\chi}^-_1\rightarrow \tilde{\chi}^0_1 W^-$  & $2.951\cdot 10^{-3}$ & $2.910\cdot 10^{-1}$ & $0.7\%$ & $-2.1\%$ & $-1.4\%$ \\\cline{2-7}
  & $\tilde{\chi}^-_2\rightarrow \tilde{\chi}^0_1 W^-$  & $5.684\cdot 10^{-1}$ & $5.552\cdot 10^{-1}$ & $0.2\%$ & $-2.5\%$ & $-2.3\%$ \\\cline{2-7}
  & $\tilde{\chi}^-_2\rightarrow \tilde{\chi}^0_2 W^-$  & $2.115$ & $2.141$ & $0.6\%$ & $0.6\%$ & $1.2\%$ \\\hline\hline

 \multirow{7}{*}{$4$} 
  & $\tilde{\chi}^0_4\rightarrow \tilde{\chi}^-_1 W^+$  & $4.719\cdot 10^{-2}$ & $5.080\cdot 10^{-2}$ & $-0.3\%$ & $7.9\%$ & $7.6\%$ \\\cline{2-7}
  & $\tilde{\chi}^0_5\rightarrow \tilde{\chi}^-_1 W^+$  & $7.442\cdot 10^{-1}$ & $7.288\cdot 10^{-1}$ & $0.2\%$ & $-2.3\%$ & $-2.1\%$ \\\cline{2-7}
  & $\tilde{\chi}^-_1\rightarrow \tilde{\chi}^0_1 W^-$  & $1.623\cdot 10^{-1}$ & $1.650\cdot 10^{-1}$ & $-0.9\%$ & $2.5\%$ & $1.6\%$  \\\cline{2-7} 
  & $\tilde{\chi}^-_2\rightarrow \tilde{\chi}^0_1 W^-$  & $2.357\cdot 10^{-1}$ & $2.291\cdot 10^{-1}$ & $0.2\%$ & $-3.0\%$ & $-2.8\%$ \\\cline{2-7}
  & $\tilde{\chi}^-_2\rightarrow \tilde{\chi}^0_2 W^-$  & $5.758\cdot 10^{-1}$ & $5.586\cdot 10^{-1}$ & $0.2\%$ & $-3.2\%$ & $-3.0\%$ \\\cline{2-7}
  & $\tilde{\chi}^-_2\rightarrow \tilde{\chi}^0_3 W^-$  & $6.024\cdot 10^{-1}$ & $5.875\cdot 10^{-1}$ & $0.2\%$ & $-2.7\%$ & $-2.5\%$ \\\cline{2-7}
  & $\tilde{\chi}^-_2\rightarrow \tilde{\chi}^0_4 W^-$  & $7.007\cdot 10^{-2}$ & $6.963\cdot 10^{-2}$ & $-0.3\%$ & $-0.3\%$ & $-0.6\%$\\\hline
\end{tabular}
\end{center}
\caption{NLO corrections for the mSUGRA benchmark scenarios; 
 $\delta$ is defined in \eqref{eq:corrfactor}.}
\label{tab:benchmarkNMSSMmSUGRA}
\end{table}
\begin{table}[]
\begin{center}
\begin{tabular}{| c || c | c | c || c | c || c |}
\hline
Sc. & Decay & $\Gamma^0$ (in GeV) & $\Gamma^1$ (in GeV) & $\delta_{1(\tilde{q},q)}$ & $\delta_2$ & $\delta_{1+2}$\\\hline\hline
 \multirow{7}{*}{$1$}
  & $\tilde{\chi}^0_4\rightarrow \tilde{\chi}^-_1 W^+$  & $2.891$ & $3.068$ & $-0.2\%$ & $6.3\%$ & $6.1\%$ \\\cline{2-7}
  & $\tilde{\chi}^0_5\rightarrow \tilde{\chi}^-_1 W^+$  & $2.943$ & $3.114$ & $-0.1\%$ & $5.9\%$ & $5.8\%$ \\\cline{2-7}
  & $\tilde{\chi}^-_1\rightarrow \tilde{\chi}^0_1 W^-$  & $1.027\cdot 10^{-2}$ & $1.028\cdot 10^{-1}$ & $0.0\%$ & $0.1\%$ & $0.1\%$ \\\cline{2-7}
  & $\tilde{\chi}^-_1\rightarrow \tilde{\chi}^0_2 W^-$  & $8.907\cdot 10^{-5}$ & $8.942\cdot 10^{-5}$ & $-0.1\%$ & $0.5\%$ & $0.4\%$ \\\cline{2-7}
  & $\tilde{\chi}^-_2\rightarrow \tilde{\chi}^0_1 W^-$  & $8.941\cdot 10^{-1}$ & $8.795\cdot 10^{-1}$ & $0.2\%$ & $-1.8\%$ & $-1.6\%$ \\\cline{2-7}
  & $\tilde{\chi}^-_2\rightarrow \tilde{\chi}^0_2 W^-$  & $3.787\cdot 10^{-3}$ & $3.697\cdot 10^{-3}$ & $-0.5\%$ & $-1.9\%$ & $-2.4\%$ \\\cline{2-7}
  & $\tilde{\chi}^-_2\rightarrow \tilde{\chi}^0_3 W^-$  & $2.962$ & $3.126$ & $-0.1\%$ & $5.6\%$ & $5.5\%$ \\\hline\hline

 \multirow{7}{*}{$2$}
  & $\tilde{\chi}^0_3\rightarrow \tilde{\chi}^-_1 W^+$  & $7.431$ & $7.241$ & $0.2\%$ & $-2.8\%$ & $-2.6\%$ \\\cline{2-7}
  & $\tilde{\chi}^0_4\rightarrow \tilde{\chi}^-_1 W^+$  & $7.442$ & $7.254$ & $0.4\%$ & $-2.9\%$ & $-2.5\%$ \\\cline{2-7}
  & $\tilde{\chi}^0_5\rightarrow \tilde{\chi}^-_1 W^+$  & $1.164\cdot 10^{-2}$ & $9.868\cdot 10^{-3}$ & $-3.4\%$ & $-11.8\%$ & $-15.2\%$ \\\cline{2-7}
  & $\tilde{\chi}^0_5\rightarrow \tilde{\chi}^-_2 W^+$  & $1.890$ & $1.851$ & $-2.0\%$ & $-0.1\%$ & $-2.1\%$ \\\cline{2-7}
  & $\tilde{\chi}^-_1\rightarrow \tilde{\chi}^0_1 W^-$  & $6.150\cdot 10^{-3}$ & $6.135\cdot 10^{-3}$ & $0.0\%$ & $-0.2\%$ & $-0.2\%$\\\cline{2-7}
  & $\tilde{\chi}^-_2\rightarrow \tilde{\chi}^0_1 W^-$  & $1.807$ & $1.708$ & $-0.7\%$ & $-4.8\%$ & $-5.5\%$ \\\cline{2-7}
  & $\tilde{\chi}^-_2\rightarrow \tilde{\chi}^0_2 W^-$  & $7.491$ & $7.235$ & $0.3\%$ & $-3.7\%$ & $-3.4\%$ \\\hline\hline

 \multirow{7}{*}{$5$}
  & $\tilde{\chi}^0_4\rightarrow \tilde{\chi}^-_1 W^+$  & $2.283$ & $2.439$ & $-0.3\%$ & $7.1\%$ & $6.8\%$ \\\cline{2-7}
  & $\tilde{\chi}^0_5\rightarrow \tilde{\chi}^-_1 W^+$  & $2.333$ & $2.485$ & $-0.1\%$ & $6.6\%$ & $6.5\%$ \\\cline{2-7}
  & $\tilde{\chi}^-_1\rightarrow \tilde{\chi}^0_1 W^-$  & $1.462\cdot 10^{-5}$ & $1.453\cdot 10^{-5}$ & $-0.3\%$ & $-0.3\%$ & $-0.6\%$\\\cline{2-7} 
  & $\tilde{\chi}^-_1\rightarrow \tilde{\chi}^0_2 W^-$  & $8.424\cdot 10^{-3}$ & $8.407\cdot 10^{-3}$ & $-0.1\%$ & $-0.1\%$ & $-0.2\%$ \\\cline{2-7}
  & $\tilde{\chi}^-_2\rightarrow \tilde{\chi}^0_1 W^-$  & $1.279\cdot 10^{-3}$ & $1.242\cdot 10^{-3}$ & $-0.6\%$ & $-2.2\%$ & $-2.9\%$ \\\cline{2-7}
  & $\tilde{\chi}^-_2\rightarrow \tilde{\chi}^0_2 W^-$  & $7.898\cdot 10^{-1}$ & $7.775\cdot 10^{-1}$ & $-0.1\%$ & $-1.5\%$ & $-1.6\%$ \\\cline{2-7}
  & $\tilde{\chi}^-_2\rightarrow \tilde{\chi}^0_3 W^-$  & $2.339$ & $2.486$ & $-0.1\%$ & $6.4\%$ & $6.3\%$ \\\hline
\end{tabular}
\end{center}
\caption{NLO corrections for the GMSB benchmark scenarios; 
 $\delta$ is defined in \eqref{eq:corrfactor}.}
\label{tab:benchmarkNMSSMGMSB}
\end{table}
We show the corresponding masses $m$, one-loop masses $m^{1L}$ and particle characters $C$
in Table~\ref{tab:benchmarkmasses}.
In Tables~\ref{tab:benchmarkNMSSMmSUGRA} and \ref{tab:benchmarkNMSSMGMSB} 
we present our results for the decay widths in the
mSUGRA and GMSB scenarios, respectively.
Tables \ref{tab:benchmarkNMSSMmSUGRA} and \ref{tab:benchmarkNMSSMGMSB}
contain beside the tree-level and one-loop corrected widths the correction factor
\begin{align}
 \delta=\frac{\Gamma^1-\Gamma^0}{\Gamma^0}\qquad,
\label{eq:corrfactor}
\end{align}
which is also split in the parts $\delta_1=\delta_{1(\tilde{q},q)}$ due to
squark and quark corrections and
$\delta_2$ containing the other contributions, which includes the hard photon emission
for comparison with \cite{Zhang:2001rd}.
As already stated, the mass corrections $m\rightarrow m^{1L}$ presented
in Table~\ref{tab:benchmarkmasses} are small, in most
cases in the per-mil range. Only the light Singlino in the mSUGRA $4$ scenario gets a large
correction of $2.6$\% from squark and quark contributions.

As expected the widths are larger in
case the neutralino involved has either large wino and/or higgsino components.
Note, that from \eqref{eq:treelevelcouplings} follows that the $W$ boson
couples either to a wino-wino or a higgsino-higgsino combination. This explains
several of at first glance surprising features, e.g.~the fact that in mSUGRA scenarios
$1$ and $3$ the width  
$\Gamma(\tilde \chi^0_5 \rightarrow \tilde \chi^+_2 W^- )$ is larger
than  $\Gamma(\tilde \chi^0_5 \rightarrow W^- \tilde \chi^+_1)$ 
 despite the smaller phase space.
Another surprise might be the difference in $\delta_2$ for the decay
$\tilde{\chi}_2^-\rightarrow \tilde{\chi}_1^0 W^-$
in  scenarios mSUGRA $1$ and mSUGRA $3$. However this can be understood
from the differences in the  Higgs sector.
We see that the corrections are in general in the order of $1$-$3\%$ but can easily
go up to $10\%$. Note, that depending on the parameters the corrections can have
both signs.

\subsection{NLO corrections in case of bino and singlino like neutralinos}

It follows from \eqref{eq:treelevelcouplings} that the partial width into a
$W$-boson vanishes in the limit that the neutralino involved is either a pure
bino or a pure singlino. This is the main reason that in Tables
 \ref{tab:benchmarkNMSSMmSUGRA} and \ref{tab:benchmarkNMSSMGMSB} 
the processes containing states which are to a large extent bino or singlino
have small widths. However, even in the limit of pure states the corresponding
couplings are induced at the one-loop level. In this section we investigate
this in more detail. We consider here a wide mass range and are aware that
neutralinos with masses above 1 TeV will hardly be produced at LHC and will
therefore most likely require a multi-TeV lepton collider such as CLIC.
All the plots and decay widths are based on tree-level
masses $m_{\tilde{\chi}_i^{\pm 0}}$ for neutralinos and charginos. 
The corresponding plots with the 
one-loop corrected masses differ only slightly and the differences
are hardly visible. 

\subsubsection{Bino decays}

\begin{figure}[t]
\includegraphics[width=0.5\textwidth]{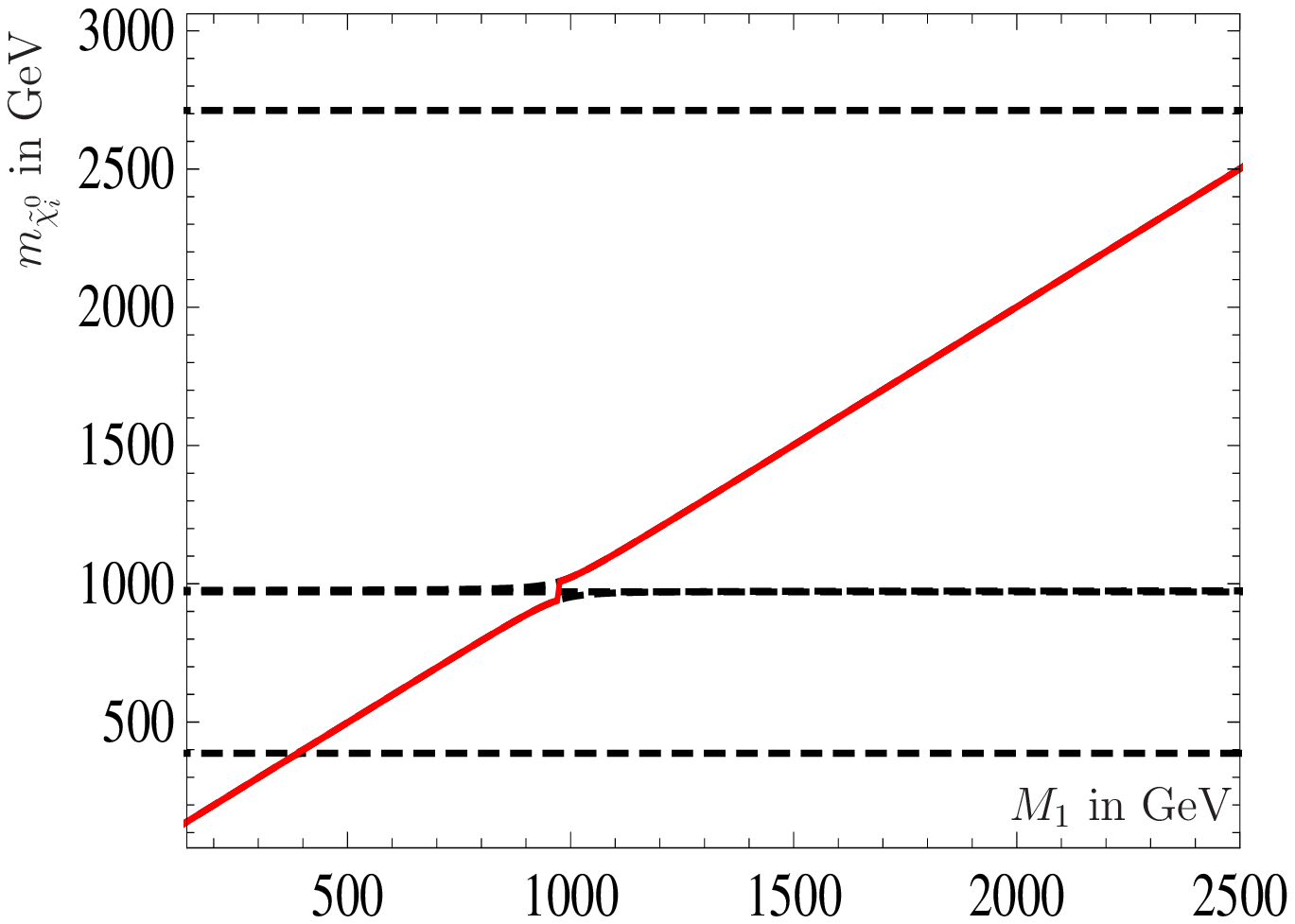}
\hfill
\includegraphics[width=0.5\textwidth]{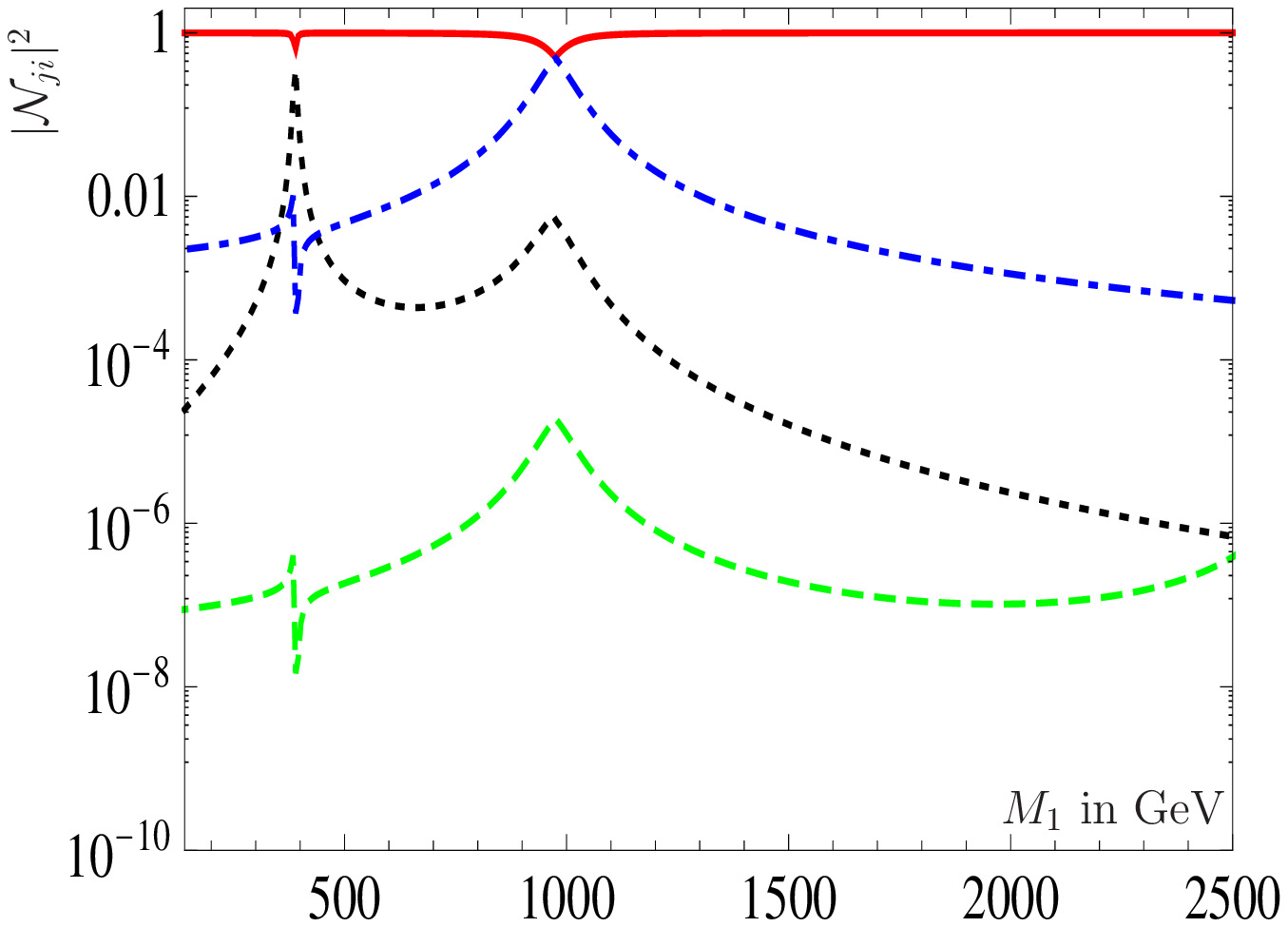}
\caption{a) (left): Neutralino masses  as a function
of $M_1$ and the other parameters according to mSUGRA $1$ apart from
$\kappa=0.14$. The red solid line marks $\tilde B$ whereas the other states
are shown with black dashed lines.
b) (right): Particle character for the bino state $\tilde{B}$ as a function of $M_1$:
red (solid): Bino character $|\Nge_{j1}|^2$,
blue (dot-dashed): Higgsino character $|\Nge_{j3}|^2+|\Nge_{j4}|^2$,
black (dotted): Wino character $|\Nge_{j2}|^2$,
green (dashed): Singlino character $|\Nge_{j5}|^2$.}
\label{fig:binodecay}
\end{figure}

We define a state denoted by $\tilde B$ to be bino-like if  $|\Nge_{j1}|^2>0.5$.
We take benchmark point mSUGRA $1$ and vary the gaugino mass $M_1$ for our
subsequent numerical investigations. Moreover, we shift $\kappa$ from $0.11$ to $0.14$
to disentangle different effects and to simplify the discussion. 
The corresponding neutralino mass spectrum  is
shown in Figure~\ref{fig:binodecay} a) as a function of $M_1$ and
the particle character of the corresponding mainly bino-like state $\tilde{B}$ is shown in
 Figure~\ref{fig:binodecay} b). Note that at the various crossings in plot a) the index
of the corresponding neutralino mass eigenstate changes. 

\begin{figure}[h]
\includegraphics[width=0.5\textwidth]{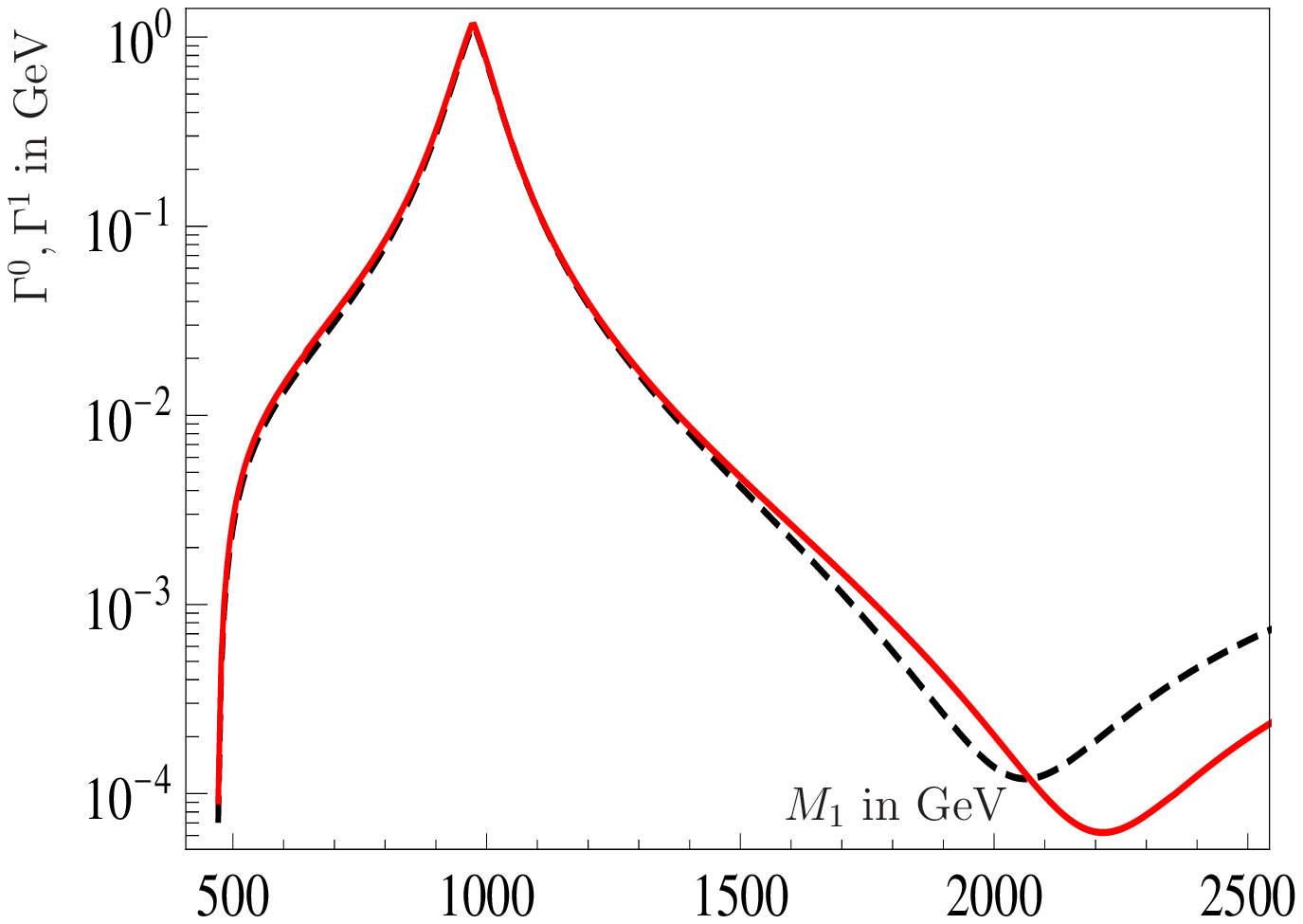}
\hfill
\includegraphics[width=0.5\textwidth]{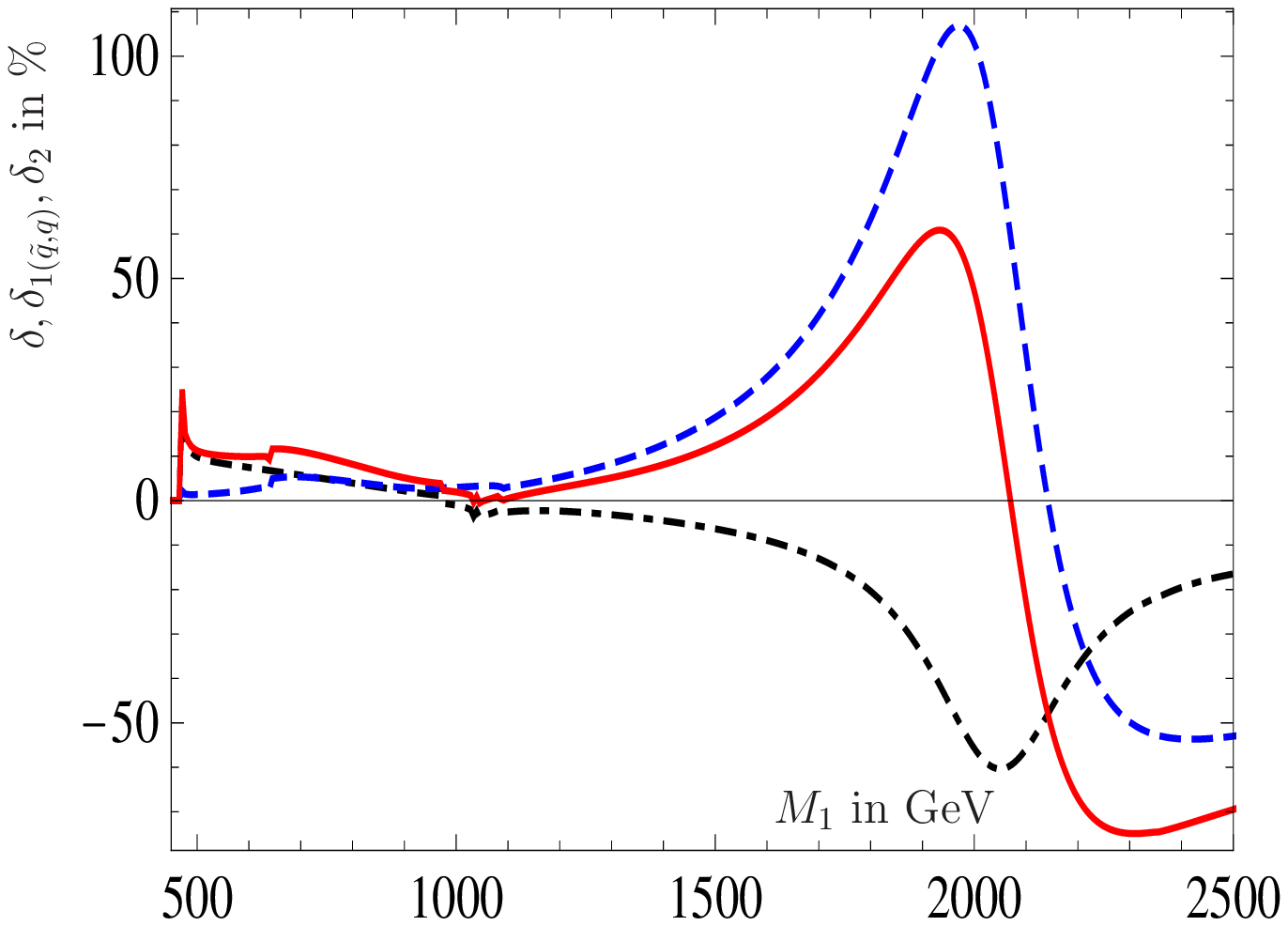}
\caption{a) (left): LO (black, dashed) and NLO (red, solid) decay widths
for $\tilde{B}\rightarrow \tilde{\chi}_{1}^- W^+$ as a function
of $M_1$ for the spectrum of Figure~\ref{fig:binodecay},
 b) (right): Correction factor $\delta$ in \% defined in \eqref{eq:corrfactor}
for $\tilde{B}\rightarrow \tilde{\chi}_{1}^- W^+$
as a function of $M_1$:
blue (dashed): Squark and quark contributions,
black (dot-dashed): Other sectors,
red (solid): Full correction.}
\label{fig:binodecay2}
\end{figure}

Figure~\ref{fig:binodecay2} a) shows the LO and NLO decay width of 
$\tilde{B}\rightarrow \tilde{\chi}_{1}^- W^+$
as a function of $M_1$. At $M_1 \simeq 1$~TeV the level crossing with the
higgsino like states occurs giving rise to the observed rise of the width
with $M_1$ and the subsequent decrease. In the later case the decrease is
strengthened by a negative interference of the higgsino and wino parts.
Moreover we note that here and also in the corresponding figures below
a Coloumb singularity occurs close to the kinematical threshold, e.g.~close
to $m_{\tilde B} = m_{\tilde{\chi}^\pm_{1}} + m_W$, which has to be resumed.
As this has not been done, we start our plots slightly above this region.

The relative size of the corrections are shown in Figure~\ref{fig:binodecay2} b)
where we also compare the third generation squark/quark contribution with the
additional ones. The kinks correspond to the level crossings in
 Figure~\ref{fig:binodecay}. Note that both parts of the correction can
be of equal importance and that in some regions of the parameter space they cancel partly
each other whereas there are also regions where they point in the same direction.
It is remarkable that the loop induced corrections can be of the same order
of magnitude as the tree-level widths. This however does not imply a break-down
of perturbation theory but is a consequence that in the limit of a pure bino
the tree-level coupling vanishes but the one-loop induced one is non-zero. 

\begin{figure}[H]
\includegraphics[width=0.5\textwidth]{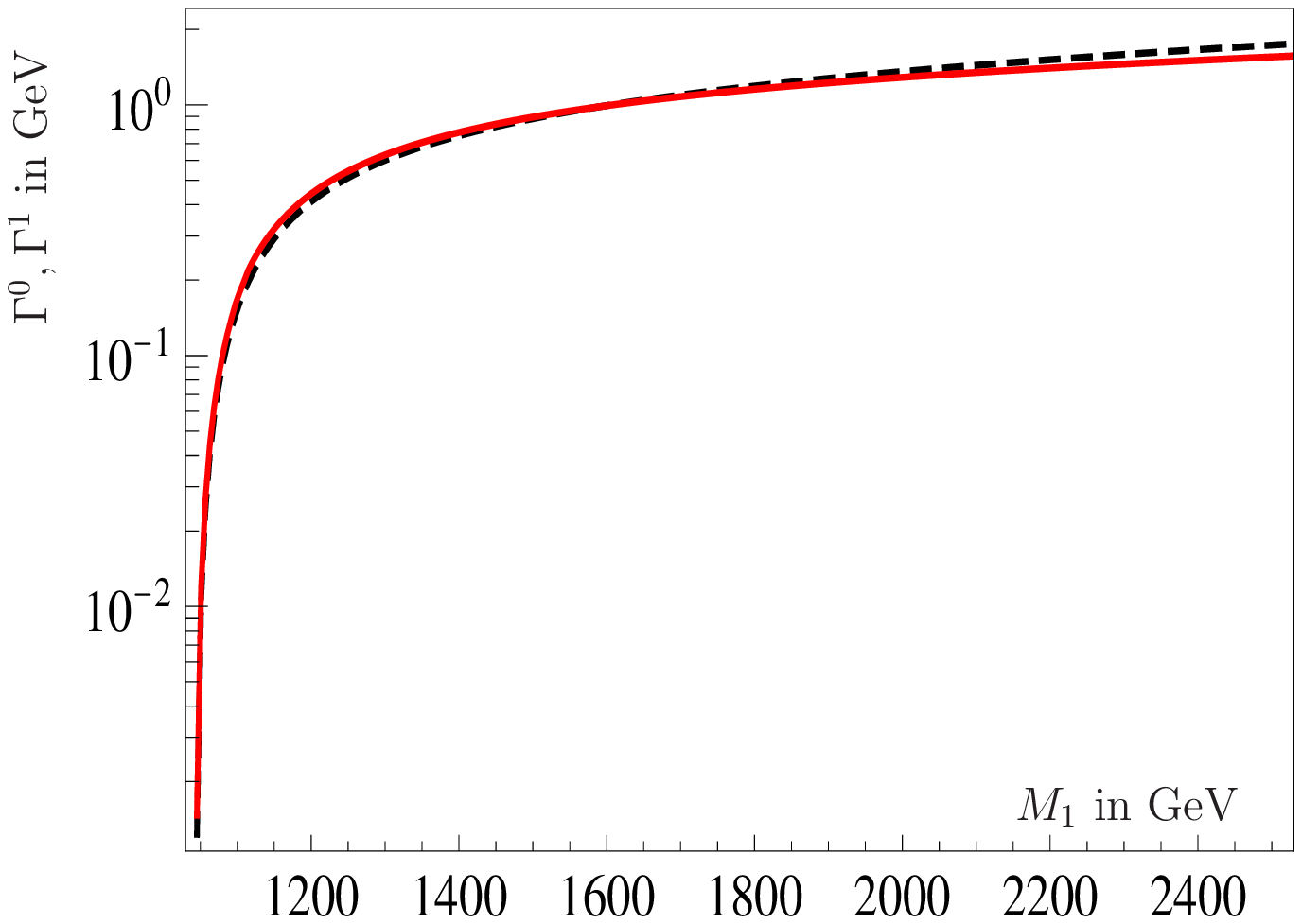}
\hfill
\includegraphics[width=0.5\textwidth]{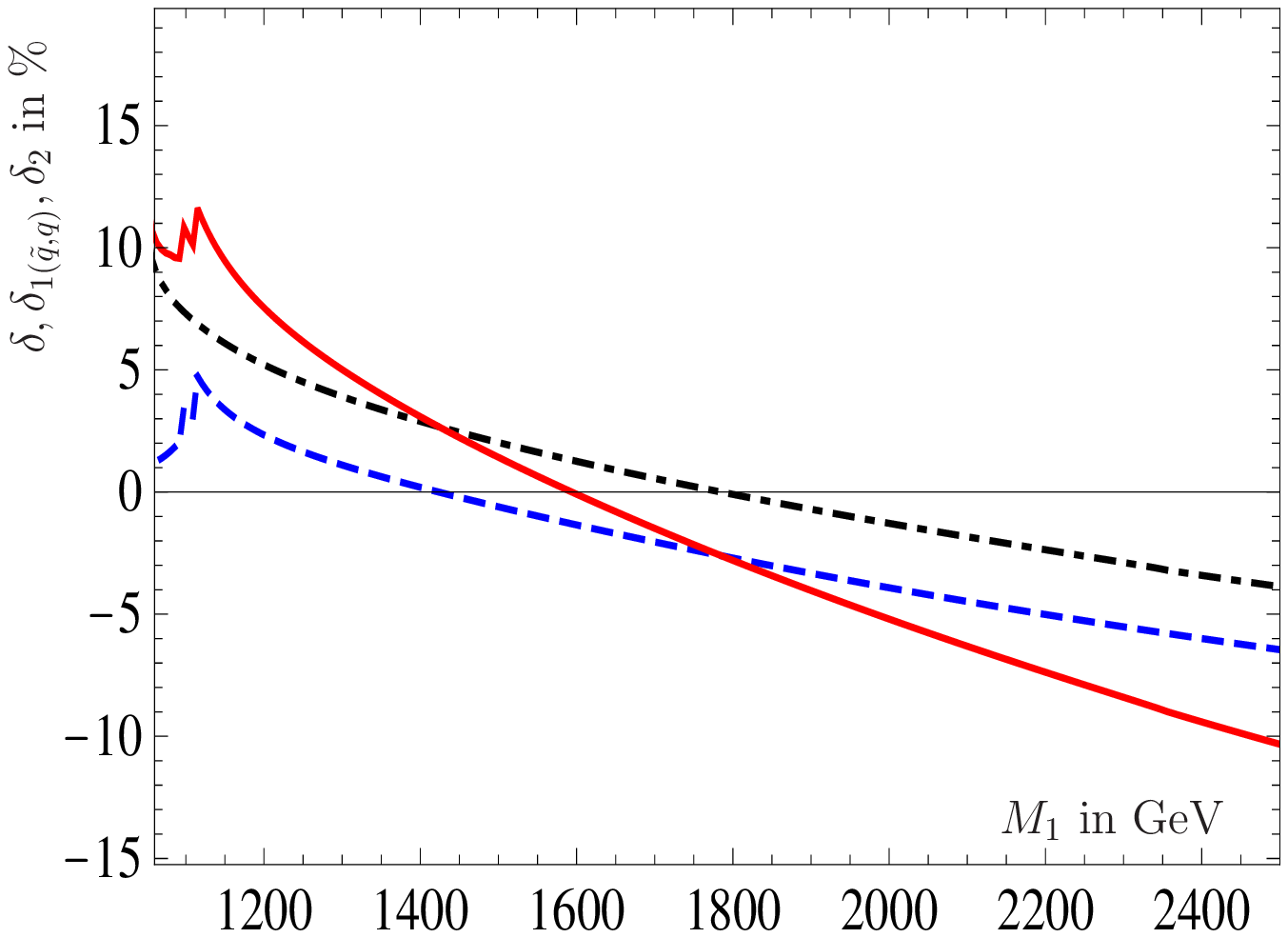}
\caption{a) (left): LO (black, dashed) and NLO (red, solid) decay widths
 for $\tilde{B}\rightarrow \tilde{\chi}_{2}^- W^+$ as a function
of $M_1$ for the spectrum of Figure~\ref{fig:binodecay},
 b) (right): Correction factor $\delta$ in \% defined in
 \eqref{eq:corrfactor} for $\tilde{B}\rightarrow \tilde{\chi}_{2}^- W^+$
as a function of $M_1$:
blue (dashed): Squark and quark contributions,
black (dot-dashed): Other sectors,
red (solid): Full correction.}
\label{fig:binodecay3}
\end{figure}

Figure~\ref{fig:binodecay3} a) shows the  widths for the decay
$\tilde{B}\rightarrow \tilde{\chi}_{2}^- W^+$
as a function of $M_1$. In contrast to the decay into $\tilde{\chi}_{1}^-$ there is
a positive interference  of the wino-wino and higgsino-higgsino components in the LO
couplings  \eqref{eq:treelevelcouplings}. Note, that the decrease of the couplings for
increasing $M_1$ is compensated by a factor $(M_1/m_W)^2$, see
Equations~\eqref{eq:neutdecaytree}--\eqref{eq:kaellenfunction} leading to a slight
increase of the width with increasing $M_1$. Also in this case the corrections
can be sizable amounting up to about $15$ percent.

\subsubsection{Singlino decays}

The  case of a singlino-like neutralino $\tilde{S}$, defined by
$|N_{j5}|^2 > 0.5$, shows similar
features to the case of a bino-like neutralino. However, there is one
important difference: In the limit of a pure singlino it only couples to
the doublet Higgs/higgsino states and the singlet Higgs boson. Hence
one expects that the squark/quark contributions to be of less importance
compared to the bino case.

\begin{figure}[H]
\includegraphics[width=0.5\textwidth]{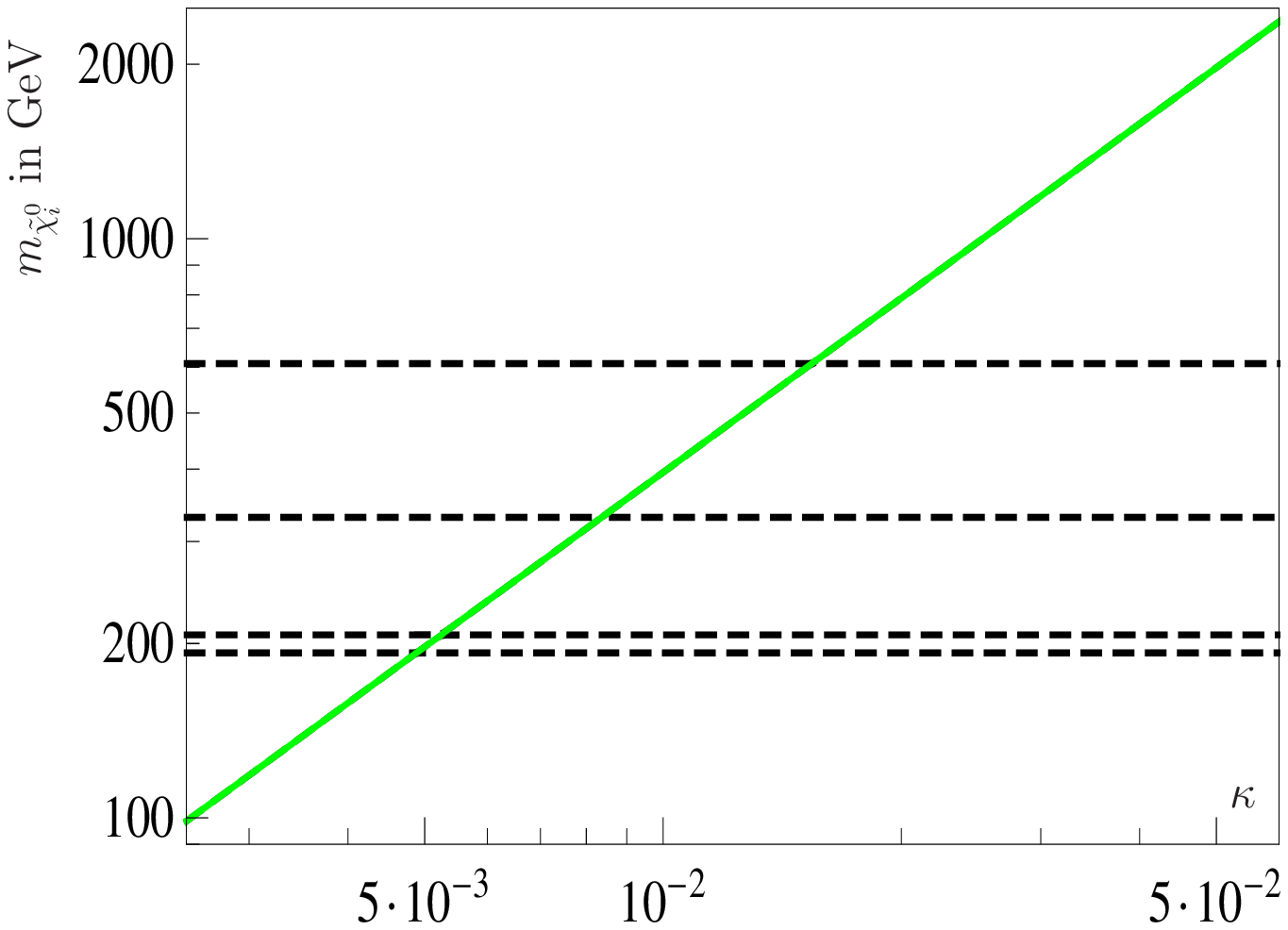}
\hfill
\includegraphics[width=0.5\textwidth]{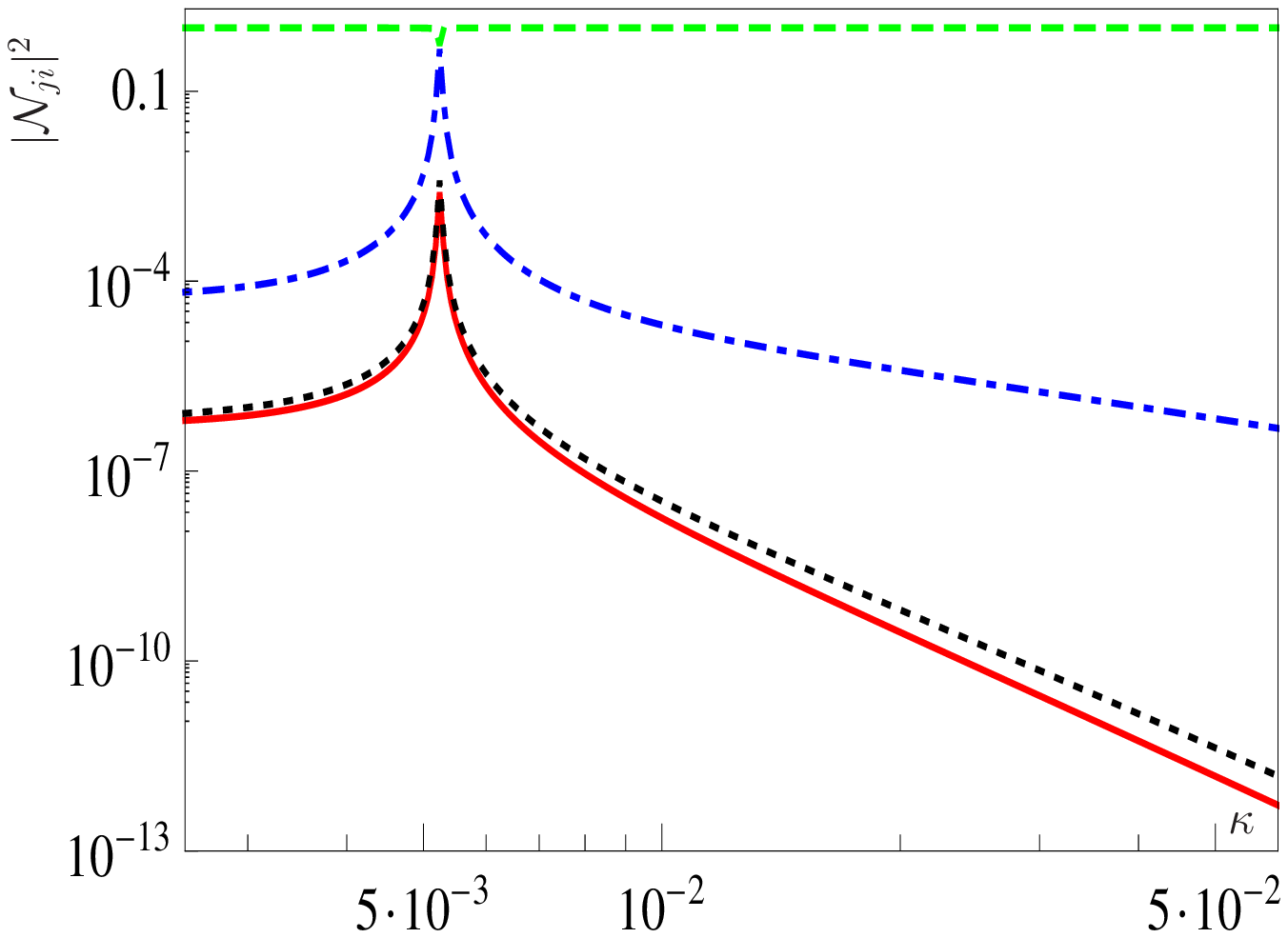}
\caption{a) (left): Neutralino masses  as a function
of $\kappa$ and the other parameters according to mSUGRA $4$ apart from
$\lambda=0.01$. The green solid line marks $\tilde S$ whereas the other states
are shown with black dashed lines.
 b) (right): Particle character for the singlino
state $\tilde{S}$ as a function of $\kappa$:
red (solid): Bino character $|\Nge_{j1}|^2$,
blue (dot-dashed): Higgsino character $|\Nge_{j3}|^2+|\Nge_{j4}|^2$,
black (dotted): Wino character $|\Nge_{j2}|^2$,
green (dashed): Singlino character $|\Nge_{j5}|^2$.}
\label{fig:singlinodecay}
\vspace{-4mm}
\end{figure}
\begin{figure}[H]
\includegraphics[width=0.5\textwidth]{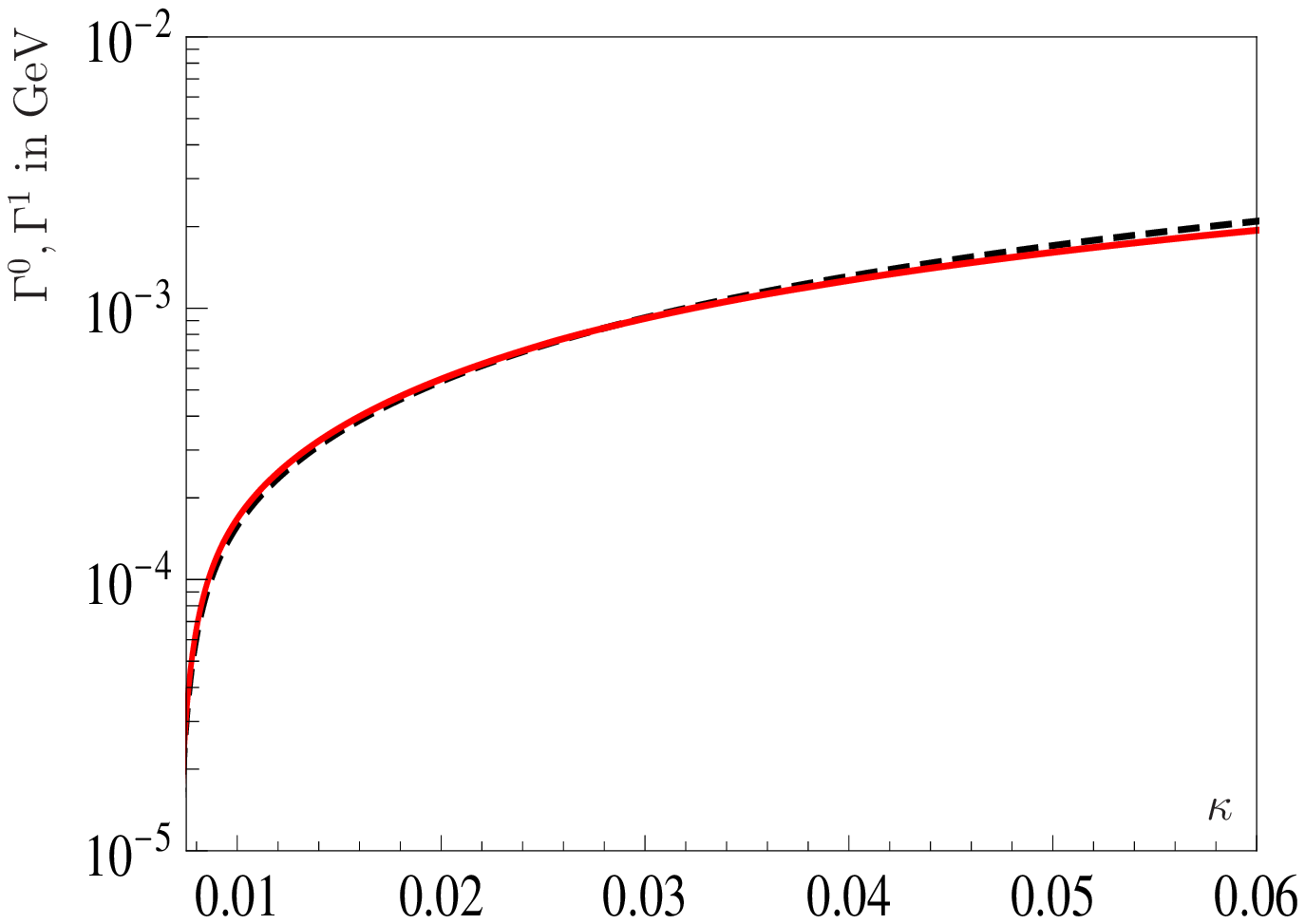}
\hfill
\includegraphics[width=0.5\textwidth]{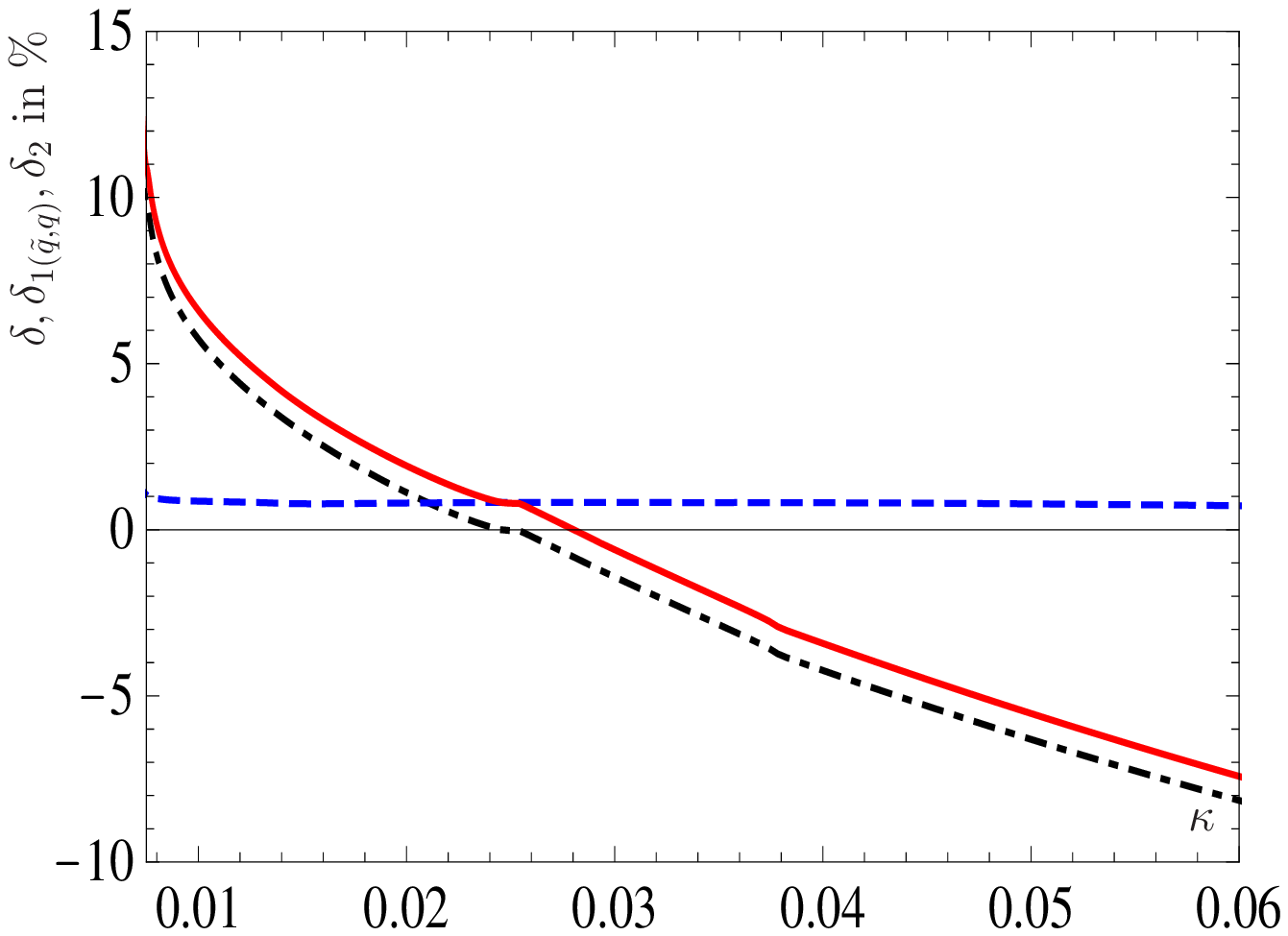}
\caption{a) (left): LO (black, dashed) and NLO (red, solid) decay widths
 for $\tilde{S}\rightarrow \tilde{\chi}_{1}^- W^+$ as a function
of $\kappa$ for the spectrum of Figure~\ref{fig:singlinodecay},
 b) (right): Correction factor $\delta$ in \% defined in
 \eqref{eq:corrfactor} for $\tilde{S}\rightarrow \tilde{\chi}_{1}^- W^+$
as a function of $\kappa$:
blue (dashed): Squark and quark contributions,
black (dot-dashed): Other sectors,
red (solid): Full correction.}
\label{fig:singlinodecay2}
\end{figure}
\begin{figure}[H]
\includegraphics[width=0.5\textwidth]{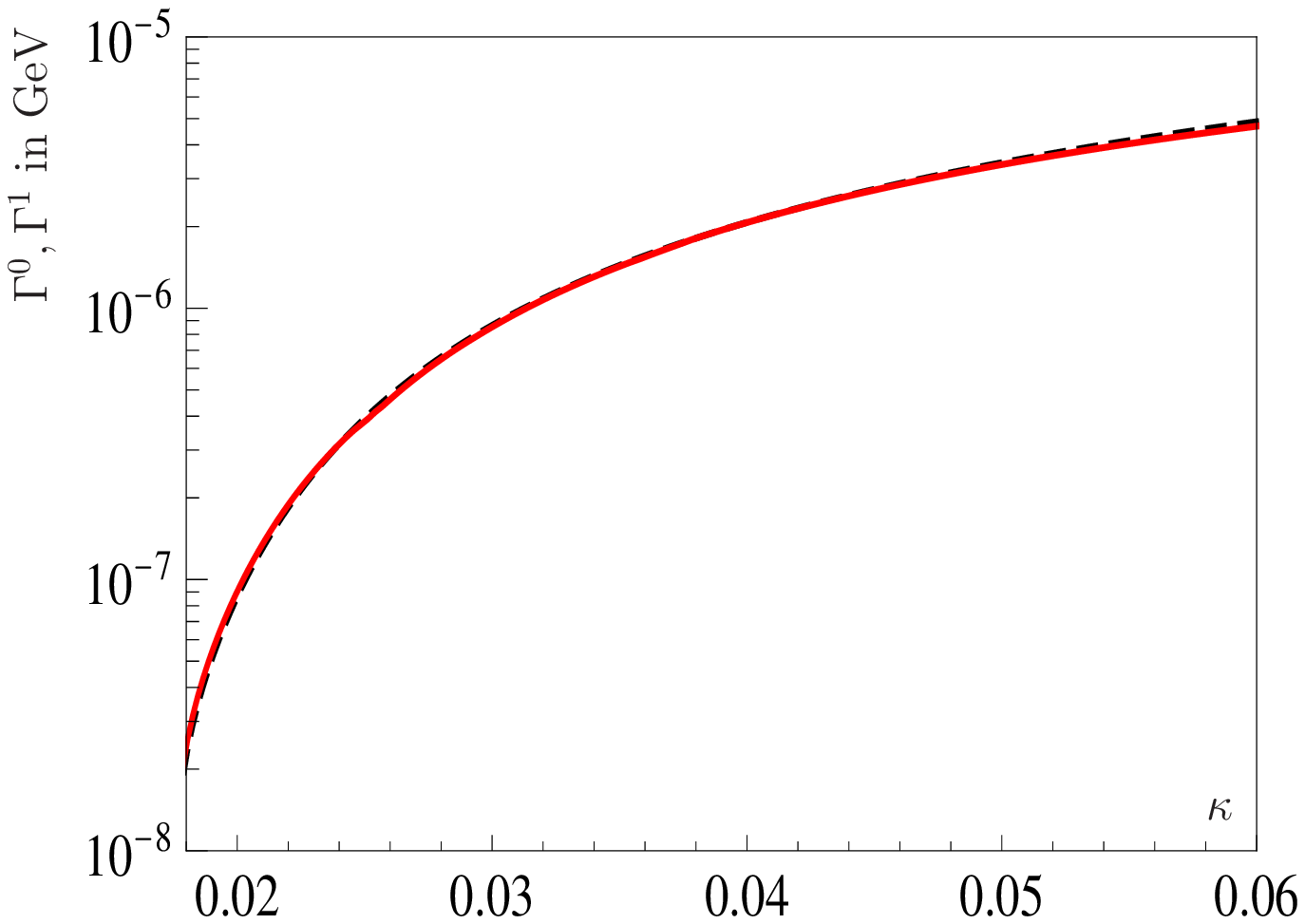}
\hfill
\includegraphics[width=0.5\textwidth]{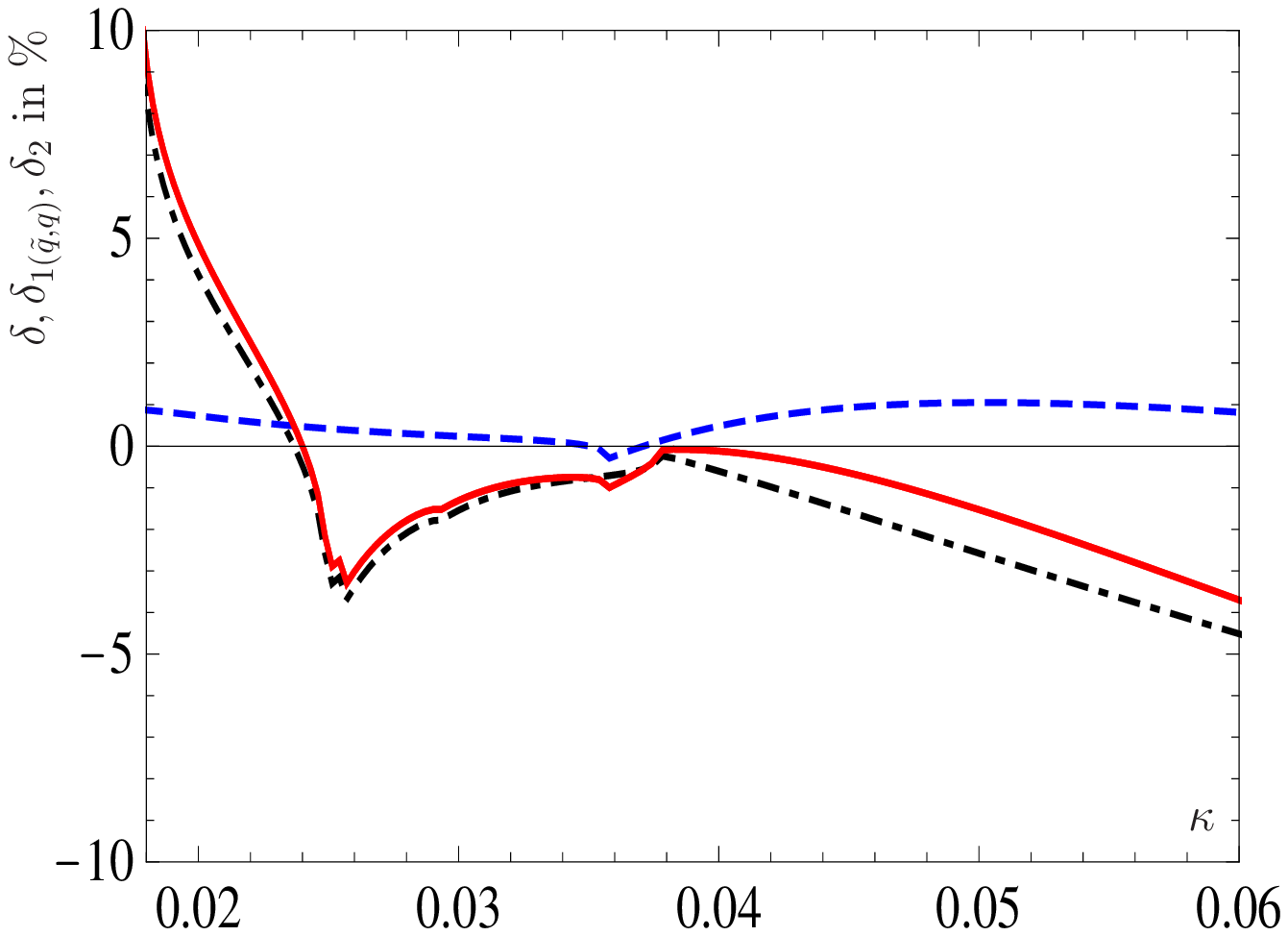}
\caption{a) (left) LO (black, dashed) and NLO (red, solid) decay widths for
 $\tilde{S}\rightarrow \tilde{\chi}_{2}^- W^+$ as a function
of $\kappa$ for the spectrum of Figure~\ref{fig:singlinodecay},
 b) (right): Correction factor $\delta$ in \% defined in
 \eqref{eq:corrfactor} for $\tilde{S}\rightarrow \tilde{\chi}_{2}^- W^+$
as a function of $\kappa$:
blue (dashed): Squark and quark contributions,
black (dot-dashed): Other sectors,
red (solid): Full correction.}
\label{fig:singlinodecay3}
\end{figure}

For the numerical investigation we choose the benchmark scenario mSUGRA~$4$, but
reduce $\lambda$ to $\lambda=0.01$ to increase the singlino character. 
We vary $\kappa$ between $2\cdot 10^{-3}$ and $6 \cdot 10^{-2}$ leading
to  singlino masses between $100$~GeV and $2.5$~TeV.
In this rather light particle spectrum the higgsino-like chargino 
has a mass of $m_{\tilde{\chi}_1^\pm}=201$~GeV and
 the wino-like chargino has a mass of $m_{\tilde{\chi}_2^\pm}=608$~GeV.
The mass spectrum of the neutralinos is given in Figure~\ref{fig:singlinodecay}~a).

In  Figure~\ref{fig:singlinodecay2} we show the details for the decay
$\tilde S \rightarrow \tilde{\chi}_1^- W^+$. Note that despite the
decrease of the coupling due to the decrease in the wino and higgsino components
with increasing $\kappa$ one gets a net increase of the width due to the
$(m_{\tilde S}/m_W)^2$ factor. In the right plot one sees clearly that
contributions of third generation squarks and quarks are less important compared to
the bino case. However, the remaining contribution can amount up to about 10 percent. 
Also the decay into the heavier chargino shows similar features as can be seen
from  Figure~\ref{fig:singlinodecay3}. The main difference is that the 
threshold effects due to on-shell intermediate states are more pronounced in this
case and are mainly caused by sleptons and Higgs bosons at $m_{\tilde{S}}\approx 1$~TeV 
and by squarks at $m_{\tilde{S}}\approx 1.6$~TeV.

\subsubsection{Chargino decays}

\begin{figure}[H]
\includegraphics[width=0.5\textwidth]{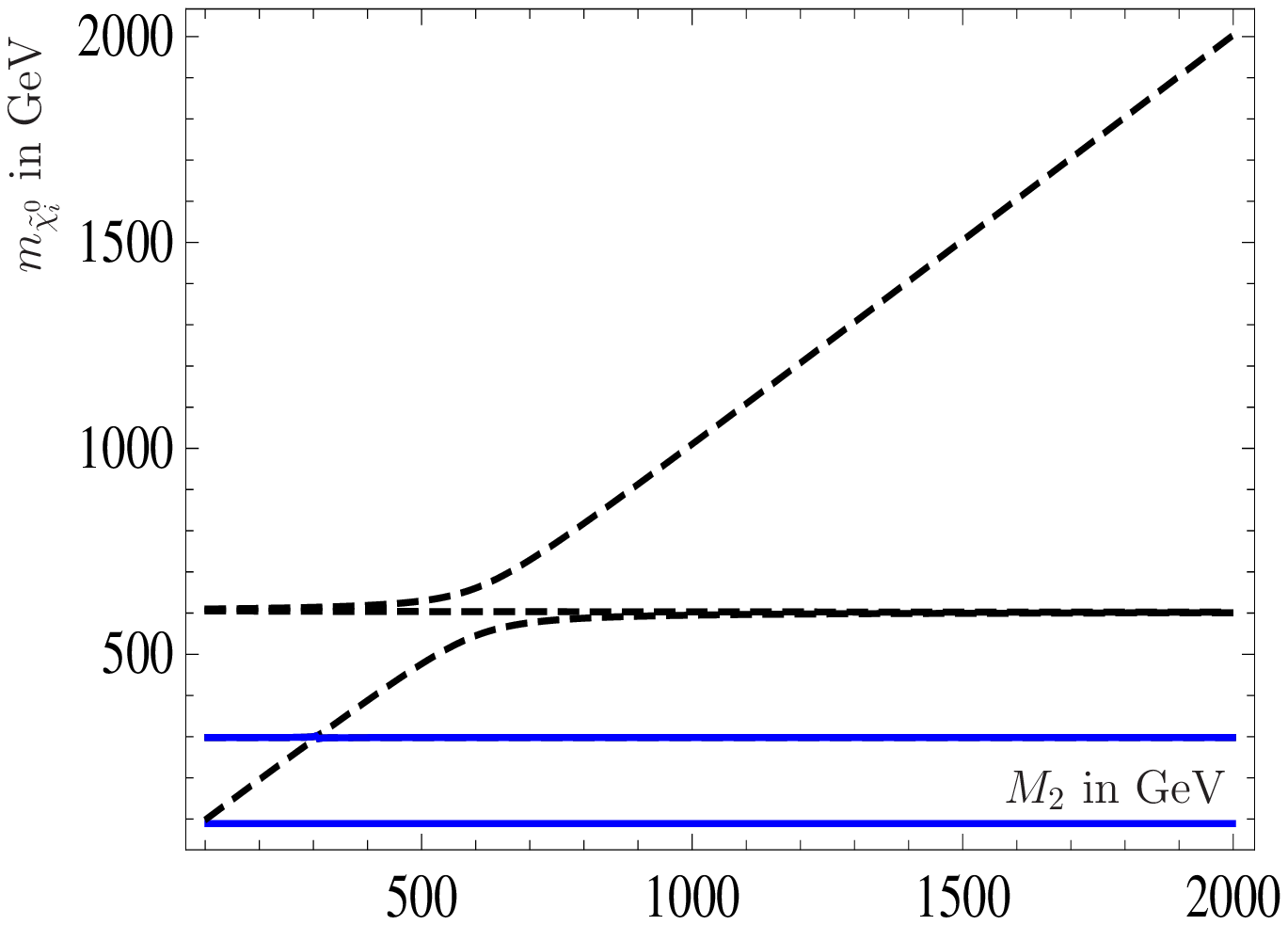}
\hfill
\includegraphics[width=0.5\textwidth]{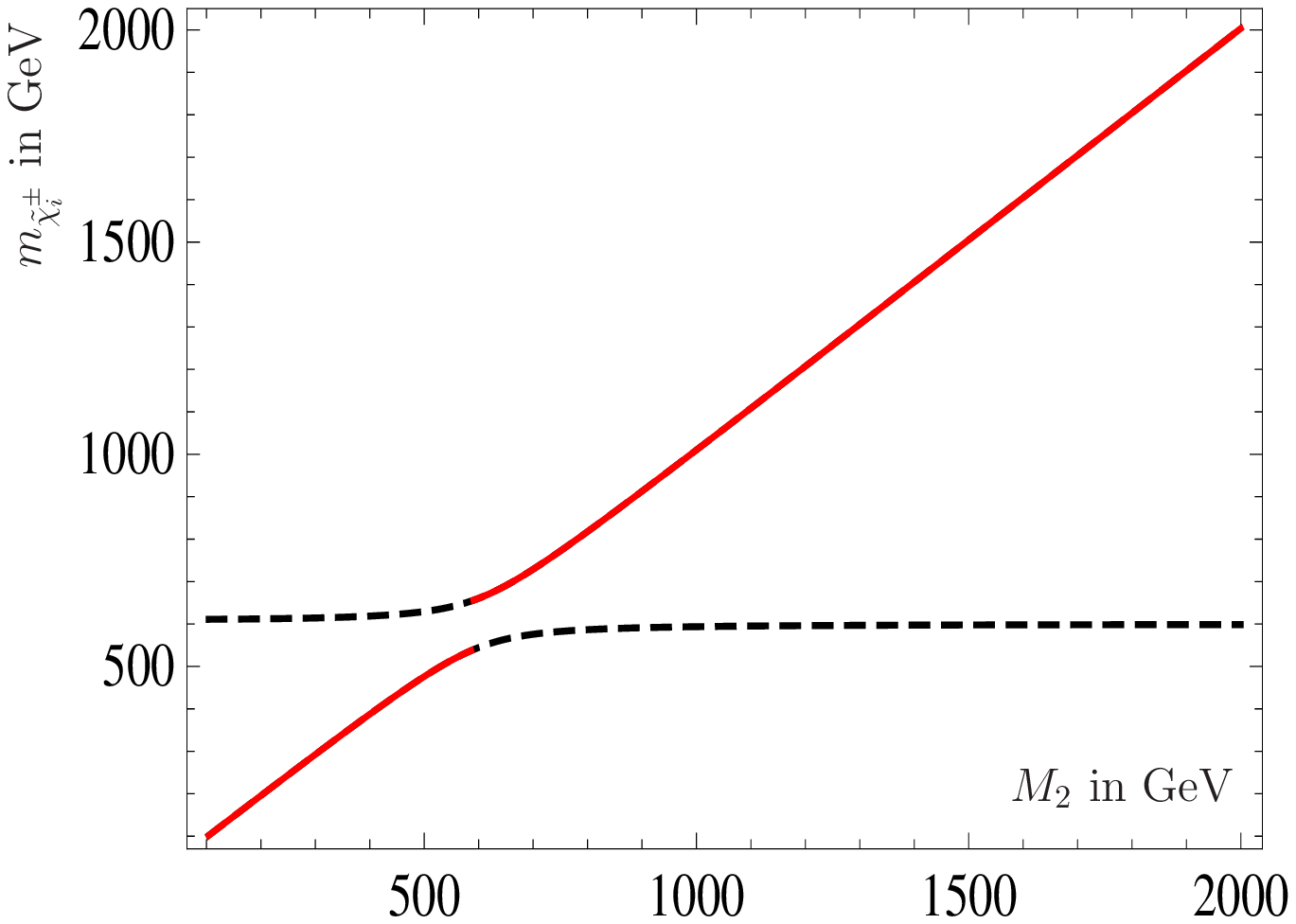}
\caption{a) (left): Neutralino masses  and b) (right): Chargino masses
as a function of $M_2$. The other parameters are as GMSB $5$ apart from
$M_1=300$~GeV and $\mu=600$~GeV. The red lines in b) correspond to the wino like
states and the two blue ones in a) to the  singlino state $\tilde{\chi}_1^0$ 
and bino state $\tilde{\chi}_2^0$.}
\label{fig:chardecay}
\end{figure}

In this  section we  present the decays of the wino-like chargino  
$\tilde{W}^+$
in a bino- or singlino-like neutralino $\tilde{\chi}_{1,2}^0$, being the 
two lightest neutralinos in the benchmark scenario GMSB $5$.
We depart from the original parameters by setting $M_1=300$~GeV
and $\mu=600$~GeV to get a rather light particle spectrum.
We vary the gaugino mass $M_2$ between $100$ and
$2000$~GeV. 
The neutralino mass spectrum  is given in Figure~\ref{fig:chardecay} a) and
the chargino mass spectrum  in Figure~\ref{fig:chardecay} b).
The two light neutralinos have a mass of $m_{\tilde{\chi}_1^0}=89$~GeV
for the singlino-like neutralino 
and $m_{\tilde{\chi}_2^0}=298$~GeV for the bino-like neutralino.

\begin{figure}[H]
\includegraphics[width=0.5\textwidth]{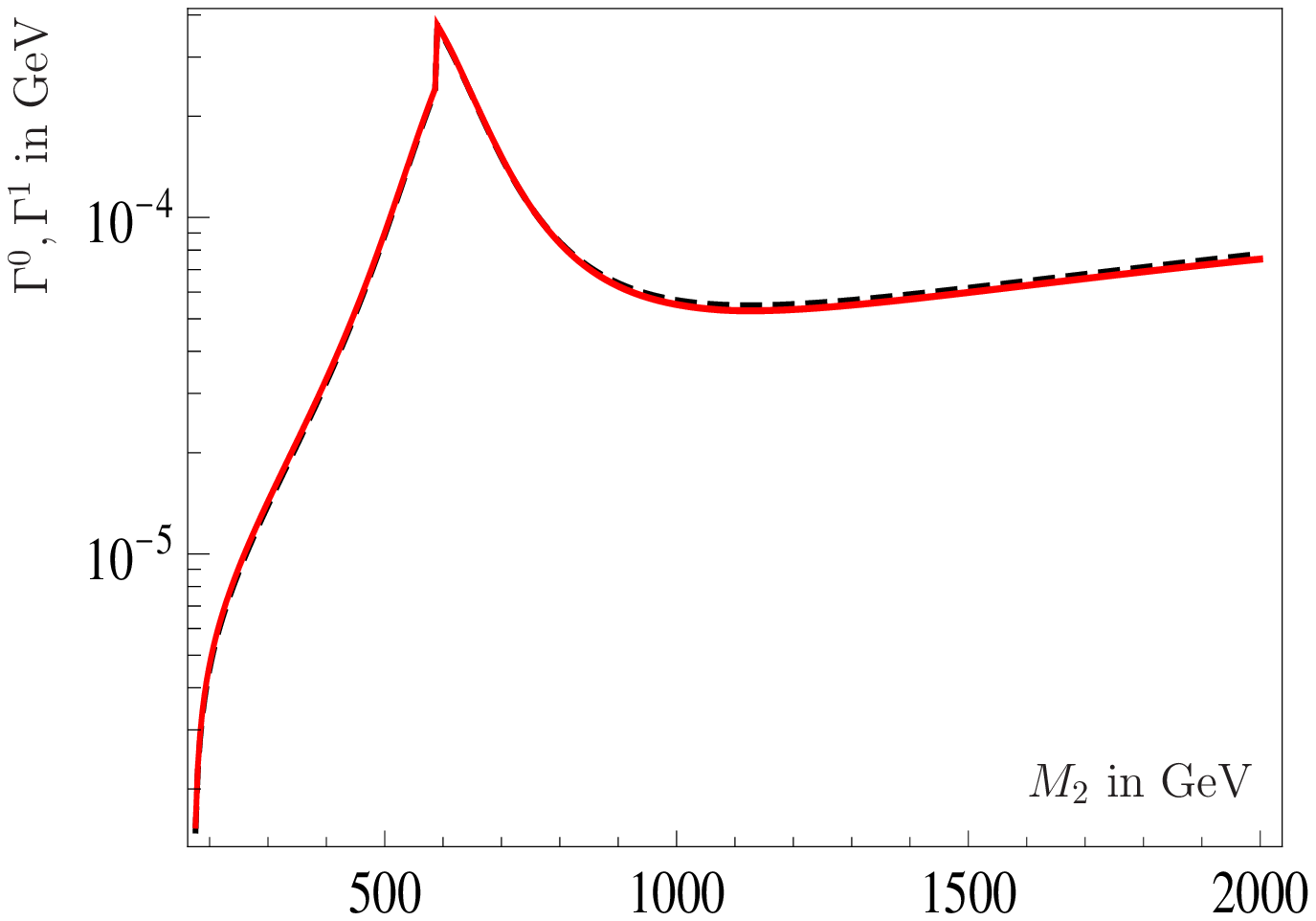}
\hfill
\includegraphics[width=0.5\textwidth]{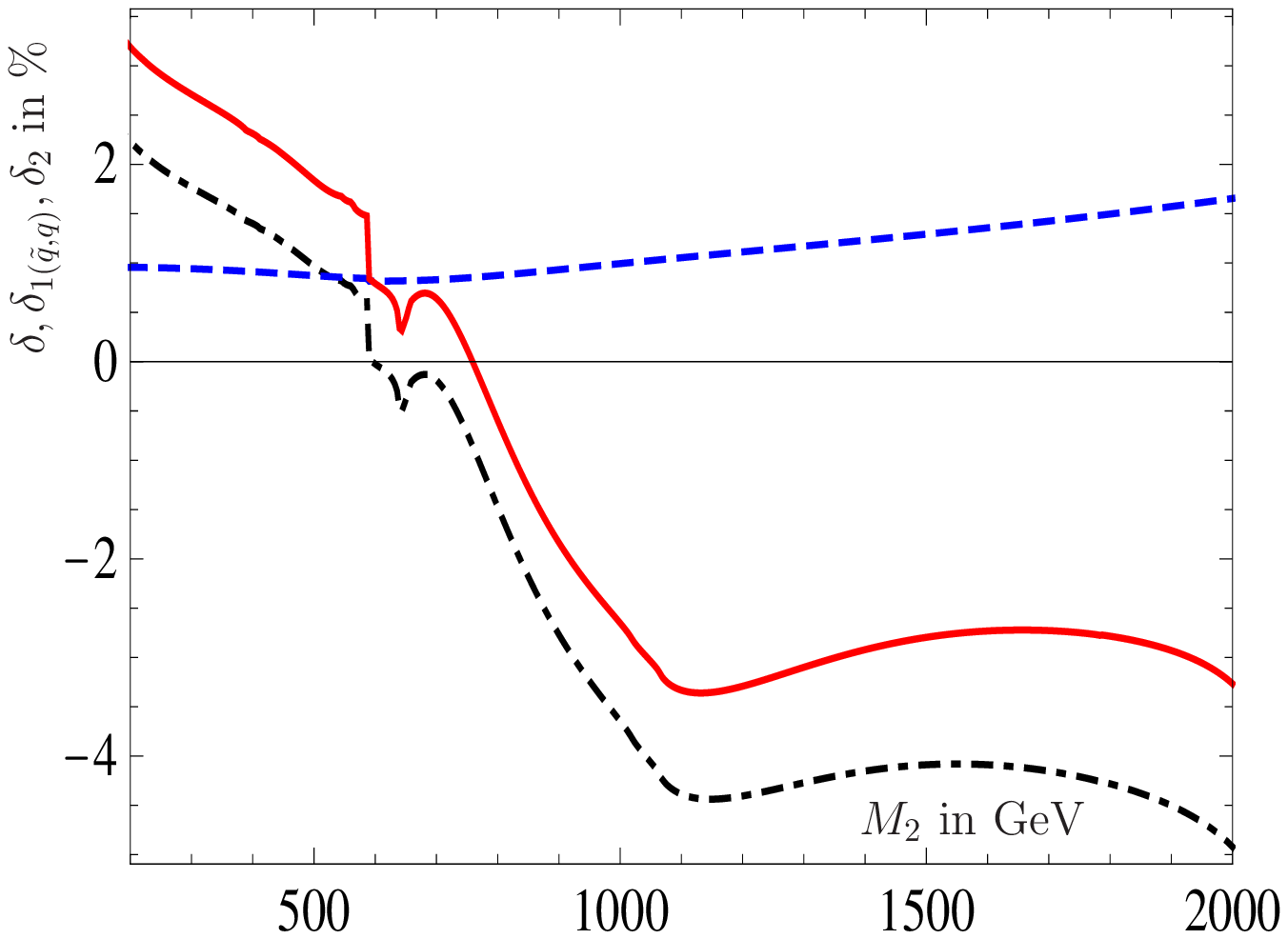}
\caption{a) (left): LO (black, dashed) and NLO (red, solid) decay widths
 for $\tilde{W}^+\rightarrow \tilde{\chi}_{1}^0 W^+$ as a function
of $M_2$  for the spectrum of Figure~\ref{fig:chardecay},
 b) (right): Correction factor $\delta$ in \% defined in
 \eqref{eq:corrfactor} for $\tilde{W}^+\rightarrow \tilde{\chi}_{1}^0 W^+$
as a function of $M_2$:
blue (dashed): Squark and quark contributions,
black (dot-dashed): Other sectors,
red (solid): Full correction.}
\label{fig:chardecay2}
\end{figure}

\begin{figure}[H]
\includegraphics[width=0.5\textwidth]{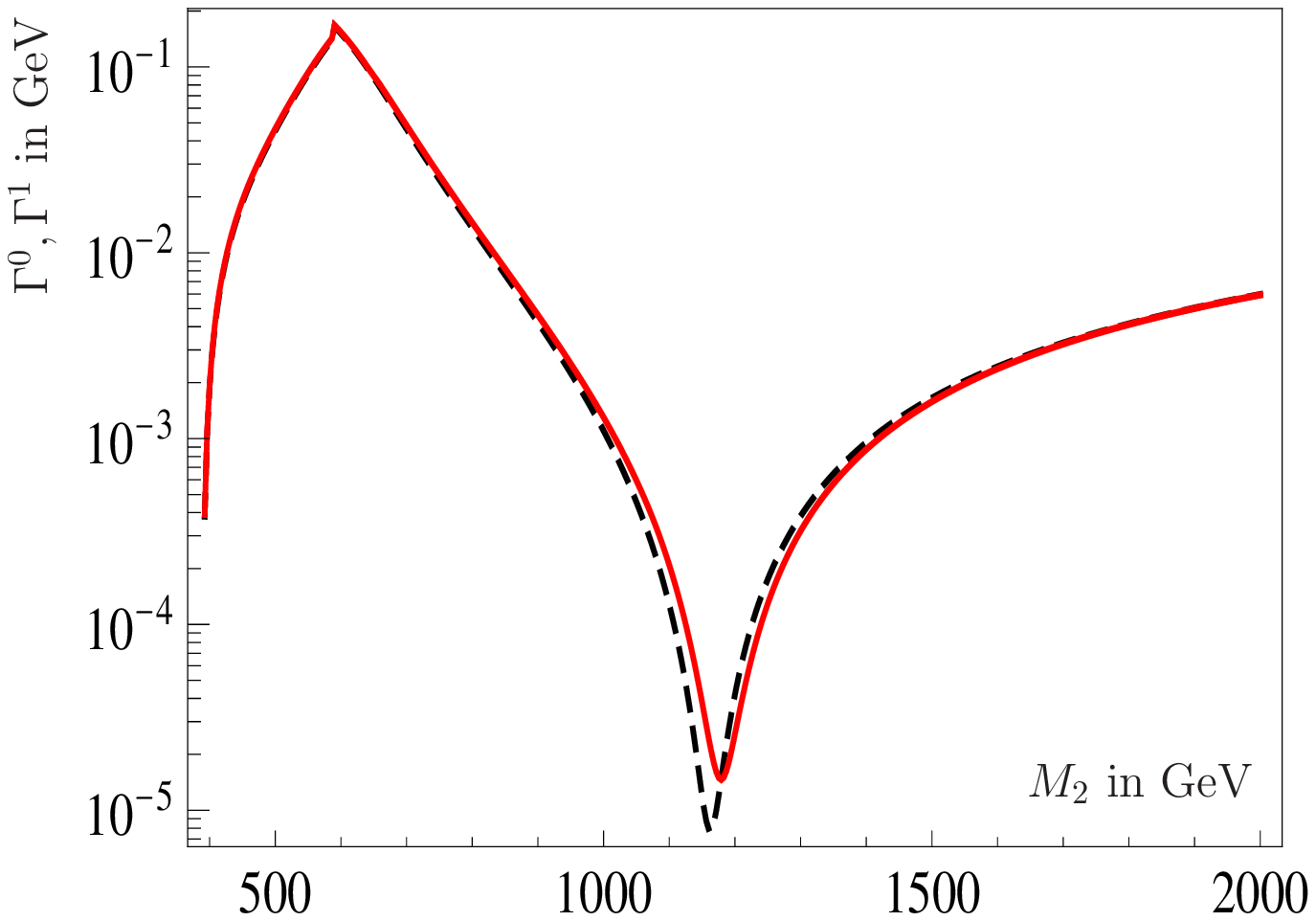}
\hfill
\includegraphics[width=0.5\textwidth]{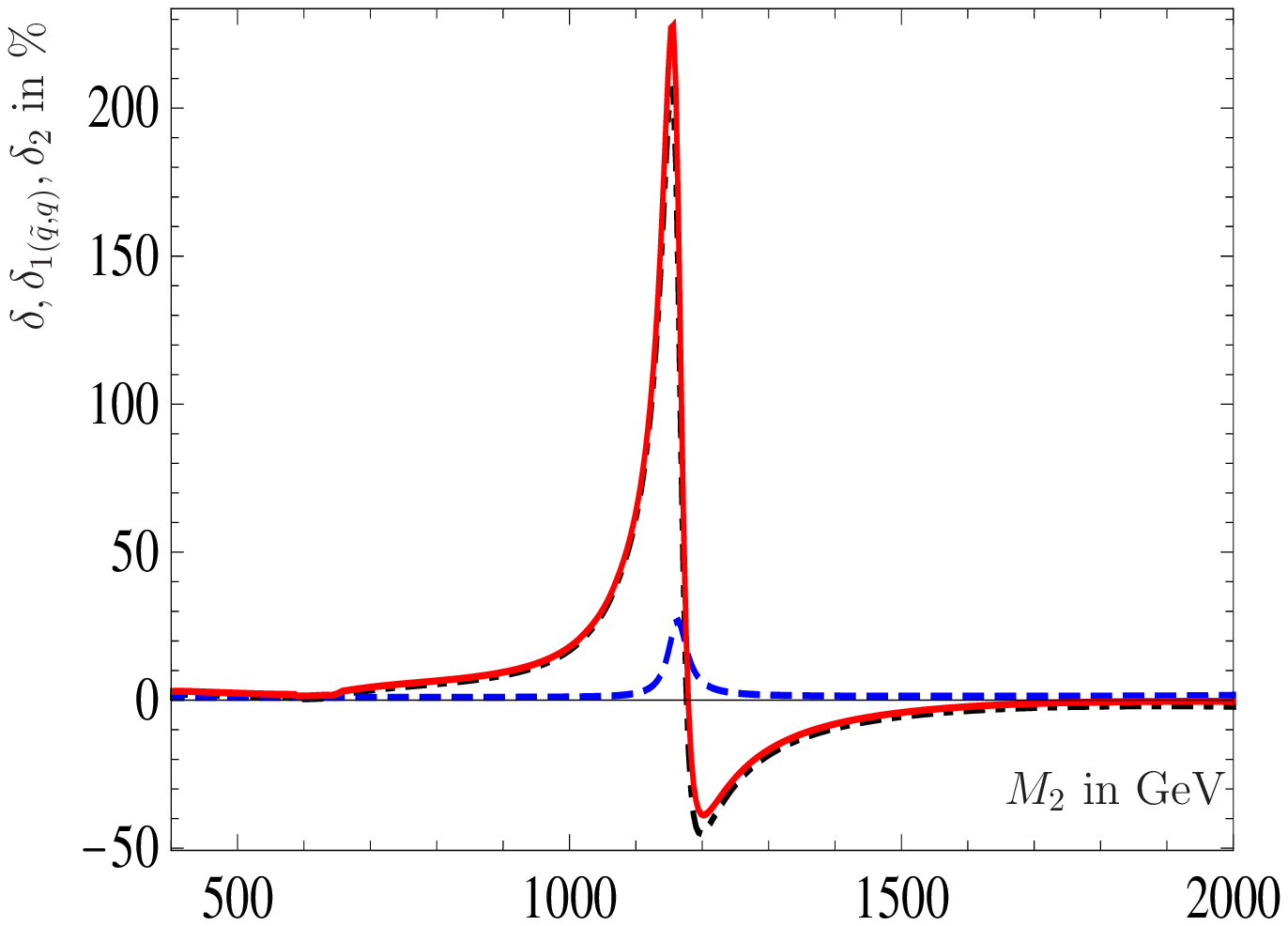}
\caption{a) (left): LO (black, dashed) and NLO (red, solid) decay widths
 for $\tilde{W}^+\rightarrow \tilde{\chi}_{2}^0 W^+$ as a function
of $M_2$ for the spectrum of Figure~\ref{fig:chardecay},
 b) (right): Correction factor $\delta$ in \% defined in
 \eqref{eq:corrfactor} for $\tilde{W}^+\rightarrow \tilde{\chi}_{2}^0 W^+$
as a function of $M_2$:
blue (dashed): Squark and quark contributions,
black (dot-dashed): Other sectors,
red (solid): Full correction.}
\label{fig:chardecay3}
\end{figure}

In Figures~\ref{fig:chardecay2} and \ref{fig:chardecay3} we show the details
of the decays $\tilde W^+ \rightarrow \tilde \chi^0_{1,2} W^+$. Note that
the peaks close to $M_2\approx 620$~GeV are due to the level crossing of the wino-like states
with the higgsino-like states. The overall features of the widths and corrections
are similar to the case of neutralino decays. In general the corrections are in the
order of a few percent except for a region close to $M_2=1.15$~TeV for the
decay $\tilde{W}^+\rightarrow \tilde{\chi}_{2}^0 W^+$ in Figure~\ref{fig:chardecay3} where the
tree-level couplings to $\tilde{\chi}^0_{2}$ nearly vanish due to a negative interference
between the wino and higgsino contributions. We find that the third generation
squark/quark contributions are in general somewhat smaller than the remaining ones.



\section{Conclusion}
\label{sec:conclusions}

In this paper we presented the complete electroweak one-loop corrections to
the partial widths for two-body decays of a chargino (neutralino)
into a $W$-boson and a neutralino (chargino). The calculations were 
done for the MSSM and the NMSSM using an  on-shell scheme and we have used
the so-called pinch technique to achieve gauge invariance.
Moreover we based the calculation of one-loop masses for neutralinos
and charginos on a reasonable renormalization of the physical parameters,
which are present on tree-level.

Whereas the mass corrections to neutralinos and charginos are typically
in the per-mil range, the corrections to the decays are
in the order of $1$-$10$ percent, in particular
if the involved neutralino is either wino or higgsino like. 
However in case of a small leading order decay width the corrections 
can be $50$ percent or larger. This is the case if either there is
a negative interference between the  wino and the higgsino part of the 
couplings or if the neutralino is either mainly bino- or singlino-like,
because for a pure bino or a pure singlino the tree-level width is identical
zero and can only be induced at the 1-loop level. Last but not least we provide
the publicly available program \prog for the numerical evaluation.

\section{Acknowledgments}

This work has been supported by the DFG, Project no. PO-1337/2-1. 
S.L.\ has been supported by the
DFG research training group GRK1147.
We thank Sven Heinemeyer, Federico von der Pahlen and Christian Schappacher for
pointing out an error in the renormalization of the electric charge
and for a detailed comparison of the Bremsstrahlung integrals.

\newpage
\begin{appendix}

\section{Formulas: Vertex corrections}
\label{sec:vertexcorr}

In this section we present the generic formulas for the next-to leading
order vertex contributions for the 't~Hooft-Feynman gauge $\xi_V=1$ ($V=W,Z$).
The formulas for $\xi_V\neq 1$ are  contained in
the program \prog \cite{Fortranonline}. The formulas for the
self-energies can be found in \cite{Pierce:1996zz}, from which also the derivatives
with respect to $p^2$ can be calculated. They are also included
in \prog \cite{Fortranonline}.

Below we present the formulas for the generic contributions
to the matrix element $M_V$ shown in Figure~\ref{fig:vertexcorr}.  In
addition we give the particle combinations to be inserted in these
diagrams for the decay $\tilde{\chi}^0\rightarrow\tilde{\chi}^- W^+$
neglecting generation indices. We use the following notation: $H$
stands for one of the $2(3)$ scalar Higgs bosons, $A$ for one of the
$2(3)$ pseudoscalar Higgs bosons including the Goldstone boson and
$H^\pm$ for one of the $2$ charged scalars including the Goldstone
boson.  The indices of couplings and  masses in the generic
formulas have to be understood in the following form: $F_i$ and $F_o$
denote the decaying and outgoing fermion, $W$ the external $W$-boson,
whereas $F$, $F_{1,2}$, $S$, $S_{1,2}$, $V$ or $V_{1,2}$ represent
possible internal fermionic, scalar or vector particles. It is understood
implicitly that one has to sum over possible flavour and generation
indices of the internal particles. All contributions have the same
generic structure:
\begin{align}
 M_V=&\frac{i}{16\pi^2}\overline{u}(p_1)\gamma^\mu\left(P_L M_1+ P_R M_2\right)u(k)\epsilon_\mu^*(p_2)\\\nonumber+&
\frac{i}{16\pi^2}\overline{u}(p_1)\left(P_L M_3^\mu+ P_R M_4^\mu\right)u(k)\epsilon_\mu^*(p_2)
\end{align}

We  start with the vertex in Figure~\ref{fig:vertexcorr} a) 
containing two internal fermions and one scalar. We use
the following abbreviations for the couplings:
\begin{align}
O_1&=O_{FFV,L}(F_2,F_1,W),\qquad O_2=O_{FFV,R}(F_2,F_1,W)\\
O_3&=O_{FFS,L}(F_1,F_i,S),\qquad O_4=O_{FFS,R}(F_1,F_i,S)\\
O_5&=O_{FFS,L}(F_o,F_2,S),\qquad O_6=O_{FFS,R}(F_o,F_2,S)
\end{align}
In the formulas below the Passarino-Veltman integrals have as
arguments: $B_0( m_W^2, m_{F_2}^2, m_{F_1}^2)$ and $C_i,C_{ij}(m_W^2, m_i^2,
m_o^2, m_{F_2}^2, m_{F_1}^2, m_S^2)$.  Possible particle insertions in
the notation $SF_1F_2$ are given by $H\tilde{\chi}^0\tilde{\chi}^\pm$,
$A\tilde{\chi}^0\tilde{\chi}^\pm$,
$H^\pm\tilde{\chi}^\pm\tilde{\chi}^0$, $\tilde{\nu}\nu l$,
$\tilde{l}l\nu$, $\tilde{u}ud$, $\tilde{d}du$. We get: 
{\allowdisplaybreaks
\begin{align}
M_{1,a}=&-O_1 \left[O_3 m_{F_1} (O_6 m_{F_2} C_0-O_5 m_o C_1)+O_4 O_6 m_{F_2} m_i (C_0+C_1)\right]\\\nonumber&
+O_1 O_5 m_o C_{2} (O_3 m_{F_1}+O_4 m_i)+O_2 \left[O_3 (O_6 (m_S^2 C_0+m_i^2 C_1+B_0-2 C_{00}\right.\\\nonumber&\left.
-m_o^2 (C_0+C_1+C_{2}))-O_5 m_{F_2} m_o (C_0+C_1+C_{2}))+O_4 O_6 m_{F_1} m_i C_1\right]\\
M_{2,a}=&O_1 \left\{O_3 O_5 m_{F_1} m_i C_1+O_4 \left[O_5 (m_S^2 C_0+m_i^2 C_1-m_o^2 (C_0+C_1+C_{2})\right.\right.\\\nonumber&\left.\left.
+B_0-2 C_{00})-O_6 m_{F_2} m_o (C_0+C_1+C_{2})\right]\right\}\\\nonumber&
+O_2 \left\{O_3 m_i (O_6 m_o C_{2}-O_5 m_{F_2} (C_0+C_1))\right.\\\nonumber&\left.
+O_4 m_{F_1} (O_6 m_o (C_1+C_{2})-O_5 m_{F_2} C_0)\right\}\\
M_{3,a}^\mu=&p_1^\mu 2\left\{O_1 O_4 \left[O_5 m_o (C_{12}+C_{2}+C_{22})-O_6 m_{F_2} (C_0+C_1+C_{2})\right]\right.\\\nonumber&\left.
+O_2 O_6 (O_4 m_{F_1} C_1-O_3 m_i C_{12})\right\}\\\nonumber&
+p_2^\mu 2\left\{O_2 O_6 \left[O_3 m_i (C_1+C_{12})+O_4 m_{F_1} C_1\right]-O_1 O_4 O_5 m_o (C_1+2 C_{12})\right\}\\
M_{4,a}^\mu=&p_1^\mu 2\left\{O_1 O_5 (O_3 m_{F_1} C_1-O_4 m_i C_{12})\right.\\\nonumber&\left.
+O_2 O_3 \left[O_6 m_o (C_{12}+C_{2}+C_{22})-O_5 m_{F_2} (C_0+C_1+C_{2})\right]\right\}\\\nonumber&
+p_2^\mu 2\left\{O_1 O_5 (O_3 m_{F_1} C_1+O_4 m_i (C_1+C_{12}))-O_2 O_3 O_6 m_o (C_1+2 C_{12})\right\}
\end{align}
}
For the vertex in Figure~\ref{fig:vertexcorr} b) we define:
\begin{align}
O_1&=O_{SSV}(S_1,S_2,V)\\
O_2&=O_{FFS,L}(F,F_i,S_1),\qquad O_3=O_{FFS,R}(F,F_i,S_1)\\
O_4&=O_{FFS,L}(F_o,F,S_2),\qquad O_5=O_{FFS,R}(F_o,F,S_2)
\end{align}
For completeness we note that in case of $O_1$ the following internal
momentum combination appears $(p_{S_1}-p_{S_2})$ over which of course
has been integrated.
The Passarino-Veltman integrals below
 have the arguments $C_i,C_{ij}(m_o^2,m_W^2,m_i^2,m_F^2,m_{S_2}^2,m_{S_1}^2)$.
Possible particle insertions in the notation $FS_1S_2$ are given by
$\tilde{\chi}^0HH^\pm$, $\tilde{\chi}^\pm H^\pm H$, $\tilde{\chi}^0AH^\pm$,
$\tilde{\chi}^\pm H^\pm A$, $\nu\tilde{\nu}\tilde{l}$, $l\tilde{l}\tilde{\nu}$,
$u\tilde{u}\tilde{d}$, $d\tilde{d}\tilde{u}$.
We obtain for the different parts:
{\allowdisplaybreaks
\begin{align}
M_{1,b}=&2 O_1 O_2 O_5 C_{00}\\
M_{2,b}=&2 O_1 O_3 O_4 C_{00}\\
M_{3,b}^\mu=&p_2^\mu O_1 \left\{O_3 (O_5 m_F (C_0+2 C_{2})-O_4 m_o (C_{1}+2 C_{12}))\right.\\\nonumber&\left.
-O_2 O_5 m_i (C_{2}+2 C_{22})\right\}-2 p_1^\mu O_1 \left\{O_2 O_5 m_i (C_{12}+C_{2}+C_{22})\right.\\\nonumber&\left.
+O_3 \left[O_4 m_o (C_{1}+C_{11}+C_{12})-O_5 m_F (C_0+C_{1}+C_{2})\right]\right\}\\
M_{4,b}^\mu=&2 p_1^\mu O_1 \left\{O_2 \left[ O_4 m_F (C_0+C_{1}+C_{2})-O_5 m_o (C_{1}+C_{11}+C_{12})\right]\right.\\\nonumber&\left.
-O_3 O_4 m_i (C_{12}+C_{2}+C_{22})\right\}+p_2^\mu O_1 \left\{O_2 \left[O_4 m_F (C_0+2 C_{2})\right.\right.\\\nonumber&\left.\left.
-O_5 m_o (C_{1}+2 C_{12})\right]-O_3 O_4 m_i (C_{2}+2 C_{22})\right\}
\end{align}
}
For the vertex in Figure~\ref{fig:vertexcorr} c) we define:
\begin{align}
O_1&=O_{FFV,L}(F_1,F_i,V),\qquad O_2=O_{FFV,R}(F_1,F_i,V)\\
O_3&=O_{FFV,L}(F_2,F_1,W),\qquad O_4=O_{FFV,R}(F_2,F_1,W)\\
O_5&=O_{FFV,L}(F_o,F_2,V),\qquad O_6=O_{FFV,R}(F_o,F_2,V)
\end{align}
The Passarino-Veltman integrals below
have as arguments $C_i,C_{ij}(m_W^2,m_i^2,m_o^2,m_{F_2}^2,m_{F_1}^2,m_V^2)$
and $B_0(m_W^2,m_{F_2}^2,m_{F_1}^2)$.
Possible particle insertions in the notation $VF_1F_2$ are given by
$Z\tilde{\chi}^0\tilde{\chi}^\pm$, $W\tilde{\chi}^\pm\tilde{\chi}^0$.
We get: 
{\allowdisplaybreaks
\begin{align}
M_{1,c}=&2 O_1 O_5 \left\{O_3 \left[m_V^2 C_0+m_i^2 (C_{1}-C_{2})-m_o^2 (C_0+C_{1}+2 C_{2})\right.\right.\\\nonumber&\left.\left.
+m_W^2 C_{2}+B_0-2 C_{00}\right]-O_4 m_{F_1} m_{F_2} C_0\right\}-2O_2 O_4 O_6 m_i m_o C_{2}\\
M_{2,c}=&2 O_2 O_6 \left\{O_4 \left[m_V^2 C_0+m_i^2 (C_{1}-C_{2})-m_o^2 (C_0+C_{1}+2 C_{2})\right.\right.\\\nonumber&\left.\left.
+m_W^2 C_{2}+B_0-2 C_{00}\right]-O_3 m_{F_1} m_{F_2} C_0\right\}-2 O_1 O_3 O_5 m_i m_o C_{2}\\
M_{3,c}^\mu=&4 p_1^\mu \left\{O_2 \left[O_3 O_5 m_{F_1} C_{2}+O_4 (O_5 m_{F_2} C_{2}+O_6 m_o (C_{2}+C_{22}))\right]\right.\\\nonumber&\left.
-O_1 O_3 O_5 m_i (C_{2}+C_{12})\right\}-4 p_2^\mu \left\{O_2 \left[O_3 O_5 m_{F_1} C_{1}+O_4 (O_5 m_{F_2} (C_0+C_{1})
\right.\right.\\\nonumber&\left.\left.
+O_6 m_o (C_{1}+C_{11}+C_{12}+C_{2}))\right]-O_1 O_3 O_5 m_i (C_{1}+C_{11})\right\}\\
M_{4,c}^\mu=&4 p_1^\mu \left\{O_1 \left[O_3 (O_5 m_o (C_{12}+C_{22})+O_6 m_{F_2} C_{2})+O_4 O_6 m_{F_1} C_{2}\right]\right.\\\nonumber&\left.
-O_2 O_4 O_6 m_i (C_{2}+C_{12})\right\}-4 p_2^\mu \left\{O_1 \left[O_3 (O_5 m_o (C_{1}+C_{11}+C_{12}+C_{2})
\right.\right.\\\nonumber&\left.\left.
+O_6 m_{F_2} (C_0+C_{1}))+O_4 O_6 m_{F_1} C_{1}\right]-O_2 O_4 O_6 m_i (C_{1}+C_{11})\right\}
\end{align}
}
For the vertex in Figure~\ref{fig:vertexcorr} d) we define:
\begin{align}
O_1&=O_{FFV,L}(F_o,F,V),\qquad O_2=O_{FFV,R}(F_o,F,V)\\
O_3&=O_{FFS,L}(F,F_i,S),\qquad O_4=O_{FFS,R}(F,F_i,S)\\
O_5&=O_{SVV}(S,W,V)
\end{align}
The Passarino-Veltman integrals below
 have as arguments $C_i,C_{ij}( m_o^2, m_W^2, m_i^2, m_F^2, m_V^2, m_S^2)$.
Possible particle insertions in the notation $FSV$ are given by
$\tilde{\chi}^\pm H^\pm\gamma$, $\tilde{\chi}^\pm H^\pm Z$,
 $\tilde{\chi}^0 H W$, $\tilde{\chi}^0 A W$. We get:
{\allowdisplaybreaks
\begin{align}
M_{1,d}=&O_5 (O_1 O_3 m_F C_0 - O_1 O_4 m_i C_2 + O_2 O_3 m_o C_1)\\
M_{2,d}=&O_5 (O_1 O_4 m_o C_1-O_2 O_3 m_i C_2+O_2 O_4 m_F C_0)\\
M_{3,d}^\mu=& 2 p_1^\mu O_1 O_4 O_5 C_1\\
M_{4,d}^\mu=& 2 p_1^\mu O_2 O_3 O_5 C_1
\end{align}
}
For the vertex  in Figure~\ref{fig:vertexcorr} e) we define:
\begin{align}
O_1&=O_{FFV,L}(F,F_i,V),\qquad O_2=O_{FFV,R}(F,F_i,V)\\
O_3&=O_{FFS,L}(F_o,F,S),\qquad O_4=O_{FFS,R}(F_o,F,S)\\
O_5&=O_{SVV}(S,W,V)
\end{align}
The Passarino-Veltman integrals, which appear in the following formulas,
 have as arguments $C_i,C_{ij}( m_i^2, m_W^2, m_o^2, m_F^2, m_V^2, m_S^2)$.
Possible particle insertions in the notation $FVS$ are given by
$\tilde{\chi}^0ZH^\pm$, $\tilde{\chi}^\pm WH$, $\tilde{\chi}^\pm WA$.
We obtain:
{\allowdisplaybreaks
\begin{align}
M_{1,e}=&O_5 (-O_1 O_3 m_o C_2+O_1 O_4 m_F C_0+O_2 O_4 m_i C_1)\\
M_{2,e}=&O_5 (O_1 O_3 m_i C_1+O_2 O_3 m_F C_0-O_2 O_4 m_o C_2)\\
M_{3,e}^\mu=& 2 (p_1^\mu+p_2^\mu) O_2 O_4 O_5 C_1\\
M_{4,e}^\mu=& 2 (p_1^\mu+p_2^\mu) O_1 O_3 O_5 C_1
\end{align}
}
For the vertex in Figure~\ref{fig:vertexcorr} f) we define:
\begin{align}
O_1&=O_{FFV,L}(F,F_i,V_1), O_2=O_{FFV,R}(F,F_i,V_1)\\
O_3&=O_{FFV,L}(F_o,F,V_2), O_4=O_{FFV,R}(F_o,F,V_2)\\
O_5&=O_{VVV}(W,V_1,V_2)
\end{align}
For completeness we note, that $O_5$ has the following internal momentum
contribution over which has been integrated: 
$((p^\mu_{V_1}-p^\mu_W)g^{\nu\sigma} + \ldots)$.
The Passarino-Veltman integrals, which appear in the following formulas,
have the arguments $C_i,C_{ij}(m_o^2,m_W^2,m_i^2,m_F^2,m_{V_2}^2,m_{V_1}^2)$
and $B_0(m_W^2,m_{V_2}^2,m_{V_1}^2)$.
Possible particle insertions in the notation $FV_1V_2$ are given by
$\tilde{\chi}^\pm W\gamma$, $\tilde{\chi}^0ZW$, $\tilde{\chi}^\pm WZ$.
We obtain:
{\allowdisplaybreaks
\begin{align}
M_{1,f}=&O_5 \left\{O_1 \left[O_3 (2 m_F^2 C_0+m_i^2 (2 C_{1}+3 C_{2})\right.\right.\\\nonumber&\left.
+m_o^2 (3 C_{1}+2 C_{2})-2 m_W^2 (C_{1}+C_{2})+2 B_0+4 C_{00})+3 O_4 m_F m_o C_0\right]\\\nonumber&\left.
+3 O_2 m_i \left[O_3 m_F C_0+O_4 m_o (C_{1}+C_{2})\right]\right\}\\
M_{2,f}=&O_5 \left\{3 O_1 m_i (O_3 m_o (C_{1}+C_{2})+O_4 m_F C_0)+O_2 \left[3 O_3 m_F m_o C_0\right.\right.\\\nonumber&
+O_4 (2 m_F^2 C_0+m_i^2 (2 C_{1}+3 C_{2})\\\nonumber&\left.\left.
+m_o^2 (3 C_{1}+2 C_{2})-2 m_W^2 (C_{1}+C_{2})+2 B_0+4 C_{00})\right]\right\}\\
M_{3,f}^\mu=&-2O_5 \left\{p_1^\mu \left[O_1 O_3 m_i (2 (C_{12}+C_{22})-C_{1})+O_2 (3 O_3 m_F (C_{1}+C_{2})\right.\right.\\\nonumber&\left.
+O_4 m_o (2 (C_{11}+C_{12})-C_{2}))\right]+p_2^\mu \left[O_1 O_3 m_i (C_{2}+2 C_{22})\right.\\\nonumber&\left.\left.
-O_2 (O_4 m_o (C_{1}-2 C_{12}+C_{2})-3 O_3 m_F C_{2})\right]\right\}\\
M_{4,f}^\mu=&-2O_5 \left\{p_1^\mu \left[O_1 (O_3 m_o (2 (C_{11}+C_{12})-C_{2})+3 O_4 m_F (C_{1}+C_{2}))\right.\right.\\\nonumber&\left.
+O_2 O_4 m_i (2 (C_{12}+C_{22})-C_{1})\right]+p_2^\mu \left[O_2 O_4 m_i (C_{2}+2 C_{22})\right.\\\nonumber&\left.\left.
-O_1 (O_3 m_o (C_{1}-2 C_{12}+C_{2})-3 O_4 m_F C_{2})\right]\right\}
\end{align}
}

\section{Hard photon emission}
\label{sec:photemission}

Here we present the formulas for the hard photon emission for the
 Feynman graphs  given  in Figures~\ref{fig:realcorrtext1}
and \ref{fig:realcorrtext2}. 
We  use the notation of \cite{Denner:1991kt} for the Bremsstrahlung integrals,
which are defined for the decay of a particle with mass $m_0$
and momentum $p_0$ into two particles with masses $m_1$ and $m_2$
and momenta $p_1$ and $p_2$ and a photon with momentum $q$ by
\begin{align}
I^{ij}_{lk}(m_0,m_1,m_2) =\frac{1}{\pi^2}\int
\frac{d^3p_1}{2p_{10}}\frac{d^3p_2}{2p_{20}}\frac{d^3q}{2q_{0}}\delta^{(4)}\left(p_0-p_1-p_2-q\right)
\frac{(\pm 2p_i q)(\pm 2p_j q)}{(\pm 2p_l q)(\pm 2p_k q)},
\end{align}
where the minus signs refer to the momentum $p_0$ of the initial particle. In this context we need the integrals
$I_{lk}^{ij}(m_i,m_o,m_W)$, allowing us to write the final result in the form
\begin{align}
 \Gamma^R = \frac{\alpha}{32 \pi^2 m_i m_W^2}\left[\left(|O_{WL}|^2+|O_{WR}|^2\right)\Omega_{1}
+ \left(O_{WL}O_{WR}^*+O_{WL}^*O_{WR}\right)\Omega_{2}\right],
\label{eq:realbremsstrahlung}
\end{align}
where we have introduced abbreviations
\begin{align}
\Omega_{1}&= Q_i^2\Omega_{1ii} + Q_o^2\Omega_{1oo}\\\nonumber
\Omega_{2}&= Q_i^2\Omega_{2ii} + Q_o^2\Omega_{2oo}
\end{align}
with the charges $Q_i$ and $Q_o$ of in- and outgoing particles being either
$0$ or $1$ depending on the process considered.
Due to the presence of left- and right-handed couplings to the $W$ boson 
in \eqref{eq:treelevellagrangian} the final result of the three-body 
does not factorize in the two-body width times a corrections factor.
The different parts are given by:
{\allowdisplaybreaks
\begin{align}
\Omega_{1ii}=&2 I (m_i^2+m_o^2+2 m_W^2)-4 \left[m_W^2 (m_i^2+m_o^2)+(m_i^2-m_o^2)^2-2 m_W^4\right] \\\nonumber
&\qquad\cdot \left[I_0+I_{00} m_i^2+m_W^2 (I_{02}+I_{22})+I_{02} (m_i-m_o)
 (m_i+m_o)+I_2\right] \\\nonumber
& + 2 I_0^2 (m_i^2+m_o^2+2 m_W^2)-8 m_W^2 (I_{22}^{01}+I_2^1)\\
\Omega_{1oo}=&
-8 I m_W^2-4 m_W^2 \left[m_o^2 (m_i^2 (2 I_{01}+2 I_{02}+I_{11}-2 I_{22})+I_1+I_2)\right.\\\nonumber
&\qquad+m_o^4 (-2 I_{01}-2 I_{02}+I_{11}+I_{22})+m_i^2 (I_1+I_2+I_{22} m_i^2)+I_1^0+2 I_{22}^{01}\\\nonumber
&\left.\qquad +4 I_2^0+2 I_2^1\right]\\\nonumber
&-8 m_W^6 (I_{01}+I_{02}-I_{22}) + 4 m_W^4 \left[-m_o^2 (I_{01}+I_{02}-2 I_{11}+I_{22})\right.\\\nonumber
&\left.\qquad+m_i^2 (3 (I_{01}+I_{02})-I_{22})+2 I_1+2 I_2\right]\\\nonumber
&-4 I_{01} m_i^6+12 I_{01} m_i^4 m_o^2-12 I_{01} m_i^2 m_o^4+4 I_{01} m_o^6-4 I_{02} m_i^6\\\nonumber
&+12 I_{02} m_i^4 m_o^2-12 I_{02} m_i^2 m_o^4+4 I_{02} m_o^6-4 I_1 m_i^4\\\nonumber
&+8 I_1 m_i^2 m_o^2-4 I_1 m_o^4-4 I_{11} m_i^4 m_o^2+8 I_{11} m_i^2 m_o^4\\\nonumber
&-4 I_{11} m_o^6-2 I_1^0 m_i^2- 2 I_1^0 m_o^2-4 I_2 m_i^4+8 I_2 m_i^2 m_o^2-4 I_2 m_o^4\\
\Omega_{2ii}=&2 m_i m_o \left[-2 I+12 m_W^2 (I_0+m_i^2 (I_{00}+I_{02})-I_{02} m_o^2+I_2)\right.\\\nonumber
&\left.\qquad +12 m_W^4 (I_{02}+I_{22})-2 I_0^2\right]\\
\Omega_{2oo}=&4 m_i m_o \left[6 m_W^2 (-m_o^2 (I_{01}+I_{02}-I_{11})+m_i^2 (I_{01}+I_{02})+I_1+I_2)\right.\\\nonumber
&\left.\qquad-6 m_W^4 (I_{01}+I_{02}-I_{22})+I_1^0\right]
\end{align}}
This result was obtained by using FeynArts \cite{Hahn:2000kx} and FormCalc \cite{Hahn:1998yk}.

\section{One- and Two-point functions} 
\label{sec:oneandtwopointfunctions}

Next we  provide some useful formulas for the  one- and two-point loop
functions  \cite{Passarino:1978jh}. As only part of them can be found in different
places in the literature we collect here the complete set, in particular the derivatives
of the two-point functions $B_{001}$ and $B_{111}$. We remind that the
 UV divergent part of the integrals can  be expressed as
\begin{align}
\Delta = \frac{1}{\epsilon}-\gamma +\ln(4\pi)\qquad,
\label{eq:UVdefinition}
\end{align}
where $\gamma\approx 0.577$ denotes Euler's constant. For the notation
of the Passarino-Veltman integrals we follow 
\cite{Hahn:1998yk,vanOldenborgh:1989wn}.

\subsection{Scalar integrals}
The one-point function $A_0$ is given by
\begin{align}
 A_0(m^2)&=m^2\left[\Delta+1+\ln\left(\frac{Q^2}{m^2}\right)\right]
\end{align}
and the two-point function $B_0$ by
\begin{align}
 B_0(p^2,m_1^2,m_2^2)=\Delta + 2 + \ln \left(\frac{Q^2}{m_1m_2}\right)+\frac{m_1^2-m_2^2}{p^2}\ln\left(\frac{m_2}{m_1}\right)
-\frac{m_1m_2}{p^2}\left(\frac{1}{r}-r\right)\ln r,
\end{align}
 where $r$ and $\tfrac{1}{r}$ denote the negative roots of the polynomial
\begin{align}\nonumber
 x^2+\frac{m_1^2+m_2^2-p^2}{m_1m_2}x+1=(x+r)\left(x+\frac{1}{r}\right)\qquad.
\end{align}
The derivative of $B_0$ with respect to $p^2$ yields:
\begin{align}\nonumber
 &\dot{B}_0(p^2,m_1^2,m_2^2):=\frac{\partial}{\partial p^2}B_0(p^2,m_1^2,m_2^2)\\
 &=-\frac{m_1^2-m_2^2}{p^4}\ln\left(\frac{m_2}{m_1}\right)
+\frac{m_1m_2}{p^4}\left(\frac{1}{r}-r\right)\ln r-\frac{1}{p^2}\left(1+\frac{r^2+1}{r^2-1}\ln r\right)
\label{eq:derB0}
\end{align}

\subsection{Tensor integrals}
Lorentz covariance in $d$ dimensions allows to decompose the
tensor integrals in terms of scalar integrals which are given by:
{\allowdisplaybreaks
\begin{align}
 A_{00}(m^2)&=\frac{1}{4}m^2A_0(m^2)+\frac{1}{8}m^4\\
 B_1(p^2,m_1^2,m_2^2)&=\frac{1}{2p^2}\left[(m_2^2-m_1^2)(B_0(p^2,m_1^2,m_2^2)-B_0(0,m_1^2,m_2^2))\right]\\\nonumber
 &\qquad\quad-\frac{1}{2}B_0(p^2,m_1^2,m_2^2)\\
 B_{00}(p^2,m_1^2,m_2^2)&=\frac{1}{6}\left[A_0(m_2^2)+(p^2-m_2^2+m_1^2)B_1(p^2,m_1^2,m_2^2) \frac{}{}\right.\\\nonumber
  &\left.\qquad +2m_1^2B_0(p^2,m_1^2,m_2^2)+m_0^2+m_1^2-\frac{1}{3}p^2\right]\\
 B_{11}(p^2,m_1^2,m_2^2)&=\frac{1}{3p^2}\left[A_0(m_2^2)-m_1^2B_0(p^2,m_1^2,m_2^2)\frac{}{}\right.\\\nonumber
  &\left.\qquad -2(p^2-m_2^2+m_1^2)B_1(p^2,m_1^2,m_2^2)+\frac{1}{6}(p^2-3m_1^2-3m_2^2)\right]\\
 B_{001}(p^2,m_1^2,m_2^2)&=\frac{1}{8}\left[2m_1^2B_1(p^2,m_1^2,m_2^2)-
 A_0(m_2^2) \frac{}{}\right.\\\nonumber
 &\left.\qquad+(p^2-m_2^2+m_1^2)B_{11}(p^2,m_1^2,m_2^2)-
 \frac{1}{6}(2m_1^2+4m_2^2-p^2)\right]\\
 B_{111}(p^2,m_1^2,m_2^2)&=-\frac{1}{4p^2}\left[A_0(m_2^2)
 +3(p^2-m_2^2+m_1^2)B_{11}(p^2,m_1^2,m_2^2) \frac{}{}\right.\\\nonumber
  &\left.\qquad +2m_1^2B_1(p^2,m_1^2,m_2^2)-
 \frac{1}{6}(2m_1^2+4m_2^2-p^2)\right]
\end{align}}

\subsection{Special cases for $B$ functions}

\begin{table}[t]
\begin{center}
\begin{tabular}{| c | c || c | c |}
\hline
PV integral & UV behaviour & PV integral & UV behaviour \\\hline\hline
$A_0$ & $m^2\Delta$ & $A_{00}$ & $\frac{1}{4} m^4\Delta$\\\hline
$B_0$ & $\Delta$ & $B_1$ & $-\frac{1}{2}\Delta$ \\\hline
$B_{00}$ & $\frac{1}{12}(3m_1^2+3m_2^2-p^2)\Delta$ & $B_{11}$ & $\frac{1}{3}\Delta$ \\\hline
$B_{001}$ & $\frac{1}{24}(-2m_1^2-4m_2^2+p^2)\Delta$ & $B_{111}$ & $-\frac{1}{4}\Delta$ \\\hline
$C_{00}$ & $\frac{1}{4}\Delta$ & $C_{001}$ & $-\frac{1}{12}\Delta$ \\\hline
$C_{002}$ & $-\frac{1}{12}\Delta$ & &\\\hline
$\dot{B}_{00}$ & $-\frac{1}{12}\Delta$ & $\dot{B}_{001}$ & $\frac{1}{24}\Delta$\\\hline
\end{tabular}
\end{center}
\caption{UV divergent parts of the Passarino-Veltman integrals.}
\label{tab:UVdivergentparts}
\end{table}

The following special cases turn out to be useful in the numerical evaluation.
Here we give only the finite parts and summarize the UV divergent parts of the 
functions appearing in the calculation in Table~\ref{tab:UVdivergentparts}.
{\allowdisplaybreaks
\begin{align}
 B_0(0,0,m^2)&=B_0(0,m^2,0)=1+\ln\left(\frac{Q^2}{m^2}\right)\\
 B_0(0,m_1^2,m_2^2)&=1+\frac{1}{m_1^2-m_2^2}\left[m_1^2\ln\left(\frac{Q^2}{m_1^2}\right)-m_2^2\ln\left(\frac{Q^2}{m_2^2}\right)\right]\\
 B_0(0,m^2,m^2)&=\ln\left(\frac{Q^2}{m^2}\right)\\
 B_0(p^2,0,0)&=2+\ln\left(\frac{Q^2}{p^2}\right)+i\pi\\
 B_0(p^2,0,m^2)&=B_0(p^2,m^2,0)=2+\ln\left(\frac{Q^2}{m^2}\right)+\frac{m^2-p^2}{p^2}\ln\left(1-\frac{p^2}{m^2}\right)\\
 B_0(p^2,m^2,m^2)&=2+\ln\left(\frac{Q^2}{m^2}\right)-\frac{m^2}{p^2}\left(\frac{1}{r}-r\right)\ln r\\
 B_0(m^2,m^2,m^2)&=2+\ln\left(\frac{Q^2}{m^2}\right)-\pi
\end{align}}

\subsection{Derivatives of the $B$ functions}

First we  present the general results for the derivatives and then
the special cases. As above we give here only the finite parts as the 
UV divergent parts are given in Table~\ref{tab:UVdivergentparts}.
$\dot{B}_0(p^2,m_1^2,m_2^2)$ is given in \eqref{eq:derB0}.
{\allowdisplaybreaks
\begin{align}\nonumber
\dot{B}_{1}(p^2,m_1^2,m_2^2)=&\frac{1}{2p^4}\left[(m_1^2-m_2^2)B_0(p^2,m_1^2,m_2^2)+(m_2^2-m_1^2)B_0(0,m_1^2,m_2^2)\right.\\
&\left.-p^2(m_1^2-m_2^2+p^2)\dot{B}_0(p^2,m_1^2,m_2^2)\frac{}{}\right]\\\nonumber
\dot{B}_{00}(p^2,m_1^2,m_2^2)=&\frac{1}{36p^4}\left[-3 (m_1^2 - m_2^2)^2 B_0(0, m_1^2, m_2^2)\right.\\\nonumber
&+ 3 (m_1^4 - 2 m_1^2 m_2^2 + m_2^4 - p^4) B_0(p^2, m_1^2, m_2^2)\\
&\left.- p^2 (3 \kappa(p^2,m_1^2,m_2^2) \dot{B}_0(p^2, m_1^2, m_2^2) + 2 p^2)\frac{}{}\right]\\\nonumber
\dot{B}_{11}(p^2,m_1^2,m_2^2)=&\frac{1}{6p^6}\left[2 (m_1-m_2) (m_1+m_2) (2 m_1^2-2 m_2^2+p^2) B_0(0,m_1^2,m_2^2)\right.\\\nonumber
&-2 (p^2 (m_1^2-2 m_2^2)+2 (m_1^2-m_2^2)^2) B_0(p^2,m_1^2,m_2^2) \\\nonumber
&+2 p^2 (p^2 (m_1^2-2 m_2^2)+(m_1^2-m_2^2)^2+p^4) \dot{B}_0(p^2,m_1^2,m_2^2) \\
&\left.-2 p^2 A_0(m_2^2)+p^2 (m_1^2+m_2^2)\right]\\\nonumber
\dot{B}_{001}(p^2,m_1^2,m_2^2)=&\frac{1}{144p^6}\left[6 (m_1^2-m_2^2) (2 (m_1^2-m_2^2)^2-p^2 (m_1^2+2 m_2^2)) B_0(0,m_1^2,m_2^2)\right.\\\nonumber
&+6(p^2 (m_1^2-m_2^2) (m_1^2+3 m_2^2)-2 (m_1^2-m_2^2)^3+p^6) B_0(p^2,m_1^2,m_2^2)\\\nonumber
&+6p^2 (m_1^2-m_2^2+p^2) \\\nonumber
&\qquad(-2 p^2 (m_1^2+m_2^2)+(m_1^2-m_2^2)^2+p^4) \dot{B}_0(p^2,m_1^2,m_2^2)\\
&\left.+6p^2 (m_2^2-m_1^2) A_0(m_2^2)+p^2 (3 m_1^4-3 m_2^4+4 p^4)\right]\\\nonumber
\dot{B}_{111}(p^2,m_1^2,m_2^2)=&\frac{1}{12p^8}\left[3 (m_1^2-m_2^2) (2 p^2 (m_1^2-2 m_2^2)\right.\\\nonumber
&\qquad+3 (m_1^2-m_2^2)^2+p^4) B_0(0,m_1^2,m_2^2)\\\nonumber
&-3 (p^4 (m_1^2-3 m_2^2)+2 p^2 (m_1^2-3 m_2^2) (m_1^2-m_2^2)\\\nonumber
&\qquad+3 (m_1^2-m_2^2)^3) B_0(p^2,m_1^2,m_2^2)\\\nonumber
&+3 p^2 (m_1^2-m_2^2+p^2) ((m_1^2-m_2^2)^2-2 m_2^2 p^2+p^4) \dot{B}_0(p^2,m_1^2,m_2^2)\\\nonumber
&-6 p^2 A_0(m_2^2) (m_1^2-m_2^2+p^2)\\
&\left.+p^2 (3 m_1^4+2 m_1^2 p^2-3 m_2^4+4 m_2^2 p^2)\right]
\end{align}}
In subsequent formulas for $p^2=0$ we use the abbreviations
\begin{align}
 \LogA=\log\left(\frac{Q^2}{m_1^2}\right), \LogB=\log\left(\frac{Q^2}{m_2^2}\right),
 \LogC=\log\left(\frac{m_2^2}{m_1^2}\right)
\end{align}
and get for $m_1\ne m_2\ne 0$
{\allowdisplaybreaks
\begin{align}
\dot{B}_0(0,m_1^2,m_2^2)&=\dfrac{1}{2(m_1^2-m_2^2)^2}
\left[m_1^2+m_2^2+\dfrac{2m_1^2m_2^2\LogC}{m_1^2-m_2^2}\right]\\
\dot{B}_1(0,m_1^2,m_2^2)&=\dfrac{1}{6(m_1^2-m_2^2)^4}\left[
-3 m_1^4 m_2^2 \left(2 \LogC+1\right)-2 m_1^6 + 6 m_1^2 m_2^4 - m_2^6\right]\\\nonumber
\dot{B}_{00}(0,m_1^2,m_2^2)&=\dfrac{1}{72(m_1^2-m_2^2)^3}\left[
-\left(6 \LogA+5\right) m_1^6+9 \left(2 \LogA+3\right) m_1^4 m_2^2\right.\\
&\quad\left.-9 \left(2 \LogB+3\right) m_1^2 m_2^4+\left(6 \LogB+5\right) m_2^6\right]\\\nonumber
\dot{B}_{11}(0,m_1^2,m_2^2)&=\dfrac{1}{24(m_1^2-m_2^2)^5}\left[
4 m_1^6 m_2^2 \left(6 \LogC+5\right)\right.\\
&\quad\left.+6 m_1^8-36 m_1^4 m_2^4+12 m_1^2 m_2^6-2 m_2^8\right]\\\nonumber
\dot{B}_{001}(0,m_1^2,m_2^2)&=\dfrac{1}{288(m_1^2-m_2^2)^4}\left[
\left(12 \LogA+13\right) m_1^8-8 \left(6 \LogA+11\right) m_1^6 m_2^2\right.\\\nonumber
&\quad+36 \left(2 \LogB+3\right) m_1^4 m_2^4-8 \left(6 \LogB+5\right) m_1^2 m_2^6\\
&\quad\left.+\left(12 \LogB+7\right) m_2^8\right]\\\nonumber
\dot{B}_{111}(0,m_1^2,m_2^2)&=\dfrac{1}{60(m_1^2-m_2^2)^6}\left[
-5 m_1^8 m_2^2 \left(12 \LogC+13\right)\right.\\
&\quad\left.-12 m_1^{10}-120 m_1^6 m_2^4-60 m_1^4 m_2^6+20 m_1^2 m_2^8-3 m_2^{10}\right]
\end{align}}
and for the remaining cases:
{\allowdisplaybreaks
\begin{align}
&\dot{B}_0(0,m^2,m^2)=-\dfrac{1}{6 m^2} & &\dot{B}_1(0,m^2,m^2)=-\dfrac{1}{12 m^2}\\
&\dot{B}_0(0,0,m^2)=\dfrac{1}{2 m^2} & &\dot{B}_1(0,0,m^2)=-\dfrac{1}{6 m^2}\\
&\dot{B}_0(0,m^2,0)=\dfrac{1}{2 m^2} & &\dot{B}_1(0,m^2,0)=\dfrac{1}{3 m^2}\\
&\dot{B}_{00}(0,m^2,m^2)=-\dfrac{1}{12}\log\left(\frac{Q^2}{m^2}\right) & &\dot{B}_{11}(0,m^2,m^2)=\dfrac{1}{20 m^2}\\
&\dot{B}_{00}(0,0,m^2)=-\dfrac{1}{72}\left[6\log\left(\frac{Q^2}{m^2}\right)+5\right] & &\dot{B}_{11}(0,0,m^2)=\dfrac{1}{12 m^2}\\
&\dot{B}_{00}(0,m^2,0)=-\dfrac{1}{72}\left[6\log\left(\frac{Q^2}{m^2}\right)+5\right] & &\dot{B}_{11}(0,m^2,0)=\dfrac{1}{4 m^2}\\
&\dot{B}_{001}(0,m^2,m^2)=\dfrac{1}{24}\log\left(\frac{Q^2}{m^2}\right) & &\dot{B}_{111}(0,m^2,m^2)=-\dfrac{1}{30m^2}\\
&\dot{B}_{001}(0,0,m^2)=\dfrac{1}{288}\left[12\log\left(\frac{Q^2}{m^2}\right)+7\right] & &\dot{B}_{111}(0,0,m^2)=-\dfrac{1}{20m^2}\\
&\dot{B}_{001}(0,m^2,0)=\dfrac{1}{288}\left[12\log\left(\frac{Q^2}{m^2}\right)+13\right] & &\dot{B}_{111}(0,m^2,0)=-\dfrac{1}{5m^2}
\end{align}}

\section{The program \prog}
\label{sec:genericfortran}

We provide for the numerical evaluation of the NLO corrections to the
decays $\tilde \chi^\pm_j \to  \tilde \chi^0_l W^\pm$ and 
$\tilde \chi^0_i \to  \tilde \chi^\mp_k W^\pm$ the program
\prog which can be obtained from \cite{Fortranonline}.
It is written in {\tt Fortran 95} and  based on {\tt SPheno}
\cite{Porod:2003um}. The program folder contains the following sub-folders:
\begin{itemize}
\item {\tt callcorrections}: routines to combine the generic routines 
contained in  {\tt corrections} with the model dependent information concerning
masses and couplings.
\item {\tt corrections}: generic NLO routines, which are 
provided in $R_\xi$-gauge and 't~Hooft-Feynman gauge, as well as the loop functions
which are not contained in the {\tt SPheno} package.
\item {\tt couplings}: NMSSM couplings which have been generated by a Mathematica code
and then were 
cross-checked with the program {\tt Sarah} \cite{Staub:2009bi}.
\item {\tt oneloop}: contains the main program {\tt CNNDecays.f90} and the main module 
for the calculation {\tt  Renormbasic.f90}. The later one can also be used to 
implement the package in other programs.
The calculation of wave-function
renormalization constants and counterterms is  included in
{\tt wavemassrenorm.f90}. The module {\tt Bremsstrahlung.f90} 
contains the calculation of the hard photon emission.
\item {\tt sphenooriginal}: necessary parts of {\tt SPheno}.
\end{itemize}
Before compiling it might be necessary to adjust the f90-compiler
and the corresponding flags in the Makefile which is placed in the main folder.
Using make the  program \prog is created which is stored in the sub-folder 
{\tt bin}. 

The input and output files are based on the SUSY Les Houches Accord (SLHA)
\cite{Skands:2003cj,Allanach:2008qq}. Concerning the input, which is expected
to be given at the electroweak scale, there
are two main differences with respect to the SLHA:
\begin{enumerate}
\item The entries of the block {\tt EXTPAR} are interpreted as effective on-shell
values for the masses and mixing entries. Therefore the entry 0 setting the scale
is ignored.
\item A new block called {\tt NLOPAR} has been created containing the information
to check for the gauge and renormalization scale independence of the results.
The program allows to use only the
UV divergent parts of the Passarino-Veltman integrals. Moreover
the divergence itself can be set to an arbitrary value. By varying
the photon mass, one can simply check the IR finiteness. In addition
the gauge parameter $\xi_V$ can be set to an arbitrary value and
it can be chosen, whether $R_\xi$-gauge or 't~Hooft-Feynman gauge
should be used for the photon, the $Z$-boson and the $W$-boson
independently. Note, that the renormalization
 scale $Q$ in {\tt NLOPAR} only affects
the scale within the Passarino-Veltman integrals and does not imply any running
of the parameters of the block {\tt EXTPAR}. Last but not least, one
can choose whether LO or NLO neutralino and chargino on-shell
masses are used for the calculation
of the processes, meaning they enter as external as well as internal
masses.
\end{enumerate}
We use for both models, the MSSM and the NMSSM, the entry 23 in the block
{\tt EXTPAR} to provide $\mu$. In case of the NMSSM it is used together with
entry 61 to calculate the singlet vacuum expectation value $v_S$. 
Below we give an example input file {\tt LesHouches.in}
based on the benchmark scenario mSUGRA $1$:\\
\lstset{basicstyle=\scriptsize, frame=shadowbox}
\begin{lstlisting}
Block MODSEL                 # Select model
 3    1                      # NMSSM
Block SMINPUTS               # Standard Model inputs
     1     1.27920000E+02   # ALPHA_EM^-1(MZ)
     2     1.16639000E-05   # GF
     3     1.17200000E-01   # ALPHA_S(MZ)
     4     9.11870000E+01   # MZ
     5     4.21400000E+00   # MB(MB)
     6     1.71400000E+02   # MTOP (POLE MASS)
     7     1.77700000E+00   # MTAU
Block MINPAR                 # Input parameters
 3   10       		     # tanb
 4   1                       # sign(mu)
BLOCK EXTPAR
     1    2.11635141E+02    # M1
     2    3.91898115E+02    # M2
     3    1.11230823E+03    # M3
    11   -1.42395369E+03    # ATOP
    12   -2.61046378E+03    # ABOT
    13   -1.77741018E+03    # ATAU
    23    9.68523016E+02    # MU
    31    3.77391179E+02    # M_eL
    32    3.77391179E+02    # M_muL
    33    3.65780798E+02    # M_tauL
    34    2.57517852E+02    # M_eR
    35    2.57517852E+02    # M_muR
    36    2.21138594E+02    # M_tauR
    41    1.02413355E+03    # M_q1L
    42    1.02413355E+03    # M_q2L
    43    8.46048325E+02    # M_q3L
    44    9.86146013E+02    # M_uR
    45    9.86146013E+02    # M_cR
    46    5.65558510E+02    # M_tR
    47    9.81632542E+02    # M_dR
    48    9.81632542E+02    # M_sR
    49    9.66133394E+02    # M_bR
    61    1.00000000E-01    # LAMBDA
    62    1.08485437E-01    # KAPPA/2
    63   -9.59966990E+02    # T_LAMBDA/LAMBDA = A_LAMBDA
    64   -1.58051889E+00    # T_KAPPA/KAPPA = A_KAPPA
BLOCK NLOPAR
     1    0		    # UV divergence: 1 = only UV div. parts
     2    0.00000000E+00    # UV divergence: Delta
     3    1.00000000E-05    # IR divergence: Photon regulator mass
     4    9.11870000E+01    # Renormalization scale: Q for NLO calc.
     5    1.00000000E+05    # Gauge dependence: Xi
     6    1		    # Xi = 1 for photon, otherwise set to 0
     7    1		    # Xi = 1 for Z, otherwise set to 0
     8    1		    # Xi = 1 for W^\pm, otherwise set to 0
     9    0		    # NLO masses for process: 0 uses LO masses, 1 uses NLO masses 
\end{lstlisting} 
\vskip 2mm
A successful run creates the output file {\tt CNNDecays.dec}.
In the example we give only the crucial information. We store
in the SLHA block {\tt MASS} the NLO masses of neutralinos
and charginos, whereas the LO masses are only part of the screen output.
In the SLHA block {\tt DECAYTREE} the LO decay widths $\Gamma^0$ in GeV are shown.
The corresponding NLO decay width $\Gamma^1$ in GeV are given in the
SLHA block {\tt DECAY}:\\
\begin{lstlisting}
# SUSY Les Houches Accord 2 - Neutralino + Chargino NLO Decays into W boson
# S. Liebler, unpublished
# in case of problems send email to sliebler@physik.uni-wuerzburg.de
# Created: 02.03.2011,  19:14
Block SPINFO            # Program information
     1   CNNDecays      # spectrum calculator
Block SMINPUTS  # SM parameters
  .......
Block MODSEL  # Model selection
 3    1       # NMSSM
Block MINPAR  # Input parameters
     3     1.00000000E+01  # tanb at m_Z   
Block EXTPAR  # 
  .......
Block gauge
  .......
Block MASS  # Mass spectrum
#   PDG code      mass          particle
  .......
   1000022     2.10611052E+02  # chi01
   1000023     3.87101530E+02  # chi02
   1000025    -9.71754981E+02  # chi03
   1000035     9.75135414E+02  # chi04
   1000045     2.10157308E+03  # S
  .......
   1000024     3.87225368E+02  # chi1+
   1000037     9.76694000E+02  # chi2+
DECAYTREE   1000045     1.68667313E-01   # S
#    BR                NDA   ID1       ID2
     9.79638284E-01    2     1000037       -24   # BR(S -> chi2+ W-)
     2.03617164E-02    2     1000024       -24   # BR(S -> chi1+ W-)
DECAY   1000045     1.64835565E-01   # S on NLO
#    BR                NDA   ID1       ID2
     9.81192113E-01    2     1000037       -24   # BR(S -> chi2+ W-)
     1.88078869E-02    2     1000024       -24   # BR(S -> chi1+ W-)
DECAYTREE   1000035     2.33679246E+00   # chi04
#    BR                NDA   ID1       ID2
     1.00000000E+00    2     1000024       -24   # BR(chi04 -> chi1+ W-)
DECAY   1000035     2.40601180E+00   # chi04 on NLO
#    BR                NDA   ID1       ID2
     1.00000000E+00    2     1000024       -24   # BR(chi04 -> chi1+ W-)
DECAYTREE   1000025     2.30613500E+00   # chi03
#    BR                NDA   ID1       ID2
     1.00000000E+00    2     1000024       -24   # BR(chi03 -> chi1+ W-)
DECAY   1000025     2.38181900E+00   # chi03 on NLO
#    BR                NDA   ID1       ID2
     1.00000000E+00    2     1000024       -24   # BR(chi03 -> chi1+ W-)
DECAYTREE   1000037     2.98577444E+00   # chi2+
#    BR                NDA   ID1       ID2
     7.89761843E-01    2     1000023        24   # BR(chi2+ -> chi02 W+)
     2.10238157E-01    2     1000022        24   # BR(chi2+ -> chi01 W+)
DECAY   1000037     3.06171556E+00   # chi2+ on NLO
#    BR                NDA   ID1       ID2
     7.93795442E-01    2     1000023        24   # BR(chi2+ -> chi02 W+)
     2.06204558E-01    2     1000022        24   # BR(chi2+ -> chi01 W+)
DECAYTREE   1000024     2.75859651E-03   # chi1+
#    BR                NDA   ID1       ID2
     1.00000000E+00    2     1000022        24   # BR(chi1+ -> chi01 W+)
DECAY   1000024     2.78599108E-03   # chi1+ on NLO
#    BR                NDA   ID1       ID2
     1.00000000E+00    2     1000022        24   # BR(chi1+ -> chi01 W+)
Block NLOPAR # Renormalization parameters
         1           0  # UV divergence: 1 = only UV div. parts
         2     0.00000000E+00  # UV divergence: Delta
         3     1.00000000E-05  # IR divergence: Photon regulator mass
         4     9.11870000E+01  # Renormalization scale: Q for NLO calc.
         5     1.00000000E+05  # Gauge dependence: Xi
         6           1  # Special choice of gauge for photon
         7           1  # Special choice of gauge for Z boson
         8           1  # Special choice of gauge for W boson
         9           0  # NLO masses used for process: 0 LO, 1 NLO masses
\end{lstlisting}

\end{appendix}

\bibliographystyle{h-physrev}

\end{document}